\documentclass[11pt,a4paper]{article}
\usepackage{jheppub,xcolor}
\pdfoutput=1
\usepackage[utf8]{inputenc}
\graphicspath{{./figs/}}

\usepackage[a4paper,top=4cm,bottom=0cm,left=4.0cm,right=-0.8cm,bindingoffset=0mm]{geometry}

\usepackage{amssymb}
\usepackage{dcolumn}	
\usepackage{bm}			
\usepackage{bbm} 		
\usepackage{multirow}
\usepackage{slashed}
\usepackage{blkarray}
\usepackage{booktabs}
\usepackage{makecell}
\usepackage{longtable}
\usepackage{placeins}
\usepackage{listings}
\usepackage{subcaption}
\usepackage{fontawesome}

\usepackage{xcolor}
\usepackage{tikz}
\usetikzlibrary{positioning}
\usetikzlibrary{shapes.geometric, arrows, fit}
\usetikzlibrary{arrows.meta}

\definecolor{codegreen}{rgb}{0,0.6,0}
\definecolor{codegray}{rgb}{0.5,0.5,0.5}
\definecolor{codepurple}{rgb}{0.58,0,0.82}
\definecolor{backcolour}{rgb}{0.95,0.95,0.92}

\lstdefinestyle{mystyle}{
    backgroundcolor=\color{backcolour},   
    commentstyle=\color{codegreen},
    keywordstyle=\color{magenta},
    numberstyle=\tiny\color{codegray},
    stringstyle=\color{codepurple},
    basicstyle=\ttfamily\footnotesize,
    breakatwhitespace=false,         
    breaklines=true,                 
    captionpos=b,
    mathescape=false,
    keepspaces=true,                 
    numbers=left,                    
    numbersep=5pt,                  
    showspaces=false,                
    showstringspaces=false,
    showtabs=false,                  
    tabsize=2
}
\definecolor{Red}{rgb}{1.,0.,0.}

\newcommand{\lag}{\mathcal{L}}

\newcommand{\mcO}{\mathcal{O}}

\newcommand{\lp}{\left(}
\newcommand{\rp}{\right)}

\newcommand{\bc}{\begin{center}}
\newcommand{\ec}{\end{center}}
\newcommand{\ba}{\begin{array}}
\newcommand{\ea}{\end{array}}

\newcommand{\be}{\begin{equation}}
\newcommand{\ee}{\end{equation}}
\renewcommand{\(}{\ensuremath{\left(}}
\renewcommand{\)}{\ensuremath{\right)}}	
\newcommand{\cQteight}{\ensuremath{c_{Qt}^{(8)}}}

\newcommand{\bea}{\begin{eqnarray}}
\newcommand{\eea}{\end{eqnarray}}
\newcommand{\bi}{\begin{itemize}}
\newcommand{\ei}{\end{itemize}}
\newcommand{\ben}{\begin{enumerate}}
\newcommand{\een}{\end{enumerate}}

\newcommand{\lc}{\left[}
\newcommand{\rc}{\right]}

\def\frac#1#2{{{#1}\over {#2}}}
\def\gsim{\mathrel{\rlap{\lower4pt\hbox{\hskip1pt$\sim$}}
    \raise1pt\hbox{$>$}}}       
\def\lsim{\mathrel{\rlap{\lower4pt\hbox{\hskip1pt$\sim$}}
    \raise1pt\hbox{$<$}}}

\newcommand{\smefit}{{\sc\small SMEFiT}}
\newcommand{\matchTOfit}{{\sc \small match2fit}}
\newcommand{\draft}[1]{}

\newcommand{\chm}{\checkmark}
\newcommand{\rchm}{{\color{red}{\chm}}}

\def\beq{\begin{equation}}
\def\eeq{\end{equation}}

\numberwithin{equation}{section}
\numberwithin{figure}{section}
\numberwithin{table}{section}

\usepackage{tabularx}
\newcolumntype{C}[1]{>{\centering\arraybackslash}p{#1}}

\definecolor{darkblue}{rgb}{0.0,0,0.5}
\definecolor{darkgreen}{rgb}{0.0,0.3,0.0}
\definecolor{redish}{rgb}{0.675,0,0.2}
\definecolor{red}{rgb}{0.8,0,0}
\definecolor{green}{rgb}{0,0.6,0}
\definecolor{bluish}{rgb}{0.2,0.2,0.675}
\definecolor{mygrey}{rgb}{0.6,0.6,0.6}

\usepackage[normalem]{ulem}
\definecolor{ao}{rgb}{0.0, 0.5, 0.0}

\numberwithin{equation}{section}
\numberwithin{figure}{section}
\numberwithin{table}{section}

\title{The automation of SMEFT-Assisted Constraints on UV-Complete Models}
\author[a,b]{Jaco~ter~Hoeve,}
\author[a,b]{Giacomo~Magni,}
\author[a,b]{Juan~Rojo,}
\author[c]{Alejo~N.~Rossia,}
\author[c]{and Eleni Vryonidou}
\affiliation[a]{Nikhef Theory Group, Science Park 105, 1098 XG Amsterdam, The Netherlands}
\affiliation[b]{Physics and Astronomy, Vrije Universiteit Amsterdam, NL-1081 HV Amsterdam, The Netherlands}
\affiliation[c]{Department of Physics and Astronomy, University of Manchester, Oxford Road, Manchester M13 9PL, United Kingdom}

\emailAdd{j.j.ter.hoeve@vu.nl}
\emailAdd{gmagni@nikhef.nl}
\emailAdd{j.rojo@vu.nl}
\emailAdd{alejo.rossia@manchester.ac.uk}
\emailAdd{eleni.vryonidou@manchester.ac.uk}

\date{\today}
\abstract{
  The ongoing Effective Field Theory (EFT) program at the LHC
  and elsewhere is motivated by streamlining the connection between experimental data and UV-complete scenarios
  of heavy new physics beyond the Standard Model (BSM).
  This connection is provided by matching relations mapping the Wilson coefficients of the EFT
  to the couplings and masses of UV-complete models.
  Building upon recent work on the automation of tree-level and one-loop  matching
  in the SMEFT, we present a novel
  strategy automating the 
  constraint-setting procedure on the parameter space
  of general heavy UV-models matched to dimension-six SMEFT operators.
  A new Mathematica package, \matchTOfit, interfaces 
  {\sc\small MatchMakerEFT}, which derives
  the matching relations for a given UV model,
  and 
  {\sc\small SMEFiT}, which provides bounds on the
  Wilson coefficients by comparing with data.
  By means of this pipeline and using both tree-level and one-loop
matching, we derive
  bounds on a wide range of single- and multi-particle extensions
of the SM from a global dataset composed
  by LHC and LEP measurements.
  Whenever possible, we benchmark our results with existing studies.
  Our framework realises one of the main objectives of the EFT program in particle physics: deploying
  the SMEFT to bypass
  the need of directly comparing the predictions of heavy UV models with experimental data.
}


\preprint{Nikhef 2023-011}

\begin{document}
\maketitle
\flushbottom

\clearpage
\section{Introduction}
\label{sec:intro}
One of the main motivations for the ongoing Standard Model Effective Field Theory (SMEFT) program in particle
physics, see~\cite{Isidori:2023pyp} for a recent review, is to streamline the connection between  experimental data and UV-complete scenarios
of new physics beyond the Standard Model (BSM) that contain new particles which are too heavy to be directly produced at available facilities.
In this paradigm, rather than comparing the predictions of specific UV-complete models
directly with data to derive information on its parameters (masses and couplings),
UV-models are first matched to the SMEFT and subsequently the resulting Wilson coefficients are constrained by means
of a global EFT analysis including a broad range of observables.

The main advantage of this approach is to bypass the need to recompute predictions for physical observables with different UV-complete models. 
The global SMEFT analysis essentially encapsulates, for a well-defined
set of assumptions, the information provided by available experimental observables, while
 the matching relations determine how this information relates
to the masses and couplings of the UV-complete model.
This feature becomes specially relevant whenever new BSM models are introduced:
one can then quantify to which extent their parameter space is constrained
by current data from a pre-existing global SMEFT analysis,
rather than having first to provide predictions for a large number of observables and then compare
those with data.

In recent years, several groups~\cite{Fuentes-Martin:2016uol,deBlas:2017xtg,DasBakshi:2018vni,Kramer:2019fwz,Gherardi:2020det,Angelescu:2020yzf,Ellis:2020ivx,Cohen:2020qvb,Fuentes-Martin:2020udw,Summ:2020dda,Carmona:2021xtq,Fuentes-Martin:2022vvu,Fuentes-Martin:2022jrf,DeAngelis:2023bmd,Banerjee:2023iiv,Chakrabortty:2023yke} have systematically studied the matching
between UV-complete models and the Wilson coefficients of the SMEFT,
with various degrees of automation and in many cases accompanied by the release of the corresponding open-source codes.
In order to realise the full potential of such EFT matching studies, it is however necessary to interface these results with global SMEFT analyses parameterizing the constraints provided by the experimental data.
Such an interface must be constructed in a manner that benefits from the automation of EFT matching tools
and that does not impose restrictions in the type of UV-models to be matched.
In particular, it must admit non-linear matching relations with 
additional constraints such as parameter positivity.
Several groups have reinterpreted global SMEFT fits in terms of matched UV models~\cite{Ellis:2020unq,Brivio:2021alv,Anisha:2021hgc,Greljo:2023adz}, but their focus is limited to pre-determined,
relatively simple models with few parameters.
No framework has been released to date that enables performing such fits with generic, user-specified, multi-particle UV-complete models.

Here we bridge this gap in the SMEFT literature by developing a framework automating the limit-setting procedure  on the parameter space
of generic UV-models which can be matched to dimension-six SMEFT operators.
This is achieved by extending {\sc\small SMEFiT}~\cite{Hartland:2019bjb,vanBeek:2019evb,Ethier:2021ydt,Ethier:2021bye,Giani:2023gfq} with the capabilities of working directly on
 the parameter space of UV-models, given arbitrary matching relations between UV couplings and EFT coefficients as an input.
To this end, we have designed an interface to the {\sc\small MatchMakerEFT} code~\cite{Carmona:2021xtq} such that for any
of the available UV-models it outputs a run card with the relevant Wilson coefficients entering
the {\sc\small SMEFiT} analysis.
This  interface, consisting of
the Mathematica package \matchTOfit~(available on \href{https://github.com/arossia94/match2fit}{ Github~\faicon{github}}),
also provides a list of the UV variables (denoted
as UV-invariants) 
that can be inferred from the data
and corresponding to specific combinations
of UV couplings and masses. 
The adopted procedure removes any limitations on the type of matching relations involved.

We exploit this new pipeline to derive bounds on a broad range
of UV-complete scenarios both at linear and quadratic order in the SMEFT expansion
from a global dataset composed by LHC and LEP measurements, using either tree-level or
one-loop matching relations.
We consider both relatively simple single-particle
extensions of the SM as well as more
complex multi-particle extensions,
in particular with a benchmark model
composed by two heavy vector-like fermions
and one heavy vector boson.
We study the stability of the fits results with respect to the order in the EFT expansion
and  the perturbative QCD accuracy for the EFT cross-sections.
  Whenever possible, we compare our results with existing SMEFT matching studies
  in the literature. 
  
  The framework presented
  in this work is made publicly available
  both 
in terms of the latest {\sc\small SMEFiT} release and
with the independent \matchTOfit~interface,
providing  a valuable
  resource for the EFT community streamlining the connection between UV-models and EFT studies.
Our work brings one step closer one of the primary goals of the SMEFT program: constraining the
parameters of general BSM Lagrangians using EFTs as a bridge between UV models and experimental data.

The structure of this paper is as follows.
  First, Sect.~\ref{sec:general_strategy}
  discusses the general strategy adopted to automate the matching between
  UV-models and the corresponding SMEFT fits.
  Sect.~\ref{sec:smefit} describes how the {\sc\small SMEFiT} framework is extended
  to enable constraint-setting
 directly in the parameter space of UV-complete models.
  The main results of this work are presented in Sect.~\ref{sec:results}, where we derive
  bounds on single- and multi-particle BSM models from a global dataset and  compare our findings with existing results.
  We conclude and summarise possible future developments in Sect.~\ref{sec:summary}.
  
  Technical details of our work are provided in the appendices.
  App.~\ref{app:ewpos} summarises the baseline
  SMEFT analysis used, in particular
  reporting on a recent update concerning
  the description of electroweak
precision observables (EWPOs) from electron-positron
colliders.
App.~\ref{app:mmeft2smefit} presents a concise description of the {\sc\small Match2Fit} package.
App.~\ref{app:logarithms} discusses the origin of the logarithms arising in the one-loop matching formulae.
Additional details on the single-particle model fits from Sect.~\ref{sec:results} are provided in App.~\ref{app:UV_details}.


\section{Tree-level and one-loop matching in  the SMEFT}
\label{sec:general_strategy}
The low-energy phenomenology of a quantum field theory can in many cases be described by an EFT with a
reduced number of dynamical degrees of freedom. 
This feature presents clear advantages in terms of BSM searches, since a general-enough EFT could describe a plethora of different UV completions. 
The success of the Standard Model in describing the physics explored at the energy scales accessible
by current experiments justifies the use of an EFT based upon it, which is known as the SMEFT~\cite{Brivio:2017vri,Isidori:2023pyp}, to search for BSM physics in a model-independent manner.
To ensure that the UV theory and the low-energy EFT predict the same observables in the energy range where the EFT is valid, the Wilson coefficients of the latter must be computed in terms of the parameters of the UV theory. 
This procedure is called matching.

Matching an EFT, such as 
the SMEFT, with the associated UV-complete model can be performed by means of two well-established techniques, as well as with another recently developed method based on on-shell amplitudes~\cite{DeAngelis:2023bmd}.
The first of these matching techniques is known as the functional method and is based on the manipulation of the path integral, the action, and the Lagrangian~\cite{Fuentes-Martin:2016uol,Ellis:2017jns,Kramer:2019fwz,Angelescu:2020yzf,Ellis:2020ivx,Summ:2020dda,Banerjee:2023iiv,Chakrabortty:2023yke}.
It requires to specify the UV-complete Lagrangian, the heavy fields, and the matching scale, 
while the EFT Lagrangian is part of the result although not necessarily in the desired basis.
The second technique is the diagrammatic method, based on equating off-shell Green's functions computed in both the EFT and the UV model, and therefore it requires the explicit form of both Lagrangians from the onset~\cite{deBlas:2017xtg,Carmona:2021xtq}.
Both methods provide the same final results and allow for both tree-level and one-loop matching computations. 
The automation of this matching procedure up to 
the one-loop level is mostly solved in the case of the diagrammatic technique~\cite{Carmona:2021xtq}, and is well advanced in the functional method case~\cite{DasBakshi:2018vni,Fuentes-Martin:2022jrf}.

Let us illustrate the core ideas underlying
this procedure by reviewing the matching to 
the SMEFT at tree and one-loop level of a 
specific benchmark UV-complete model.
This is taken to be the single-particle
extension of the SM resulting from adding a new heavy scalar boson, $\phi$. This scalar transforms under the SM gauge group in the same manner as the Higgs boson, i.e. $\phi\sim\left(1,2\right)_{1/2}$, where we denote the irreducible representations under the SM gauge group as $(\text{SU}(3)_{\text{c}},\text{SU}(2)_{\text{L}})_{\text{U}(1)_{\text{Y}}}$. 
Following the notation of~\cite{deBlas:2017xtg}, the Lagrangian of this model reads
\begin{equation}
\begin{split}
 \lag_{\rm UV}= &\lag_{\rm SM} + |D_{\mu}\phi|^2 - m_{\phi}^2 \phi^{\dagger}\phi - \left((y_{\phi}^{e})_{ij}\, \phi^{\dagger} \bar{e}_{R}^{i} \ell_{L}^{j} + (y_{\phi}^{d})_{ij}\, \phi^{\dagger} \bar{d}_{R}^{i} q_{L}^{j} \right.\\
	&\left. + (y_{\phi}^{u})_{ij} \, \phi^{\dagger} i\sigma_2  \bar q_{L}^{T,i} u_{R}^{j} + \lambda_{\phi}\, \phi^{\dagger}\varphi |\varphi|^2+\text{h.c.}\right) - {\rm scalar~potential}\, ,
\end{split}	
\label{eq:Lag_UV_full_Varphi_model}
\end{equation}
with $\lag_{\rm SM}$ being the SM
Lagrangian and  $\varphi$ the SM Higgs
doublet.
We do not write down explicitly the complete form of the scalar potential in Eq.~(\ref{eq:Lag_UV_full_Varphi_model}), of which $\lambda_{\phi}\, \phi^{\dagger}\varphi |\varphi|^2$ is one of the components, since it has no further effect on the matching outcome as long as it leads to an expectation value satisfying  $\langle\phi\rangle=0$,
such that $m_\phi^2>0$ corresponds to the pole mass. 

This heavy doublet $\phi$ interacts with the SM fields via the Yukawa couplings $(y_{\phi}^{u,d,e})_{ij}$, the scalar coupling $\lambda_\phi$, and the electroweak gauge couplings.
In the following, we consider as  ``UV couplings''
exclusively those couplings between UV and SM particles that are not gauge couplings.
The model described by Eq.~(\ref{eq:Lag_UV_full_Varphi_model}) corresponds to the two-Higgs doublet model (2HDM) in the decoupling limit~\cite{Gunion:2002zf,Bernon:2015qea}.
For simplicity, we assume that all the couplings between the SM and the heavy particle are real
and satisfy $(y_{\phi}^{\psi})_{ij} = \delta_{i,3}\delta_{j,3}\, (y_{\phi}^{\psi})_{33}$ for $\psi=u,\,d,\,e$, and the only SM Yukawa couplings that we consider as non-vanishing are the ones of the third-generation fermions. 
 
\subsection{Tree-level matching}
\label{subsec:tree_level_matching}

The matching of UV-complete models to dimension-6 SMEFT operators at tree level has been fully tackled in~\cite{deBlas:2017xtg}, which considers all possible UV-completions  with particles of spin up to $s=1$  generating non-trivial Wilson coefficients. 
These results can be reproduced with the automated codes {\sc\small MatchingTools}~\cite{Criado:2017khh} and {\sc\small MatchMakerEFT}~\cite{Carmona:2021xtq}
based on the diagramatic approach.
At tree level, the diagrammatic method requires computing the tree-level Feynman diagrams contributing to multi-point Green's functions with only light external particles. 
Then, the covariant propagators $\Delta_i$ must be expanded to a given order in 
inverse powers of the heavy masses. 
The computation of the Feynman diagrams in the EFT can be performed in a user-defined operator basis. 

\begin{table}[t]
    \centering
      \renewcommand{\arraystretch}{1.60}
    \begin{tabular}{c|c}
\toprule
        $\qquad$ Operator type $\qquad$ &  Generated operators\\
         \midrule
      $X^3$                & $\overline\mcO_{W}$ \\
      $\varphi^6$          & {\color{blue}$\overline\mcO_{\varphi}$} \\
      $\varphi^4 D^2$      & $\overline\mcO_{\varphi \square}$,$\overline\mcO_{\varphi D}$\\
      $\psi^2\varphi^3$    & {\color{blue}$\overline{\mcO}_{u\varphi}$}, {$\overline{\textcolor{blue}{\mcO}}_{\textcolor{blue}{d\varphi}}$}, {$\overline{\textcolor{blue}{\mcO}}_{\textcolor{blue}{e\varphi}}$} \\
      $X^2\varphi^2$       & -\\
      $\psi^2 X \varphi$   & $\overline\mcO_{dB}$, $\overline\mcO_{dG}$, $\overline\mcO_{dW}$, $\overline\mcO_{uB}$, $\overline\mcO_{uG}$, $\overline\mcO_{uW}$, $\mcO_{eB}$, $\mcO_{eW}$ \\
      $\psi^2 \varphi^2 D$ & $\overline\mcO_{\varphi u}$, $\overline\mcO_{\varphi d}$, $\overline\mcO_{\varphi e}$, $\overline\mcO_{\varphi \ell}^{(1)}$, $\overline\mcO_{\varphi \ell}^{(3)}$,  $\overline\mcO_{\varphi q}^{(1)}$, $\overline\mcO_{\varphi q}^{(3)}$, $\overline\mcO_{\varphi ud}$\\
      \midrule
      \multirow{3}{*}{$\psi^4$}             & {\color{blue}$\overline\mcO_{qu}^{(1)}$}, {\color{blue}$\overline\mcO_{qu}^{(8)}$}, {$\overline{\textcolor{blue}{\mcO}}_{\textcolor{blue}{\ell e}}$}, {$\overline{\textcolor{blue}{\mcO}}\textcolor{blue}{_{\ell equ}^{(1)}}$}, {$\overline{\textcolor{blue}{\mcO}}\textcolor{blue}{_{qd}^{(1)}}$}, {$\overline{\textcolor{blue}{\mcO}}\textcolor{blue}{_{qd}^{(8)}}$}, {$\overline{\textcolor{blue}{\mcO}}\textcolor{blue}{_{quqd}^{(1)}}$}, {\color{blue}$\mcO_{\ell edq}$}, \\
                           & $\overline\mcO_{uu}$, $\overline\mcO_{dd}$, $\overline\mcO_{\ell\ell}$, $\overline\mcO_{\ell d}$ $\overline\mcO_{ee}$, $\overline\mcO_{ed}$, $\overline\mcO_{eu}$, $\overline\mcO_{\ell q}^{(1)}$, $\overline\mcO_{\ell q}^{(3)}$,\\
                           & $\overline\mcO_{\ell u}$, $\overline\mcO_{qe}$, $\overline\mcO_{qq}^{(1)}$, $\overline\mcO_{qq}^{(3)}$, $\overline\mcO_{quqd}^{(8)}$, $\overline\mcO_{ud}^{(1)}$, $\overline\mcO_{ud}^{(8)}$, $\mcO_{\ell equ}^{(3)}$.   \\
      \bottomrule
    \end{tabular}
    \caption{SMEFT dimension-6 operators in the Warsaw basis generated by the integration of the heavy doublet scalar $\phi$
    from the Lagrangian defined in
    Eq.~(\ref{eq:Lag_UV_full_Varphi_model}).
    The operators generated at tree level are highlighted in blue, while the operators that appear at one-loop level are in black. 
    We assume that the non-vanishing couplings of this heavy particle with the SM are $\lambda_\phi$ and $(y_{\phi}^{\psi})_{ij} = \delta_{i,3}\delta_{j,3}\, (y_{\phi}^{\psi})_{33}$ for $\psi=u,\,d,\,e$.
    The operators with a bar over their name survive the further constraint of $(y_{\phi}^{e})_{33}=(y_{\phi}^{d})_{33}=0$, required to fulfil the {\sc SMEFiT} flavour assumption at tree level.
    We take all the SM Yukawa couplings and masses to be zero, except those for the third-generation fermions.}
    \label{tab:one_loop_matching_phi}
\end{table}

The  outcome of matching the model defined by the Lagrangian in Eq.~\eqref{eq:Lag_UV_full_Varphi_model} to 
the SMEFT at tree level is provided in~\cite{deBlas:2017xtg}.
Table~\ref{tab:one_loop_matching_phi} summarizes the  dimension-6 operators in the Warsaw basis~\cite{Grzadkowski:2010es}
generated by both tree-level (in blue)
and one-loop (in black) matching.
A representative subset of the resulting tree-level matching expressions is given by
\begin{equation}
	\begin{alignedat}{2}
	    \frac{\lp c_{qd}^{(1)}\rp_{3333}}{\Lambda^2}=-\frac{\(y_{\phi}^{d}\)_{33}^{2}}{6\,m_\phi^2}, \quad & \frac{\(c_{qd}^{(8)}\)_{3333}}{\Lambda^2} = -\frac{\(y_{\phi}^{d}\)_{33}^{2}}{ m_{\phi}^2}, \quad & \frac{\(c_{d\varphi}\)_{33}}{\Lambda^2}= \frac{\lambda_{\phi}\,\left(y_{\phi}^{d}\right)_{33}}{m_\phi^2}, \quad & \frac{c_{\varphi}}{\Lambda^2} =  \frac{\lambda_{\phi}^2}{ m_{\phi}^2}, \\
          \frac{\(c_{qu}^{(1)}\)_{3333}}{\Lambda^2} =-\frac{\(y_{\phi}^{u}\)_{33}^2}{6\, m_{\phi}^2}, \quad & \frac{\(c_{qu}^{(8)}\)_{3333}}{\Lambda^2} = -\frac{\(y_{\phi}^{u}\)_{33}^2}{ m_{\phi}^2}, \quad & \frac{\(c_{u \varphi}\)_{33}}{\Lambda^2}=-\frac{\lambda_{\phi}\,\left(y_{\phi}^{u}\right)_{33}}{m_\phi^2}, \quad & \frac{\(c_{\varphi q}^{(3)}\)_{33}}{\Lambda^2} = 0.
     \end{alignedat}
     \label{eq:matching_result_example_tree}
\end{equation}
Eq.~(\ref{eq:matching_result_example_tree}) showcases the type of constraints on the EFT coefficients that tree-level matching can generate.
First of all, the relations between UV couplings
and Wilson coefficients will be in general non-linear.
Second, some coefficients such as $\left(c_{\varphi q}^{(3)}\right)_{33}$ 
are set to zero by the matching relations.
Third, other coefficients acquire a well-defined
sign, such as $c_\varphi$ ($\left(c_{qd}^{(1)}\right)_{3333}$) which becomes positive-definite (negative-definite) after matching. 
Fourth, several EFT coefficients become
related among them by means of both linear and non-linear relations such as
\begin{align}
    \left(c_{qu}^{(8)}\right)_{3333} = & 6\,\left(c_{qu}^{(1)}\right)_{3333}, \label{eq:rel_UV_tree_1}\\
    \frac{\lp c_{qd}^{(1)}\rp_{3333}}{\lp c_{qu}^{(1)} \rp _{3333}} = & \lp \frac{\lp c_{d\varphi}\rp_{33}}{\lp c_{u\varphi}\rp_{33}}\rp^2.\label{eq:rel_UV_tree_2}
\end{align}
These relations must be taken
into account when performing the EFT fit
to the experimental data, using the dedicated techniques
discussed in Sect.~\ref{sec:smefit}
and App.~\ref{app:ewpos}.

When considering multi-particle UV scenarios, rather
than single-particle extensions such 
as the model defined by Eq.~\eqref{eq:Lag_UV_full_Varphi_model},
non-vanishing EFT coefficients generally consist of the sum of several rational terms. 
For example, assume that one adds
to the model of Eq.~\eqref{eq:Lag_UV_full_Varphi_model}
a second heavy scalar with gauge charges $\Phi\sim\left(8,2\right)_{1/2}$ and with mass $m_{\Phi}$ which couples to the SM  fields by means of
\begin{equation}
    \lag_{\rm UV}\supset - \left(y_{\Phi}^{qu}\right)_{ij}\,\Phi^{A\dagger}\,i\sigma_2\,\bar{q}_{L,i}^{T} T^{A} u_{R,j}+\text{h.c.} \, .
\label{eq:Lag_UV_add_1part}
\end{equation}
Integrating out this additional heavy
scalar field modifies
two of the tree-level matching relations listed in Eq.~\eqref{eq:matching_result_example_tree}
as follows
\begin{equation}
    \frac{\lp c_{qu}^{(1)} \rp_{3333}}{\Lambda^2} =-\frac{\(y_{\phi}^{u}\)_{33}^2}{6\, m_{\phi}^2}-\frac{2\left(y_{\Phi}^{qu}\right)_{33}^2}{9\,m_{\Phi}^2}\,, \quad \frac{\lp c_{qu}^{(8)} \rp_{3333}}{\Lambda^2} = -\frac{\lp y_{\phi}^{u}\rp_{33}^2}{ m_{\phi}^2}+\frac{\left(y_{\Phi}^{qu}\right)_{33}^2}{6\,m_{\Phi}^2}\,.
\end{equation}
Hence, the simple linear
relation Eq.~\eqref{eq:rel_UV_tree_1} is not valid anymore, while Eq.~\eqref{eq:rel_UV_tree_2} now becomes
\begin{equation}
    \frac{\lp c_{qd}^{(1)}\rp_{3333}}{\frac{1}{9} \lp c_{qu}^{(1)} \rp_{3333} +\frac{4}{27} \lp c_{qu}^{(8)} \rp_{3333} } = \lp \frac{\lp c_{d\varphi}\rp_{33}}{\lp c_{u\varphi}\rp_{33}}\rp^2,
\end{equation}
which shows that multiplicative relations might involve non-trivial linear combinations of the
Wilson coefficients.
In this context, we observe that many of the conditions on the EFT coefficients imposed by assuming a certain UV completion are non-linear and hence the resulting posterior distributions inferred from the data will in general be non-Gaussian. 

The tree-level matching results discussed up to now do not comply with the flavour symmetry adopted by the 
current {\sc SMEFiT} analysis, namely U$(2)_{q}\times$U$(2)_{u}\times$U$(3)_{d}\times(\text{U}(1)_{\ell}\times\text{U}(1)_{e})^3$. 
This would cause ambiguities at the moment of performing the fit, since for example {\sc\small SMEFiT} assumes that the coefficient $\lp c_{qd}^{(1)}\rp_{33ii}$ has the same value for $i=1,\,2,\,3$, while the matching result instead gives a non-vanishing coefficient only for $i=3$. 
In this specific case, the appropriate flavour symmetry 
used at the EFT fit level can be respected after tree-level matching by further imposing
\be
\left( y_{\phi}^{e} \right)_{33} = \left(y_{\phi}^{d}\right)_{33}=0 \, ,
\ee
and leaving $\lambda_\phi$ and $\left(y_{\phi}^{u}\right)_{33}$ as the only non-vanishing UV couplings, as we will assume in the rest of this work unless otherwise specified. Notice that this implies that the heavy new particle interacts only with the Higgs boson and the top quark, a common situation in well-motivated UV models. 
The operators that remain after this additional restriction
is imposed are indicated with a bar in Table~\ref{tab:one_loop_matching_phi}. 

Once this flavour symmetry is imposed, one can map unambiguously the naming of the operators and EFT coefficients 
provided by the {\sc\small MatchMakerEFT} output
to the ones defined in {\sc\small SMEFiT}.
The non-vanishing coefficients after integrating out the heavy scalar $\phi$ at tree level are then
\begin{equation}
    c_{\varphi}=c_{\varphi},\quad c_{t\varphi}=\lp c_{u\varphi}\rp_{33}, \quad c_{Qt}^{(1)} = \lp c_{qu}^{(1)}\rp_{3333}, \quad c_{Qt}^{(8)} = \lp c_{qu}^{(8)}\rp_{3333},
\end{equation}
where in the l.h.s. of the equalities we use the {\sc SMEFiT} convention and on the r.h.s the Warsaw convention adopted in the matching output~\cite{deBlas:2017xtg,Carmona:2021xtq}.

\subsection{One-loop matching}
\label{subsec:oneloop}

Extending the diagrammatic matching technique to the one-loop case is conceptually straightforward, and requires the computation of one-loop diagrams in the UV model with off-shell external light particles and at least one heavy-particle propagator inside the loop.
From the EFT side, diagrammatic matching
at one-loop involves the calculation of the diagrams with the so-called Green's basis, which includes also those operators that are redundant by equatoions of motion (EoMs). 
The dimension-6  and dimension-8 Green's bases in the
SMEFT have been computed in ~\cite{Gherardi:2020det,Chala:2021cgt,Ren:2022tvi}. 
Further technicalities such as evanescent operators acquire special relevance at this order~\cite{Fuentes-Martin:2022vvu}.
The automation of one-loop matching with the diagrammatic technique is provided by  {\sc\small MatchMakerEFT}~\cite{Carmona:2021xtq} for any renormalisable UV model with heavy scalar bosons and spin-$1/2$ fermions.
The equivalent automation
applicable to models containing heavy spin-1 bosons is work in progress.

In this work we use {\sc\small MatchMakerEFT} (v1.1.3) to evaluate the one-loop matching of 
a selected UV model. 
When applied to the Lagrangian of Eq.~\eqref{eq:Lag_UV_full_Varphi_model} with the {\sc\small SMEFiT} flavour assumptions, this procedure generates  
additional non-vanishing operators in comparison
to those arising at tree level, as indicated in Table~\ref{tab:one_loop_matching_phi}, where we also show operators generated in the more general case of 
\be
\left(y_{\phi}^{\psi}\right)_{ij} = \delta_{i,3}\delta_{j,3}\, \left(y_{\phi}^{\psi}\right)_{33} \, , \qquad \psi =e,u,d \, .
\ee
Notice that setting $\left(y_{\phi}^{e}\right)_{33}=\left(y_{\phi}^{d}\right)_{33}=0$ has a smaller impact on the number of loop-generated operators than for the tree-level case. 
The reason is that, in many cases, the operators are still generated via SM-gauge couplings so the EFT coefficients take the generic form $c\sim g_{\text{SM}}^4/16\pi^2$.
These contributions are typically flavour-universal and if they break the desired flavour symmetry, the breaking is suppressed by the SM gauge couplings and/or the loop factor.
For this reason, in this work we only enforce the {\sc SMEFiT} flavour symmetry for tree-level matching.

An example of the results of one-loop matching corrections to
the EFT coefficients is provided for 
$c_{Qt}^{(8)}$, for which the tree-level
matching relation in Eq.~\eqref{eq:matching_result_example_tree}
is now extended as follows
\begin{eqnarray} 
\nonumber
\label{eq:oneloop_logs_example}
    \frac{c_{Qt}^{(8)}}{\Lambda^2} &=& -\frac{\(y_{\phi}^{u}\)_{33}^{2}}{m_\phi^2 } - \lc \frac{25 g_1^2}{1152\pi^2}+\frac{3g_2^2}{128\pi^2} - \frac{3 \(y_{t}^{\rm SM}\)^2}{16 \pi^2} + \frac{g_3^2}{16\pi^2}\left(1-\log \lp \frac{m_\phi^2}{\mu^2}\rp\right) \rc \frac{\(y_{\phi}^{u}\)_{33}^{2}}{m_\phi^2} \\&&+ \frac{3}{64\pi^2} \lc 1-2\log \lp \frac{m_\phi^2}{\mu^2}\rp \rc \frac{\(y_{\phi}^{u}\)_{33}^4}{m_{\phi}^2} \, ,
\end{eqnarray}
with $\mu$ being the matching scale,
$g_i$ the SM gauge coupling constants,
and $y_{t}^{\rm SM}$ the top Yukawa coupling
in the SM. 
To estimate of the numerical impact of
loop corrections to matching 
in the specific case of Eq.~(\ref{eq:oneloop_logs_example}), one can substitute the corresponding values of the SM couplings.
One finds that the term proportional to $\left(y_{\phi}^{u}\right)_{33}^2$
receives a correction at the few-percent
level from one-loop matching effects. 

The logarithms in the matching scale $\mu$
appearing in Eq.~(\ref{eq:oneloop_logs_example}) are  generated by the running of the couplings and Wilson
coefficients between the heavy particle mass $m_\phi$ and $\mu$, as further discussed in App.~\ref{app:logarithms}. 
Since here we neglect RGE effects~\cite{Aoude:2022aro}, we simplify the matching expressions by choosing the scale $\mu$ to be equal to the mass of the integrated-out UV heavy field such that the logarithms vanish.

As compared to the tree-level matching relations, common features of the one-loop contributions are the appearance of terms proportional to the UV couplings at the fourth order and the presence
of the SM gauge couplings. 
Here we assume that the latter have fixed numerical
values determined by other measurements i.e. 
the PDG averages~\cite{ParticleDataGroup:2022pth}.
This implies that at the fit level some 
of the EFT coefficients 
are specified entirely by the mass of
the UV heavy particle, such as for the 
dimension-six 
triple SU$(2)$ field strength tensor operator
for the considered model
\begin{equation}
\label{eq:cwww_matching}
    \frac{c_{WWW}}{\Lambda^2}=\frac{g_2^3}{5760\pi^2\, m_{\phi}^2} \, .
\end{equation} 
While one may expect that one-loop matching
relations such as Eq.~(\ref{eq:cwww_matching}),
which depend only on the gauge couplings of the heavy particle and its mass,
provide useful sensitivity to the heavy mass
$m_\phi$,  we have verified that
this 
is in practice very weak and
hence not competitive.

It is also possible to find EFT coefficients that are matched to the sum of a piece depending only on SM couplings and the UV mass and a piece proportional to the UV couplings, e.g.,
\begin{equation}
    \frac{c_{\varphi Q}^{(3)}}{\Lambda^2} = -\frac{g_2^4}{3840\pi^2 m_\phi^2} - \frac{\left(y_{t}^{\rm SM}\right)^2 \left(y_{\phi}^{u}\right)_{33}^2}{192\pi^2 m_\phi^2} + \frac{ g_2^2 \left(y_{\phi}^{u}\right)_{33}^2}{1152\pi^2 m_\phi^2}.
\end{equation} 
This kind of relations could in principle favour non-vanishing UV couplings even when the EFT
coefficients are very tightly constrained, provided that the gauge-coupling term is of similar size to the other terms in
the matching relation.

As for tree-level matching,
exemplified by Eqns.~(\ref{eq:rel_UV_tree_1})
and~(\ref{eq:rel_UV_tree_2}), at the one-loop
level one can also automatically
evaluate linear relations among the 
Wilson coefficients, though the analogous
result for non-linear relations  is
more challenging.
Nevertheless, as discussed in Sect.~\ref{sec:smefit}, in this work we carry out the fit directly at the level of UV couplings and hence the constraints between different coefficients are provided implicitly by the matching relations, rather than directly as explicit restrictions in a fit of EFT coefficients.

\subsection{UV invariants and SMEFT matching}
\label{subsec:uv_invariants}

The inverse problem to matching lies at the heart of the interpretation of the SMEFT global fit results
in terms of UV couplings
and masses.
This inverse problem is known to be plagued by degeneracies since many UV completions could yield the same EFT coefficients up to a fixed mass dimension~\cite{Zhang:2021eeo}.
Matching relations define a mapping $f$ from $U$, the parameter space spanned by the UV couplings $\mathbf{g}$,  to $W$, the space spanned by the Wilson coefficients $\mathbf{c}$,
\begin{equation}
    f: U \rightarrow W \, .
\end{equation}
The previous discussion indicates
that the matching relation $f$
associated to UV-models such as the one defined by Eq.~\eqref{eq:Lag_UV_full_Varphi_model} is in general non-injective
and hence non-invertible.
Therefore, even choosing a particular UV model does not lift completely these degeneracies.
They might be partially or fully removed by either matching at higher loop orders or considering higher-dimensional operators.
In particular, dimension-8 operators offer advantages to disentangle UV models~\cite{Zhang:2021eeo}, but their study is beyond the scope of this work.

Since the fit is, at best, only sensitive to $f(\mathbf{g})$, one can only meaningfully discriminate UV parameters $\mathbf{g}$ that map to different points
in the EFT parameter space $W$ under 
the matching relation $f$. 
Thus, we define ``UV invariants'' as those combinations of UV parameters such that $f$ remains invariant under a mapping $h$, defined as
\begin{equation}
    h: U \rightarrow I,
\end{equation}
such that $f(h(\mathbf{g})) = f(h(\mathbf{g}'))$. We denote by $I$ the space of UV invariants that is now bijective with $W$ under $f$ and that contains all the information that we can extract about the UV couplings by measuring $f$ from experimental data in the global EFT fit.

To illustrate the role of UV invariants, we consider again the tree-level matching relations 
for our benchmark model Eq.~(\ref{eq:Lag_UV_full_Varphi_model}) given by
Eq.~\eqref{eq:matching_result_example_tree}. 
Expressing the UV couplings in terms of the 
EFT coefficients leads to two different solutions,
\begin{equation}
\label{eq:uvinvariants_1}
	\big(y_{\phi}^{d}\big)_{33} = -  \sqrt{- \(c_{qd}^{(8)}\)_{3333} } \frac{m_\phi}{\Lambda}, \quad \big(y_{\phi}^{u} \big)_{33} = -\frac{c_{t\varphi}}{c_{b\varphi}}\sqrt{- \(c_{qd}^{(8)}\)_{3333} } \frac{m_\phi}{\Lambda},\quad \lambda_{\phi}=\frac{c_{b\varphi}}{\sqrt{-\( c_{qd}^{(8)}\)_{3333}}}\frac{m_\phi}{\Lambda},
\end{equation}
and
\begin{equation}
\label{eq:uvinvariants_2}
	\big(y_{\phi}^{d}\big)_{33} =  \sqrt{- \(c_{qd}^{(8)}\)_{3333} } \frac{m_\phi}{\Lambda}, \quad \big(y_{\phi}^{u}\big)_{33} = \frac{c_{t\varphi}}{c_{b\varphi}}\sqrt{- \(c_{qd}^{(8)}\)_{3333} } \frac{m_\phi}{\Lambda},\quad \lambda_{\phi} = - \frac{c_{b\varphi}}{\sqrt{ \( c_{qd}^{(8)}\)_{3333} }}\frac{m_\phi}{\Lambda},
\end{equation}
where we have omitted the flavour indices of the coefficients for clarity.
The resulting sign ambiguity stems from the sensitivity to the sign of only two products of the three UV couplings. 
Other non-vanishing EFT coefficients do not enter these solutions since they are related to the present ones via linear or non-linear relations.

Eqns.~(\ref{eq:uvinvariants_1})
and~(\ref{eq:uvinvariants_2}) hence indicate
that the EFT fit is not sensitive to
the sign of these UV couplings for this specific
model.
The sought-for mapping $h$ between the original UV couplings
$(\big(y_{\phi}^{d}\big)_{33}, \big(y_{\phi}^{u}\big)_{33}, \lambda_\phi)$
and the UV invariants which can be meaningfully
constrained by the global EFT fit is given by
\begin{equation}
\label{eq:example_UVinvariants}
    h: (\big(y_{\phi}^{d}\big)_{33}, \left(y_{\phi}^{u}\right)_{33}, \lambda_\phi) \mapsto \left( |\left(y_{\phi}^{u}\right)_{33}|,\, \lambda_\phi \,\mathrm{sgn}\lp \left(y_{\phi}^{u}\right)_{33}\rp,\, \left(y_{\phi}^{d}\right)_{33}\,\mathrm{sgn}\lp \lambda_\phi\rp \right),
\end{equation}
with $\mathrm{sgn}(x)$ being the sign function. Notice the degree of arbitrariness present in this construction, since for example one could have  also chosen $\left( |\lambda_\phi|,\, \left(y_{\phi}^{u}\right)_{33} \,\mathrm{sgn}\lp \lambda_\phi\rp\right)$ instead of the first two invariants
of Eq.~(\ref{eq:example_UVinvariants}).
This simple example displays only a sign ambiguity, but one could be unable to distinguish two UV couplings altogether since, e.g. they always appear multiplying each other.
The  \matchTOfit~package automates the computation of the transformations $h$ defining the
UV-invariants for the models at the tree-level matching level, 
see App.~\ref{app:mmeft2smefit} for more details.

\begin{figure}[t]
    \centering
    \includegraphics[width=\linewidth]{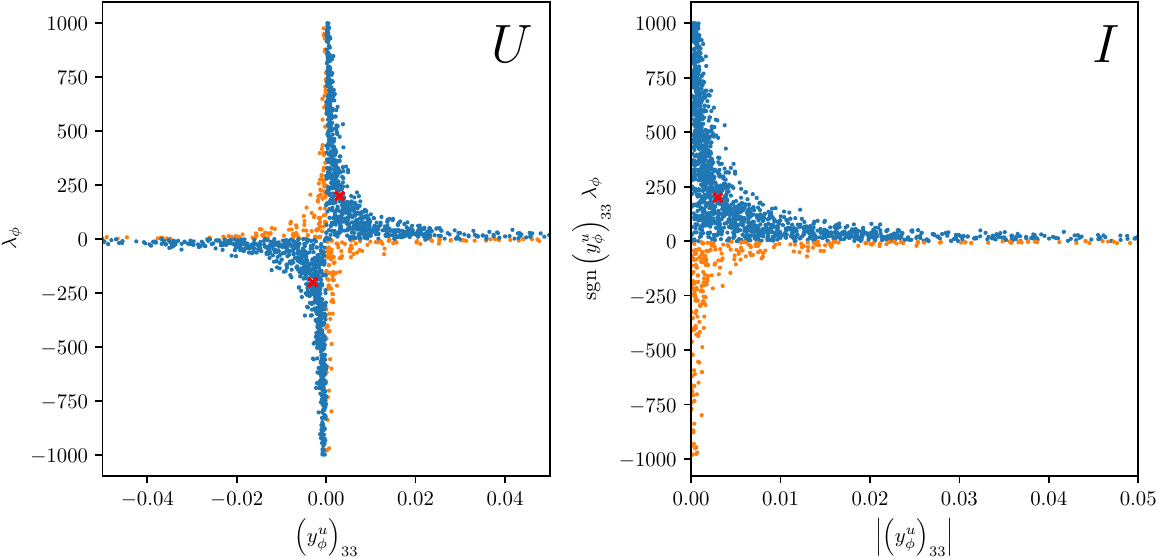}
    \caption{Left: marginalised posterior distributions in the space $U$ of UV parameters
$\lp \big(y_{\phi}^{u}\big)_{33}, \lambda_\phi\rp$
in the heavy scalar doublet model given by Eq.~\eqref{eq:Lag_UV_full_Varphi_model}
fitted to the data according to the
procedure of Sect.~\ref{sec:smefit}.
Right: the same results represented in the space $I$ of UV invariants
$
\left( |\big(y_{\phi}^{u}\big)_{33}|,\, \lambda_\phi \,\mathrm{sgn}( \big(y_{\phi}^{u}\big)_{33} ) \right).
$
The red points indicate two different
sets of UV couplings in $U$ that
are mapped to the same point in $I$
upon the transformation $h$.
Blue (orange) points indicate positive
(negative) values of the
UV-invariant $\lambda_\phi \,\mathrm{sgn}( \big(y_{\phi}^{u}\big)_{33} )$.
}
    \label{fig:uv_invariant_explained}
\end{figure}

Furthermore, one can also illustrate the concept
of UV-invariants by fitting
 the heavy scalar doublet model of Eq.~\eqref{eq:Lag_UV_full_Varphi_model}
 to the experimental data
by following the procedure which will be outlined
in Sect.~\ref{sec:smefit}.
Fig.~\ref{fig:uv_invariant_explained} shows the 
resulting 
 marginalised posterior distributions in the space $U$ of UV parameters
$( \big(y_{\phi}^{u}\big)_{33}, \lambda_\phi)$
and in the space $I$ of UV invariants
$
( |\big(y_{\phi}^{u}\big)_{33}|,\, \lambda_\phi \,\mathrm{sgn}\left( \big( y_{\phi}^{u} \big)_{33} \right) 
$.
The red points indicate two different
sets of UV couplings in $U$,
$\mathbf{g}\ne\mathbf{g}'$, that
are mapped to the same point in $I$
upon the transformation $h$.
Presenting results at the level of UV-invariants
has the benefit of making explicit the symmetries
and relations between UV-couplings that
may be hidden otherwise if one presents
results in the UV parameters space $U$. It is worth stressing that UV invariants do not necessarily correspond to combinations of UV parameters that one can constrain in a fit. Rather, they represent what can be said about the UV couplings from given values for the Wilson coefficients (WCs) and hence serve to map out the UV parameter space such that no redundant information is shown.

\section{Implementation in {\sc\small SMEFiT}}
\label{sec:smefit}
Here we describe how the {\sc\small SMEFiT} global analysis
framework~\cite{Hartland:2019bjb,vanBeek:2019evb,Ethier:2021bye,Ethier:2021ydt,Giani:2023gfq} has been extended to operate, in addition to at the level
of Wilson coefficients, directly at the level of the parameters
of UV-complete models.
We also  present the baseline SMEFT analysis
which will be used in Sect.~\ref{sec:results}
to constrain these UV parameters,
reporting on a number of improvements as compared
to its most recent version presented in~\cite{Giani:2023gfq}, in particular concerning the implementation
of precision electroweak observables (EWPOs) from the LEP legacy measurements.
App.~\ref{app:ewpos} provides additional details about the {\sc\small SMEFiT} functionalities and this baseline EFT global fit.

Assume a UV-complete model defined by the Lagrangian $\mathcal{L}_{\rm UV}(\boldsymbol{g})$
which contains $n_{\rm uv}$ free parameters $\boldsymbol{g}$.
Provided that this model has the SM as its low-energy limit with linearly realised electroweak symmetry breaking, one can derive matching relations between the SMEFT coefficients and the UV couplings of the form $\boldsymbol{c} = \boldsymbol{f}(\boldsymbol{g},\mu)$ for a given choice of the matching scale $\mu$ as discussed in Sect.~\ref{sec:general_strategy}.
Once these matching conditions are evaluated, 
the EFT cross-sections $\sigma(\boldsymbol{c})$
entering the fit can be expressed in terms of the UV couplings and masses $\sigma(\boldsymbol{f}(\boldsymbol{g},\mu))$.
By doing so, one ends up with the likelihood function $L$
now expressed in terms of UV couplings, $L(\boldsymbol{g})$.
Bayesian sampling techniques can now be applied directly to $L(\boldsymbol{g})$, assuming
a given prior distribution of the UV coupling space, in  the same manner
as for the fit of EFT coefficients.

The current release of {\sc\small SMEFiT} enables the user to impose these matching conditions $\boldsymbol{c} = \boldsymbol{f}(\boldsymbol{g},\mu)$ via run cards thanks to its support for a wide class of different constraints on the fit parameters, see App.~\ref{app:ewpos}.
The code applies the required substitutions on the 
theoretical predictions for the observables
entering the fit automatically.
Additionally, the availability of Bayesian sampling means that the functional relationship between the likelihood function $L$ and the fitted parameters 
$\boldsymbol{g}$ is unrestricted.

In order to carry out parameter inference directly at the level of UV couplings within {\sc\small SMEFiT}, three main ingredients are required:

\begin{itemize}

\item First, the matching relations $\boldsymbol{f}$ between the parameters of
the UV Lagrangian $\mathcal{L}_{\rm UV}(\boldsymbol{g})$ and the EFT Wilson coefficients, $\boldsymbol{c} = \boldsymbol{f}(\boldsymbol{g})$ in the
Warsaw basis used in the fit.
As discussed in Sect.~\ref{sec:general_strategy},
this step can be achieved automatically 
both for tree-level and for one-loop matching
relations
by using {\sc\small MatchMakerEFT}~\cite{Carmona:2021xtq}.
Other matching frameworks may also be used
in this step.

\item Second, the conversion between the output of the matching code, {\sc\small MatchMakerEFT} in our case,
and the input required for the {\sc\small SMEFiT} run cards specifying the
dependence of the Wilson coefficients $\boldsymbol{c}$
on the UV couplings $\boldsymbol{g}$ such that
the replacement $\sigma(\boldsymbol{c}) \to \sigma (\boldsymbol{f}(\boldsymbol{g}))$ on the physical observables entering the fit can be implemented.
This step has been automated by modifying accordingly \smefit~and by the release of the new Mathematica package \matchTOfit. The latter translates the output of the matching code, {\sc\small MatchMakerEFT} in our case, to an expression that can be understood by \smefit~and evaluates numerically all those parameters that are not to be fitted. More details on the use of \matchTOfit~can be found in App.~\ref{app:mmeft2smefit}.
%
The automation supports one-loop matching results, reaching the same level as state-of-the-art matching tools.
{\sc\small SMEFiT} then performs the replacements $\sigma(\boldsymbol{c}) \to \sigma (\boldsymbol{f}(\boldsymbol{g}))$ specified by the runcards.

\item Third, a choice of prior volume
in the space $U$ spanned by the UV
couplings $\boldsymbol{g}$ entering the fit.
In this work, we assume a flat prior on the UV parameters $\boldsymbol{g}$ and verify
that results are stable with respect to variations of this prior volume.
We note that, for the typical (polynomial) matching relations between UV couplings and EFT coefficients, this choice of prior implies non-trivial forms for the priors on the space of the latter.
This observation is an important motivation
to support the choice of fitting directly at the level of UV couplings.

\end{itemize}

Once these ingredients are provided,  {\sc\small SMEFiT} performs the global fit by comparing
EFT theory predictions with experimental data
and returning the posterior probability distributions
on the space of UV couplings $\boldsymbol{g}$
or any combination thereof. 
Fig.~\ref{fig:pipeline} displays a schematic representation of the pipeline adopted in this work
to map in an automated manner the parameter space of UV-complete models using the SMEFT as a bridge  to the data and based on the combination
of three tools: {\sc\small MatchMakerEFT}
to derive the matching relations,
\matchTOfit~to transform its output
into the {\sc\small SMEFiT}-compliant format,
and {\sc\small SMEFiT} to infer from
the data bounds on the UV coupling space.

\tikzstyle{io} = [trapezium, trapezium left angle=70, trapezium right angle=110, minimum width=3cm, minimum height=1cm, text centered, draw=black, fill=blue!30]
\tikzstyle{decision} = [diamond, minimum width=3cm, minimum height=1cm, text centered, draw=black, fill=green!30]
\tikzstyle{arrow} = [thick,->,>=stealth]
\tikzstyle{dottedline} = [thick, dotted]
\tikzstyle{process} = [rectangle, minimum width=2cm, minimum height=1cm, text centered, draw=black, fill=orange!30, rounded corners, text width=3cm]
\tikzstyle{smallbox} = [rectangle, minimum width=1cm, minimum height=1cm, text centered, draw=black, fill=orange!30, rounded corners, text width=1cm]
\tikzstyle{startstop} = [rectangle, minimum width=2cm, minimum height=1cm, text centered, draw=black, fill=blue!30, rounded corners, text width=6cm]
\tikzstyle{textbox} = [rectangle, minimum width=2cm, minimum height=1cm, text centered, rounded corners, text width=3cm]

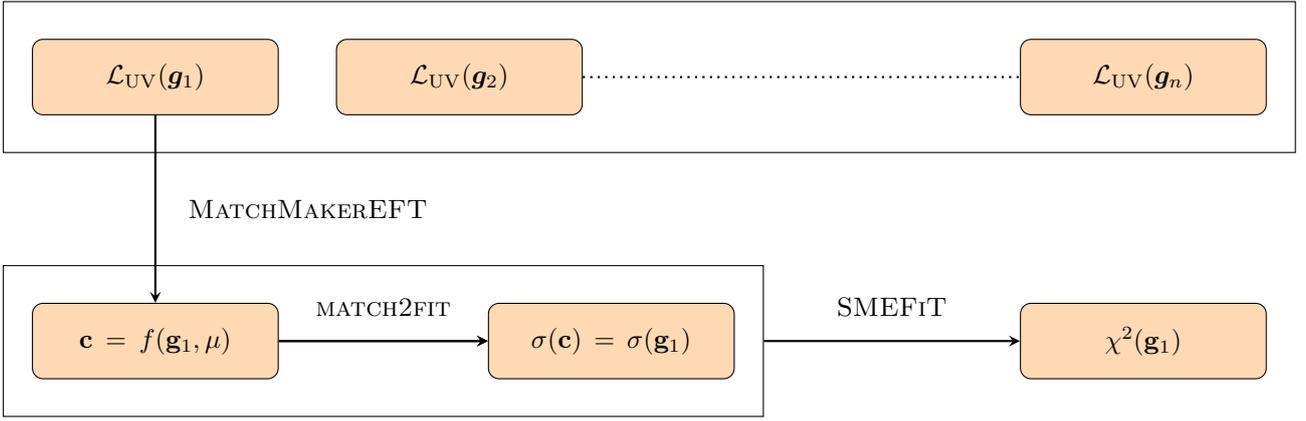
\begin{figure}[t]
\small
    \centering
    \begin{tikzpicture}[node distance=2cm]

\node (UV1) [process] {$\mathcal{L}_{\rm UV}(\boldsymbol{g}_1)$};
\node (UV2) [process, right of=UV1, xshift=2cm] {$\mathcal{L}_{\rm UV}(\boldsymbol{g}_2)$};
\node (UVn) [process, right of=UV2, xshift=7cm] {$\mathcal{L}_{\rm UV}(\boldsymbol{g}_n)$};

\node[fit=(UV1) (UVn),draw,minimum height=2cm, minimum width = 17cm] (Input) {};

\node (match) [process, below of=UV1, yshift=-1.5cm] {$\mathbf{c} = f(\mathbf{g}_1, \mu)$};

\node (xsec) [process, right of=match, xshift=4.0cm] {$\sigma(\mathbf{c})=\sigma(\mathbf{g}_1)$};

\node (chi2) [process, right of=xsec,  xshift=5cm] {$\chi^2(\mathbf{g}_1)$};

\node[fit=(match) (xsec),draw,minimum height=2cm, minimum width = 10cm] (match2fit) {};

\draw [dottedline] (UV2.east) --(UVn.west);
\draw [arrow] (UV1) -- (match) node[midway, right, xshift=0.3cm] {\sc\small MatchMakerEFT};
\draw [arrow] (match) -- (xsec);
\draw [arrow] (match) -- (xsec) node[midway, above,yshift=0.2cm] {\sc\small match2fit};
\draw [arrow] (match2fit) -- (chi2)  node[midway, above, yshift=0.2cm]{\sc\small SMEFiT} ;

\end{tikzpicture}
    \caption{Schematic representation of the pipeline adopted in this work
    to map the parameter space of UV-complete models using the SMEFT as a bridge
    to the data.
    The starting point is a UV-Lagrangian containing a number of free parameters
    $\boldsymbol{g}$ such as its masses and coupling constants,
    for which a flat prior is assumed.
    We then determine the matching relations between the UV parameters
     $\boldsymbol{g}$ and the corresponding EFT coefficients  $\boldsymbol{c}$ at
     the matching scale $\mu$
     using {\sc\small MatchMakerEFT}.
     Then the \matchTOfit~interface
     enables expressing
     cross-sections for
     processes entering the EFT
     in terms of the  UV-parameters $\boldsymbol{g}$.
     Finally, these UV-parameters
     are constrained from the data using the sampling methods of {\sc\small SMEFiT}
     applied to the figure of merit $\chi^2(\boldsymbol{g})$ evaluated
     on a global dataset.}
    \label{fig:pipeline}
\end{figure}


The UV-invariants introduced in Sect.~\ref{subsec:uv_invariants} are used only after the fit for plotting purposes, even if their functional form can be computed beforehand with \matchTOfit. 
Plotting the posteriors of UV-invariants from the results of the fit and the output of \matchTOfit~is also automated by \smefit.
Additionally, we note here that
depending on the specific UV-complete model, one might be able to constrain only the absolute value of certain UV parameters or of their product.
For this reason, here we will display results mostly at the level of UV-invariants $\boldsymbol{I}_{\rm UV}(\boldsymbol{g})$  
 determined from the matching relations.
We will find that  certain UV-invariants  may nevertheless remain unconstrained, due to the lack of sensitivity to specific EFT
coefficients in the fitted data. 
In any case, the user can easily plot and study
arbitrary combinations
of the fitted UV parameters.

The baseline global {\sc\small SMEFiT} analysis
adopted in this work to constrain
the parameter space of UV-complete models
is based on the analysis presented in~\cite{Giani:2023gfq}, 
which in turn updated the global SMEFT analysis
of Higgs, top quark, and diboson data from~\cite{Ethier:2021bye}.
In comparison with~\cite{Giani:2023gfq},
 a significant upgrade is the new implementation
 of the information provided by the legacy EWPOs~\cite{ALEPH:2005ab}
from LEP and SLC in
the Wilson coefficients.
Previously, the constraints provided by these EWPOs in the SMEFT parameter space were accounted in an approximate manner, assuming EWPOs measured with vanishing uncertainties.\footnote{
For the purposes of benchmarking with the
ATLAS EFT analysis of~\cite{ATL-PHYS-PUB-2022-037}, in~\cite{Giani:2023gfq} also
fits with EWPOs were considered, but these were based on
the exact same theory inputs as in the ATLAS paper.
}

As a consequence of this improvement,
the present global SMEFT analysis
considers $14$ additional Wilson coefficients $\boldsymbol{c}$ with respect to the 
basis used in~\cite{Ethier:2021bye,Giani:2023gfq}, leading to a total of 50 independent parameters constrained from the data.
Since these new degrees of freedom are constrained not only by the EWPOs, but also by LHC  observables, we have recomputed all EFT cross-sections for those processes where such operators enter.
As in the original analysis~\cite{Ethier:2021bye}, we use {\sc\small mg5\_aMC@NLO}~\cite{Alwall:2014hca}
interfaced to SMEFT@NLO~\cite{Degrande:2020evl} to evaluate linear and quadratic EFT corrections which account up to NLO QCD perturbative corrections, whenever available.
A dedicated description of this new
implementation of the EWPOs in {\sc\small SMEFiT}
and of their phenomenological implications for present and future colliders will be discussed
in a forthcoming publication~\cite{ewpos}.

\begin{figure}[t]
    \centering
    \includegraphics[width=0.99\linewidth] {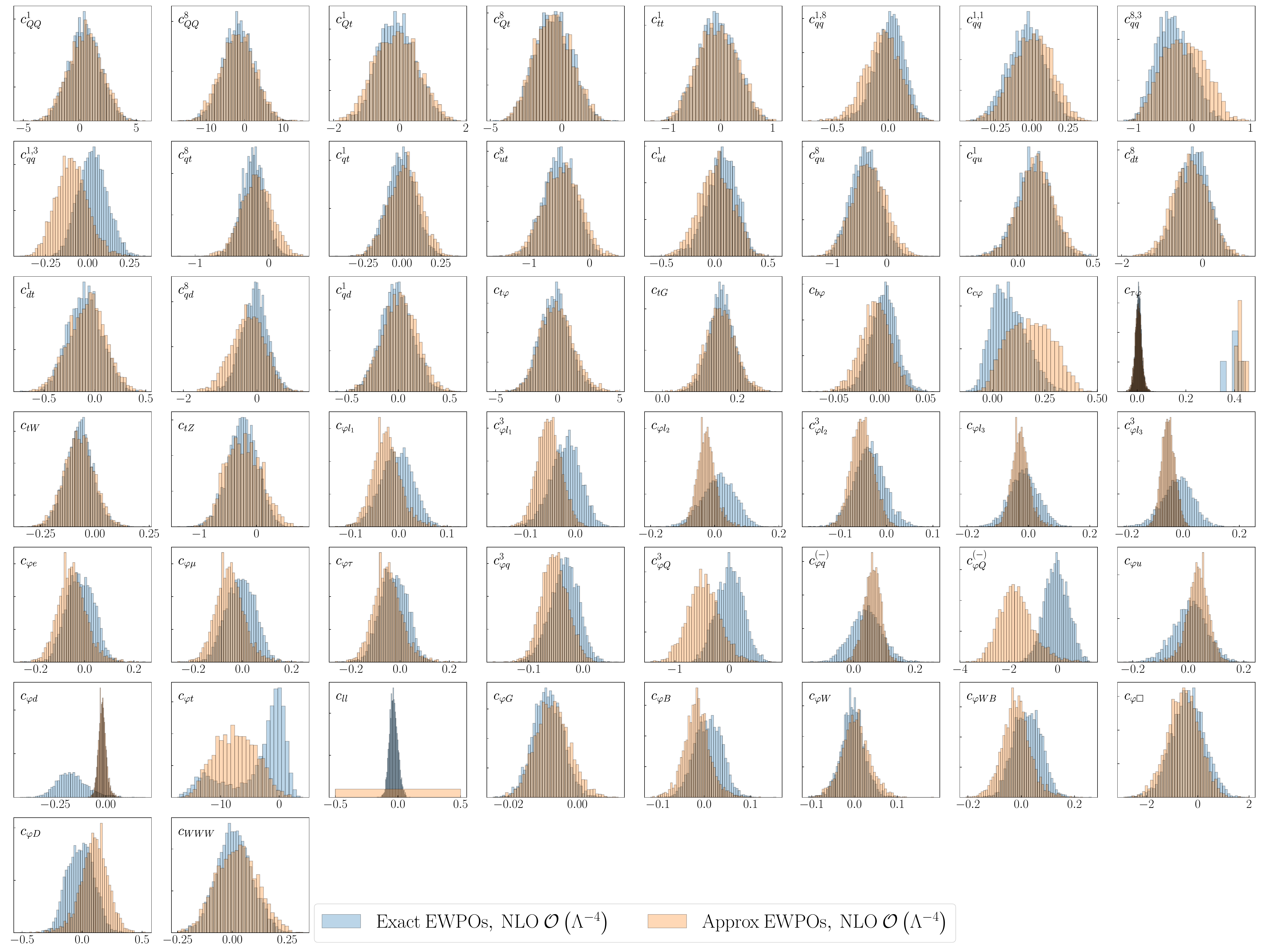}
    \caption{Marginalised single-parameter
    posterior probability
    distributions in the global SMEFT fit of~\cite{Giani:2023gfq}, based on an approximate implementation of the EWPOs, compared with the baseline fit adopted in this work and implementing the EWPOs constraints in a exact manner.
    Note that the results labelled as ``Approx
    EWPOs'' differ slightly from those
    in \cite{Giani:2023gfq} due
    to the use of updated theory
    predictions for LHC processes, see text.
    In both cases, the fits include
    quadratic EFT effects and NLO QCD
    corrections to the EFT cross-sections.
    \label{fig:post_smefit30_vs_smefit20-label}
    }
\end{figure}

The impact of this new exact EWPOs implementation is illustrated in 
Fig.~\ref{fig:post_smefit30_vs_smefit20-label},
comparing the marginalised one-parameter
posterior 
    distributions obtained in the global SMEFT fit of~\cite{Giani:2023gfq} and based
    on the approximate EWPOs implementation, with the baseline fit
    adopted in this work where EWPOs are taken into account in a exact manner.
    Note that the results labelled as ``Approx
    EWPOs'' differ slightly from those
    in \cite{Giani:2023gfq} due to
    the use of updated theory
    predictions. 
    In both cases, the fits include
    quadratic EFT effects and NLO QCD
    corrections to the EFT cross-sections.

In general, there is good agreement
at the level of Wilson coefficients between results based on the approximate and the exact
implementation of the EWPOs, with some noticeable differences.
While the posterior distributions
for most operators are only moderately
affected by the EWPOs, their
impact is visible
for some of the coefficients
affecting electroweak interactions
such as
$c_{\varphi t}$ (where a new double-peak structure is revealed), $\lp c_{\ell\ell} \rp_{1111}$ (which was
previously unconstrained),
$c_{\varphi d}$, $c_{\varphi Q}^{(3)}$,
and $c_{\varphi Q}^{(-)}$.
Nevertheless, differences between exact
and approximate implementation of EWPOs, which will be further studied in~\cite{ewpos},
are well contained within the respective
95\% CL intervals. 

\section{Results}
\label{sec:results}
We now present the main results of this work, namely
the constraints on the parameter space of
a  broad range  of UV-complete models obtained using the {\sc\small SMEFiT} global analysis
integrated with the pipeline described in Sects.~\ref{sec:general_strategy} and~\ref{sec:smefit}.
We discuss the following results in turn:
one-particle models matched at tree level,
multi-particle models also matched at tree level,
and one-particle models matched at 
the one-loop level.
For each category of models, we study both the impact of linear and quadratic corrections in the SMEFT expansion as well as of the QCD accuracy. Whenever possible, we provide comparisons with related studies to validate our results. 

The UV models considered in this work are composed of one or more of the heavy BSM particles listed in Table~\ref{tab:UV_particles} and
classified in terms of their spin
as scalar, fermion, and vector
particles.
 For each particle, we indicate the irreducible representation of the SM gauge group under which they transform.
These particles can couple linearly to the SM fields and hence generate dimension-6 SMEFT operators after being integrated out at tree level. 
The complete tree-level matching results for these particles were computed in~\cite{deBlas:2017xtg}, from where we adopt the notation.
The only  exception is the heavy scalar field $\varphi$, which we rename $\phi$ to be consistent with the convention used in Sect.~\ref{sec:general_strategy}.

\begin{table}[t]
    \centering
     \renewcommand{\arraystretch}{1.40}
\begin{tabular}{|c|c||c|c||c|c|}
\hline
\multicolumn{2}{|c||}{ $\qquad\qquad$Scalars$\qquad\qquad$} & \multicolumn{2}{c||}{$\qquad\qquad$Fermions$\qquad\qquad$} & \multicolumn{2}{c|}{$\qquad\qquad$Vectors$\qquad\qquad$} \\
\hline
Particle & Irrep & Particle & Irrep & Particle & Irrep \\
\hline
$\mathcal{S}$   & $\(1,1\)_{0\phantom{-/3}}$          & $N$        & $\left(1,1 \right)_{0\phantom{-/3}}$ & $\mathcal{B}$   & $\left(1,1\right)_{0}$ \\
$\mathcal{S}_1$ & $\(1,1\)_{1\phantom{-/3}}$  & $E$        & $\left(1,1\right)_{-1\phantom{/3}}$  & $\mathcal{B}_1$ & $\left(1,1\right)_{1}$ \\
$\phi$          & $\(1,2\)_{1/2\phantom{-}}$       & $\Delta_1$ & $\left(1,2\right)_{-1/2}$            & $\mathcal{W}$   & $\left(1,3\right)_{0}$ \\
$\Xi$           & $\(1,3\)_{0\phantom{-/3}}$  & $\Delta_3$ & $\left(1,2\right)_{-3/2}$            & $\mathcal{W}_1$ & $\left(1,3\right)_{1}$ \\
$\Xi_1$         & $\(1,3\)_{1\phantom{-/3}}$   & $\Sigma$   & $\left(1,3\right)_{0\phantom{-/3}}$  & $\mathcal{G}$   & $\left(8,1\right)_{0}$ \\
$\omega_1$      & $\left(3,1\right)_{-1/3}$    & $\Sigma_1$ & $\left(1,3\right)_{-1\phantom{/3}}$  & $\mathcal{H}$   & $\left(8,3\right)_{0}$ \\
$\omega_4$      & $\left(3,1\right)_{-4/3}$   & $U$        & $\left(3,1\right)_{2/3\phantom{-}}$  & $\mathcal{Q}_5$ & $\left(8,3\right)_{0}$ \\
$\zeta$         & $\left(3,3\right)_{-1/3}$      & $D$        & $\left(3,1\right)_{-1/3}$            & $\mathcal{Y}_5$ & $\left(\bar{6},2 \right)_{-5/6}$ \\
$\Omega_1$      & $\left(6,1\right)_{1/3\phantom{-}}$ & $Q_1$      & $\left(3,2\right)_{1/6\phantom{-}}$  &                 & \\ 
$\Omega_4$      & $\left(6,1\right)_{4/3\phantom{-}}$ & $Q_7$      & $\left(3,2\right)_{7/6\phantom{-}}$  &                 & \\
$\Upsilon$      & $\left(6,3\right)_{1/3\phantom{-}}$ & $T_1$      & $\left(3,3\right)_{-1/3}$            &                 & \\
$\Phi$          & $\left(8,2\right)_{1/2\phantom{-}}$             & $T_2$      & $\left(3,3\right)_{2/3\phantom{-}}$  &                 & \\ 
                &                                     & $Q_5$      & $\left(3,2\right)_{-5/6\phantom{-}}$ &                 & \\ 
\hline
\end{tabular}
    \caption{Heavy BSM particles entering the UV-complete models considered in this work. 
    For each particle, we indicate the irreducible representation  of the SM gauge group under which they transform, with notation $(\text{SU}(3)_{\text{c}},\text{SU}(2)_{\text{L}})_{\text{U}(1)_{\text{Y}}}$.}
    \label{tab:UV_particles}
\end{table}

Concerning one-particle models, we include each of the particles listed
in Table~\ref{tab:UV_particles} one at the time, and then impose restrictions on the couplings between the UV heavy fields  and the SM particles such that these models satisfy the {\sc\small SMEFiT} flavour assumption after matching at tree level. 
A discussion on the UV couplings allowed for each of the considered models under such restriction can be found in App.~\ref{app:UV_details}. 
We have discarded from our analysis those heavy particles considered in~\cite{deBlas:2017xtg} for which we could not find a set of restrictions on their UV couplings such that they obeyed the baseline EFT fit flavour assumptions. 

With regard to multi-particle models, we consider  combinations of the one-particle models
mentioned above without additional assumptions on their UV couplings unless specified. 
For illustration purposes,
we will present results for the model that results from adding the custodially-symmetric models with vector-like fermions and an SU$(2)_L$ triplet vector boson, presented in~\cite{Marzocca:2020jze}. 
We analyse the cases of both degenerate and different heavy particle masses.

An overview of which Wilson coefficients are generated by each UV particle at tree level is provided in Table~\ref{tab:heavy_scalar_vboson_sens} for heavy scalars and vector bosons and in Table~\ref{tab:heavy_vfermion_sens} for heavy vector-like fermions.
Black ticks indicate the EFT coefficients probed when applying the {\sc\small SMEFiT} flavour restrictions, while red ones indicate those probed only when the flavour-universal UV couplings assumption of 
the {\sc\small FitMaker} analysis~\cite{Ellis:2020unq} is used.
Notice that some operators are not generated by any of the models considered in this work, e.g.~four-fermion operators with 
two-light and two-heavy quarks
or with purely light-quark ones.
This is due, in part, to the correlations among operators imposed by UV models and the restrictions imposed by the assumed flavour symmetry.  
We recall that, in general, one-loop
matching will introduce more coefficients
as compared to the tree-level ones
listed in Tables~\ref{tab:heavy_scalar_vboson_sens} and~\ref{tab:heavy_vfermion_sens}

Finally, we will consider the model composed of a heavy scalar doublet $\phi$ matched to the SMEFT at the one-loop level. 
In this case, we impose the {\sc\small SMEFiT} flavour restriction only at tree level. 
The operators generated by this model at one-loop were already discussed in Sect.~\ref{subsec:oneloop}.

In the following, confidence level (CL)
intervals for the UV model parameters are evaluated as highest density intervals (HDI or HPDI)~\cite{Box1993}. The HDI is defined as the interval such that all values inside the HDI have higher probability than any value outside. In contrast to equal-tailed intervals (ETI) which are based on quantiles, HDI does not suffer from the property that some value inside the interval might have lower probability than the ones outside the interval. 
Whenever the lower bound is two orders of magnitude smaller than the width of the CL interval, we round it to zero.

\renewcommand{\arraystretch}{1.4}
\begin{table}[t]
    \centering
    \footnotesize
    \begin{tabular}
    {c|c|c|c|c|c|c|c|c|c|c|c|c||c|c|c|c|c|c|c|c|c}
    \toprule
& \multicolumn{12}{c||}{\bf Heavy Scalars}
& \multicolumn{9}{c}{\bf Heavy Vector Bosons} \\
    \midrule
        &${\color{red}\mathcal{S}}$ 
        & $\mathcal{S}_1$ 
        & $\color{red}{\phi}$ & ${\color{red}\Xi}$ & $\Xi_1$ & $\omega_1$ & $\omega_4$ & $\zeta$ & $\Omega_1$&
    $\Omega_4$&
    $\Upsilon$&
    $\Phi$&
      $\mathcal{B}$&
    ${\color{red}\mathcal{B}_1}$&
    $\mathcal{W}$&
    ${\color{red}\mathcal{W}_1}$&
    $\mathcal{G}$&
    $\mathcal{H}$&
    $\mathcal{Q}_5$&
    $\mathcal{Y}_5$&
    ${\color{red}\mathcal{B}_1\mathcal{B}}$ \\
    \hline
        $c_{\varphi \Box}$           & \chm      & &       & \chm & \chm &      &      &      &      &      &      &      & \chm & \chm & \chm & \chm &      &      &      &      &  \rchm   \\
        $c_{\varphi D}$              &           & &       & \chm & \chm &      &      &      &      &      &      &      & \chm & \chm &      & \chm &      &      &      &      &          \\
        $c_{\tau \varphi}$           &           & & \rchm & \chm & \chm &      &      &      &      &      &      &      &      & \chm & \chm & \chm &      &      &      &      & \rchm    \\
        $c_{b \varphi}$              &           & & \rchm & \chm & \chm &      &      &      &      &      &      &      &      & \chm & \chm & \chm &      &      &      &      & \rchm    \\
        $c_{t \varphi}$              &           & & \chm  & \chm & \chm &      &      &      &      &      &      &      &      & \chm & \chm & \chm &      &      &      &      & \rchm    \\
        $c_{\varphi l_{1,2,3}}^{(3)}$&           & &       &      &      &      &      &      &      &      &      &      &      &      & \chm &      &      &      &      &      &      \\
        $c_{\varphi l_{1,2,3}}^{(1)}$&           & &       &      &      &      &      &      &      &      &      &      & \chm &      &      &      &      &      &      &      &      \\
        $c_{\varphi (e, \mu, \tau)}$ &           & &       &      &      &      &      &      &      &      &      &      & \chm &      &      &      &      &      &      &      &      \\
        $c_{\varphi Q}^{(3)}$        &           & &       &      &      &      &      &      &      &      &      &      &      &      & \chm &      &      &      &      &      &      \\
        $c_{\varphi Q}^{(-)}$        &           & &       &      &      &      &      &      &      &      &      &      & \chm &      & \chm &      &      &      &      &      &      \\
        $c_{\varphi q}^{(3)}$        &           & &       &      &      &      &      &      &      &      &      &      &      &      &      &      &      &      &      &      &      \\
        $c_{\varphi t}$              &           & &       &      &      &      &      &      &      &      &      &      & \chm &      &      &      &      &      &      &      &      \\
        $c_{\ell\ell}$               &           & \chm &       &      &      &      &      &      &      &      &      &      &      &      & \chm &      &      &      &      &      &      \\
        $c_{Qt}^{1}$                 &           & & \chm  &      &      &      &      &      &      &      &      & \chm & \chm &      &      &      &      &      & \chm & \chm &      \\
        $c_{Qt}^{8}$                 &           & & \chm  &      &      &      &      &      &      &      &      & \chm &      &      &      &      & \chm &      & \chm & \chm &      \\
        $c_{QQ}^{1}$                 &           & &       &      &      & \chm &      & \chm & \chm &      & \chm &      & \chm &      & \chm &      &      & \chm &      &      &      \\
        $c_{QQ}^{8}$                 &           & &       &      &      & \chm &      & \chm & \chm &      & \chm &      &      &      & \chm &      & \chm & \chm &      &      &      \\
        $c_{tt}^{1}$                 &           & &       &      &      &      & \chm &      &      & \chm &      &      & \chm &      &      &      & \chm &      &      &      &      \\
        $c_{qd}^{(1)}$\color{red}{$^\dagger$}&          & & \rchm &      &      &      &      &      &      &      &      &      &      &      &      &      &      &      &      &      &      \\
        $c_{qd}^{(8)}$\color{red}{$^\dagger$}&          & & \rchm &      &      &      &      &      &      &      &      &      &      &      &      &      &      &      &      &      &      \\
        $c_{{\ell \ell}_{1111}}$    &            & &       &      &      &      &      &      &      &      &      &      & \chm &      & \chm &      &      &      &      &      &      \\
        \bottomrule
    \end{tabular}
    \caption{The Wilson coefficients, in the 
    {\sc SMEFiT} fitting basis, generated at tree level
    by the heavy scalar and vector boson particles
    whose quantum numbers are listed in Table~\ref{tab:UV_particles}.
   Particles highlighted in red corresponds to one-particle models also considered in the {\sc\small FitMaker} analysis~\cite{Ellis:2020unq}. 
    Red check marks indicate operators relevant for the flavor assumptions of~\cite{Ellis:2020unq}. 
    Coefficients marked with {\color{red}{$^\dagger$}} correspond to operators generated by the heavy scalar model $\phi$ in~\cite{Ellis:2020unq} restricted to the third quark generation.
    In our analysis, we take such a flavour-universal restriction exclusively when comparing our results with those from~\cite{Ellis:2020unq}.}
    \label{tab:heavy_scalar_vboson_sens}
\end{table}

\renewcommand{\arraystretch}{1.3}
\begin{table}[t]
    \centering
    \begin{tabular}{c|c|c|c|c|c|c|c|c|c|c|c|c}
    \toprule
& \multicolumn{12}{c}{\bf Heavy Fermions} \\
    \midrule
&${\color{red}N}$&
    ${\color{red}E}$&
    ${\color{red}\Delta_{1}}$,
    ${\color{red}\Delta_{3}}$&
    ${\color{red}\Sigma}$, ${\color{red}\Sigma_{1}}$&
    ${\color{red}U}$&
    ${\color{red}D}$&
    $Q_1$&
    ${\color{red}Q_5}$&
    ${\color{red}Q_7}$&
    ${\color{red}T_1}$,
    ${\color{red}T_2}$ &
    ${\color{red}T}$ &
    ${\color{red} Q_{17}}$\\
    \hline
   $\qquad$ $c_{\varphi \ell_1}^{(3)}$  \color{red}{$^\dagger$}  $\qquad$   &{\rchm} &\rchm&     &{\rchm}&       &       &       &       &       & & &\\
    $c_{\varphi \ell_1}$ \color{red}{$^\dagger$}           &{\rchm} &\rchm&     &{\rchm}&       &       &       &       &       & & &\\
    $c_{\varphi \ell_2}^{(3)}$ \color{red}{$^\dagger$}     &{\rchm} &\rchm&     &{\rchm}&       &       &       &       &       & & &\\
    $c_{\varphi \ell_2}$ \color{red}{$^\dagger$}           &{\rchm} &\rchm&     &{\rchm}&       &       &       &       &       & & &\\
    $c_{\varphi \ell_3}^{(3)}$ \color{red}{$^\dagger$}     &\chm    &\chm &     &\chm   &       &       &       &       &       & & &\\
    $c_{\varphi \ell_3}$ \color{red}{$^\dagger$}           &\chm    &\chm &     &\chm   &       &       &       &       &       & & &\\
    $c_{\varphi \tau}$ \color{red}{$^\dagger$}             &        &     &\chm &       &       &       &       &       &       & & &\\
    $c_{\varphi (e, \mu)}$ \color{red}{$^\dagger$}         &        &     &\rchm&       &       &       &       &       &       & & &\\
    $c_{\tau \varphi}$                                     &        &\chm &\chm &\chm   &       &       &       &       &       & & &\\
    $c_{\varphi Q}^{(3)}$ \color{red}{$^\dagger$}                                &        &     &     &       &\chm   &\chm   &       &       & &\chm   & \rchm &\\
    $c_{\varphi Q}^{(-)}$ \color{red}{$^\dagger$}                                 &        &     &     &       &\chm   &       &       &       & &\chm   & \rchm & \\
    $c_{\varphi t}$                                        &        &     &     &       &       &       &\chm   &   &  \chm     & & &\\
    $c_{t \varphi}$                                        &        &     &     &       &\chm   &       &\chm   &   &\chm   &\chm & \rchm & \rchm \\
    $c_{b \varphi}$                                        &        &     &     &       &      &\chm   &       & \chm      & & \chm & &\\
    $c_{\varphi q}^{(3)}$ \color{red}{$^\dagger$}                                 &        &     &     &       &{\rchm}&{\rchm}&       &       & &{\rchm}& & \\
    $c_{\varphi q}^{(-)}$ \color{red}{$^\dagger$}                                &        &     &     &       &{\rchm}&       &       &       & &{\rchm}& & \\
    $c_{\varphi d_i}$                                        &        &     &     &       &       &       &       &\rchm &       & & & \\
    $c_{\varphi u_i}$                                        &        &     &     &       &       &       &       & &\rchm &       & & \\
    \bottomrule
    \end{tabular}
    \caption{
    Same as Table~\ref{tab:heavy_scalar_vboson_sens}
for the heavy fermions.
Coefficients labelled with{ \color{red}{$^\dagger$}} correspond
to operators probed in~\cite{Ellis:2020unq} in a flavour-universal way, i.e. it is assumed that $c_{\varphi \ell_{1,2,3}}^{(3)}=c_{\varphi \ell}^{(3)}$ and likewise for the others. 
In our analysis, we take such a flavour-universal restriction exclusively
when benchmarking our results against those from~\cite{Ellis:2020unq}.
    }
    \label{tab:heavy_vfermion_sens}
\end{table}

\subsection{One-particle models matched at tree level}
\label{subsec:heavy-scalar}

Here we present results obtained from the tree-level matching of one-particle extensions of
the SM.
Motivated by the discussion in Sect.~\ref{subsec:uv_invariants},
we present results at the level of UV invariants
since these are the combinations of UV couplings
that are one-to-one with the Wilson coefficients under the matching relations.
We study in turn the one-particle models
containing heavy scalars,
fermions, and vector bosons
listed in Tables~\ref{tab:heavy_scalar_vboson_sens}
and~\ref{tab:heavy_vfermion_sens}.
We also present a comparison
with the one-particle models considered
in the {\sc\small FitMaker} analysis~\cite{Ellis:2020unq}. 

\paragraph{Heavy scalars.}
The upper part of Table~\ref{tab:cl-heavy-scalar-fermion} shows the $95\%$ CL intervals obtained for the UV invariants  associated to the one-particle
heavy scalar models considered
in this work and listed in Table~\ref{tab:UV_particles}.
We compare results obtained with
different settings of the global SMEFT fit:
linear and quadratic level in the EFT expansion
and either LO or NLO QCD accuracy
for the EFT cross-sections.
We exclude from Table~\ref{tab:cl-heavy-scalar-fermion} the results of heavy scalar model $\phi$ that is characterised by two UV invariants.
In all cases, we assume that the mass of the heavy particle is $m_{\rm UV}=1$~TeV, and hence
all the UV invariants shown are dimensionless.
The resulting bounds for these heavy scalar models are well below the naive perturbative limit $g\lesssim 4\pi$ in most cases, and only for a few models in linear EFT fits the bounds are close to saturate the perturbative unitary condition, $g \lesssim \sqrt{8\pi}$~\cite{DiLuzio:2016sur}.

\begin{table}[htbp]
\begin{center}
\scriptsize
\renewcommand{\arraystretch}{1.6}\begin{tabular}{c|c|c|c|c|c}
Model & UV invariants & LO $\mathcal{O}\left(\Lambda^{-2}\right)$ &LO $\mathcal{O}\left(\Lambda^{-4}\right)$ & NLO $\mathcal{O}\left(\Lambda^{-2}\right)$ & NLO $\mathcal{O}\left(\Lambda^{-4}\right)$ \\
\toprule
$\mathcal{S}$ & $\left|\kappa_{\mathcal{S}}\right|$ & [0, 1.4] & [0, 1.4] & [0, 1.5] & [0, 1.4] \\
$\mathcal{S}_1$ & $\left(y_{\mathcal{S}_1}\right)_{12} \left(y_{\mathcal{S}_1}\right)_{21}$ & [-0.041, 0.0018] & [-0.040, 0.0042] & [-0.042, 0.0027] & [-0.042, 0.0030] \\
$\Xi$ & $\left|\kappa_{\Xi}\right|$ & [0, 0.067] & [0, 0.069] & [0, 0.069] & [0, 0.069] \\
$\Xi_1$ & $\left|\kappa_{\Xi1}\right|$ & [0, 0.049] & [0, 0.049] & [0, 0.049] & [0, 0.048] \\
$\omega_1$ & $\left|\left(y_{\omega_1}^{qq}\right)_{33}\right|$ & [0, 5.0] & [0, 1.6] & [0, 5.2] & [0, 1.7] \\
$\omega_4$ & $\left|\left(y_{\omega_4}^{uu}\right)_{33}\right|$ & [0.027, 3.6] & [0.021, 1.1] & [0, 3.1] & [0.043, 1.0] \\
$\zeta$ & $\left|\left(y_{\zeta}^{qq}\right)_{33}\right|$ & [0.11, 3.7] & [0.011, 1.0] & [0.14, 3.3] & [0.034, 0.99] \\
$\Omega_1$ & $\left|\left(y_{\Omega_1}^{qq}\right)_{33}\right|$ & [0.021, 4.4] & [0, 1.5] & [0, 4.0] & [0.031, 1.4] \\
$\Omega_4$ & $\left|\left(y_{\Omega_4}\right)_{33}\right|$ & [0.099, 5.165] & [0.059, 1.553] & [0, 4.4] & [0.037, 1.4] \\
$\Upsilon$ & $\left|\left(y_{\Upsilon}\right)_{33}\right|$ & [0, 3.4] & [0, 1.1] & [0, 3.0] & [0.027, 1.0] \\
$\Phi$ & $\left|\left(y_{\Phi}^{qu}\right)_{33}\right|$ & [0.14, 11] & [0, 2.9] & [0.018, 9.8] & [0.014, 2.6] \\
\bottomrule
\toprule
$N$ & $\left|\left(\lambda_N^e\right)_3\right|$ & [0, 0.47] & [0, 0.47] & [0, 0.47] & [0, 0.48] \\
$E$ & $\left|\left(\lambda_E\right)_3\right|$ & [0, 0.24] & [0, 0.25] & [0, 0.25] & [0, 0.25] \\
$\Delta_1$ & $\left|\left(\lambda_{\Delta_1}\right)_3\right|$ & [0, 0.21] & [0, 0.20] & [0, 0.21] & [0, 0.20] \\
$\Delta_3$ & $\left|\left(\lambda_{\Delta_3}\right)_3\right|$ & [0, 0.26] & [0, 0.27] & [0.0015, 0.26] & [0, 0.27] \\
$\Sigma$ & $\left|\left(\lambda_{\Sigma}\right)_3\right|$ & [0, 0.29] & [0, 0.28] & [0, 0.28] & [0, 0.29] \\
$\Sigma_1$ & $\left|\left(\lambda_{\Sigma_1}\right)_3\right|$ & [0, 0.42] & [0, 0.42] & [0, 0.43] & [0, 0.42] \\
$U$ & $\left|\left(\lambda_{U}\right)_3\right|$ & [0, 0.84] & [0, 0.85] & [0, 0.82] & [0, 0.84] \\
$D$ & $\left|\left(\lambda_{D}\right)_3\right|$ & [0, 0.23] & [0, 0.24] & [0, 0.24] & [0, 0.23] \\
$Q_1$ & $\left|\left(\lambda_{\mathcal{Q}_1}^u\right)_3\right|$ & [0, 0.94] & [0, 0.95] & [0, 0.93] & [0, 0.92] \\
$Q_7$ & $\left|\left(\lambda_{\mathcal{Q}_7}\right)_3\right|$ & [0, 0.95] & [0, 0.93] & [0, 0.91] & [0 0.91] \\
$T_1$ & $\left|\left(\lambda_{T_1}\right)_3\right|$ & [0, 0.46] & [0, 0.46] & [0, 0.45] & [0, 0.47] \\
$T_2$ & $\left|\left(\lambda_{T_2}\right)_3\right|$ & [0, 0.39] & [0, 0.38] & [0, 0.38] & [0, 0.38] \\
\bottomrule
\end{tabular}
\end{center}
\caption{The 95\% CL intervals for the UV invariants 
relevant for the heavy scalar (upper part) and heavy fermion (lower part of the table) 
one-particle models matched at tree-level.
We quote
the 95\% CL upper limit 
and the lower limit 
is rounded to 0 whenever it is two orders of magnitudes smaller than the total CL width. For each model we compare results
obtained at the linear and quadratic level
in the EFT expansion and using either LO
or NLO perturbative QCD corrections to
the EFT cross-sections.
In all cases, we set the mass of the heavy particle to $m_{\rm UV}=1$ TeV.
Note that the model with heavy scalar  $\phi$ is considered separately in Table~\ref{tab:cl-phi}, given that it is parameterised in terms of multiple UV invariants.}
\label{tab:cl-heavy-scalar-fermion}
\end{table}


Comparing the impact of linear versus quadratic corrections, we notice a significant improvement in sensitivity for the heavy scalar models $\omega_1$, $\omega_4$, $\zeta$, $\Omega_1$, $\Omega_4$, $\Upsilon$ and $\Phi$. 
This can be explained by the fact that they generate four-fermion operators, as these are characterised by large quadratic corrections.
Indeed, four-heavy operators, constrained
in the EFT fit by $t\bar{t}t\bar{t}$
and $t\bar{t}b\bar{b}$ cross-section data, have limited
sensitivity in the linear EFT fit.
In the remaining models, the impact of
quadratic EFT corrections
on the UV-invariant bounds is small.
Considering the impact of the QCD perturbative accuracy
used for the EFT cross-sections, one notices a moderate improvement of $\sim 10\%$ for models that are sensitive to four-fermion operators with the exception of $\omega_1$. 

\begin{figure}[t]
    \centering
    \includegraphics[width=\linewidth]{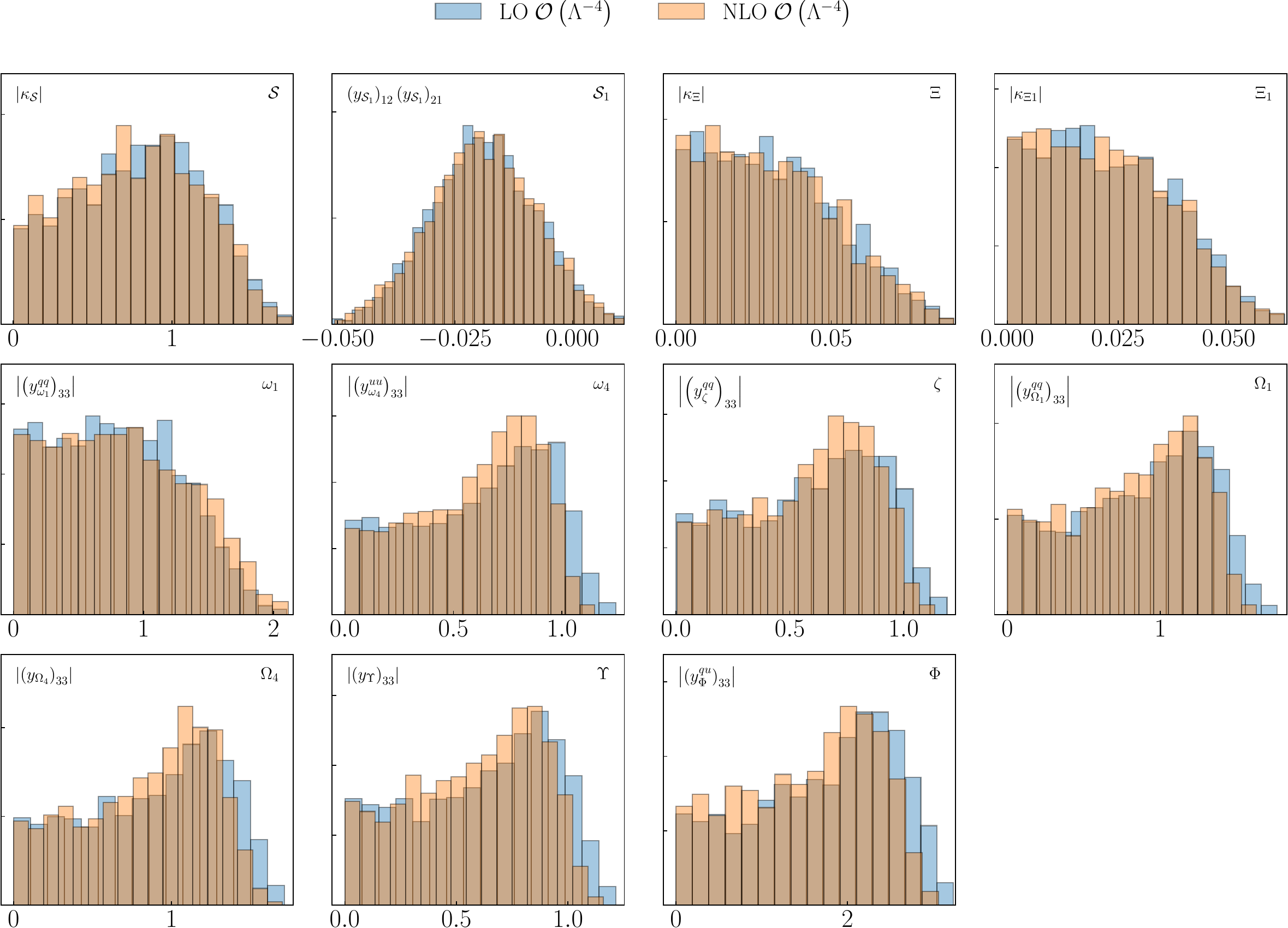}
    \caption{Same as Fig.~\ref{fig:heavy_scalar_nlo}, now comparing the baseline results,
    based on NLO QCD cross-sections
    for the EFT cross-sections, with
    the fit variant restricted to LO
    accuracy. }
    \label{fig:heavy_scalar_nlo}
\end{figure}

The posterior distributions 
associated to the heavy scalar models
listed in Table~\ref{tab:cl-heavy-scalar-fermion} are shown in Figs.~\ref{fig:heavy_scalar_quad}
and~\ref{fig:heavy_scalar_nlo}, comparing
results with different EFT expansion order
and QCD accuracy respectively.
To facilitate visualisation of the bulk region,
the distributions are
cut after the distribution has dropped to $5\%$ of its maximum, though
this choice is independent from the calculation
of the CL bounds.
One can see how
using quadratic EFT corrections (NLO QCD cross-sections)
improve significantly (moderately) the bounds on models that are sensitive to four-fermion operators for the reasons mentioned above.
The bounds most affected by quadratic EFT corrections are also the most susceptible to be modified when including dimension-8 effects.
In a few cases, one can observe a worsening of the bound when including quadratic corrections, which are due to numerical accuracy limitations.
For all models considered, the posterior
distributions indicate agreement
with the SM, namely vanishing UV model couplings.

Along the same lines,
from Figs.~\ref{fig:heavy_scalar_quad}
and~\ref{fig:heavy_scalar_nlo} one
also observes that the posterior distributions of the absolute values of UV couplings tend to exhibit a most likely value (mode) away from zero. 
This feature is not incompatible with a posterior distribution on the EFT coefficient space that favours the SM case.
The reason is that when transforming the probability density function from one space to the other, one must consider the Jacobian factor that depends on the functional relation between EFT coefficients and UV couplings. 
For most matching relations, this Jacobian factor can generate these peaks away from zero in the UV coupling space even when the EFT fit favours the SM solution.

To illustrate this somewhat counter-intuitive result, consider a toy model for the probability density of a (positive-definite) Wilson coefficient $c$,
\be
P(c) = \frac{2}{\sqrt{\pi}}e^{-c^2} \, ,\qquad \int_{0}^\infty dc\,P(c) = 1 \, ,
\ee
where the underlying law is the SM.
For a typical matching condition of the form $c=g^2$, (see i.e. the heavy
scalar model of Eq.~(\ref{eq:matching_result_example_tree})),
the transformed probability distribution for
the ``UV-invariant'' $|g|$ is
\be
P(|g|) = \frac{4}{\sqrt{\pi}}|g|e^{-|g|^4} \, , \qquad \int_{0}^\infty d|g| \,P(|g|) = 1 \, ,
\ee
which is maximised by $g=1/\sqrt{2}\ne 0$.
Hence posterior distributions in the UV
coupling space favouring non-zero values
do not (necessarily) indicate preference
for BSM solutions in the fit. On the other hand, posteriors on the UV that are approximately constant near zero correspond to posteriors in the WC space that diverge towards zero.

One also observes from Table~\ref{tab:cl-heavy-scalar-fermion} that two of the considered 
heavy scalar models, specifically those containing the
 $\Xi$ and $\Xi_1$ particles, lead to bounds which are at least two orders of magnitude more stringent than for the rest.
 These two
 models generate the same operators with slightly different matching coefficients,
 and as indicated in Table~\ref{tab:heavy_scalar_vboson_sens} they are the only scalar particles that generate the Wilson
coefficient $c_{\varphi D}$, which is  
strongly constrained by the EWPOs. 
Therefore, one concludes that the heavy scalar models
with the best constraints are those
whose generated operators are sensitive
to the high-precision electron-positron collider data.

\begin{table}[t]
\small
\begin{center}
\renewcommand{\arraystretch}{1.5}\begin{tabular}{c|c|c|c|c|c}
\toprule
Model & UV invariants & LO $\mathcal{O}\left(\Lambda^{-2}\right)$ &LO $\mathcal{O}\left(\Lambda^{-4}\right)$ & NLO $\mathcal{O}\left(\Lambda^{-2}\right)$ & NLO $\mathcal{O}\left(\Lambda^{-4}\right)$ \\
\midrule
\multirow{2}{*}{$\phi$ (tree-level)} & $\left|\lambda_{\phi}\right|$ & [0, $8.2\cdot 10^{2}$] & [0, $7.4\cdot 10^{2}$] & [0, $8.0\cdot 10^{2}$] & [0, $7.9\cdot 10^{2}$] \\
 & $\mathrm{sgn}(\lambda_{\phi})\left(y_{\phi}^u\right)_{33}$ & [-0.11, 1.0] & [-0.20, 2.1] & [-0.19, 0.62] & [-0.18, 1.7]  \\
 \midrule
\multirow{2}{*}{$\phi$ (one-loop)} & $\left|\lambda_{\phi}\right|$ & [0, 7.6] & [0, 7.6] & [0, 7.6] & [0, 7.1] \\
 & $\mathrm{sgn}(\lambda_{\phi})\left(y_{\phi}^u\right)_{33}$ & [-0.81, 2.8] & [-1.2, 2.3] & [-0.80, 2.2] & [-0.87, 2.1]  \\
\bottomrule
\end{tabular}
\end{center}
\caption{Same as Table~\ref{tab:cl-heavy-scalar-fermion} for the heavy scalar $\phi$ model,
which has associated two independent
UV-invariants.
See Fig.~\ref{fig:tree_vs_loop} for
the associated posterior distributions.
}
\label{tab:cl-phi}
\end{table}

\begin{figure}[t]
    \centering
    \includegraphics[width=\linewidth]{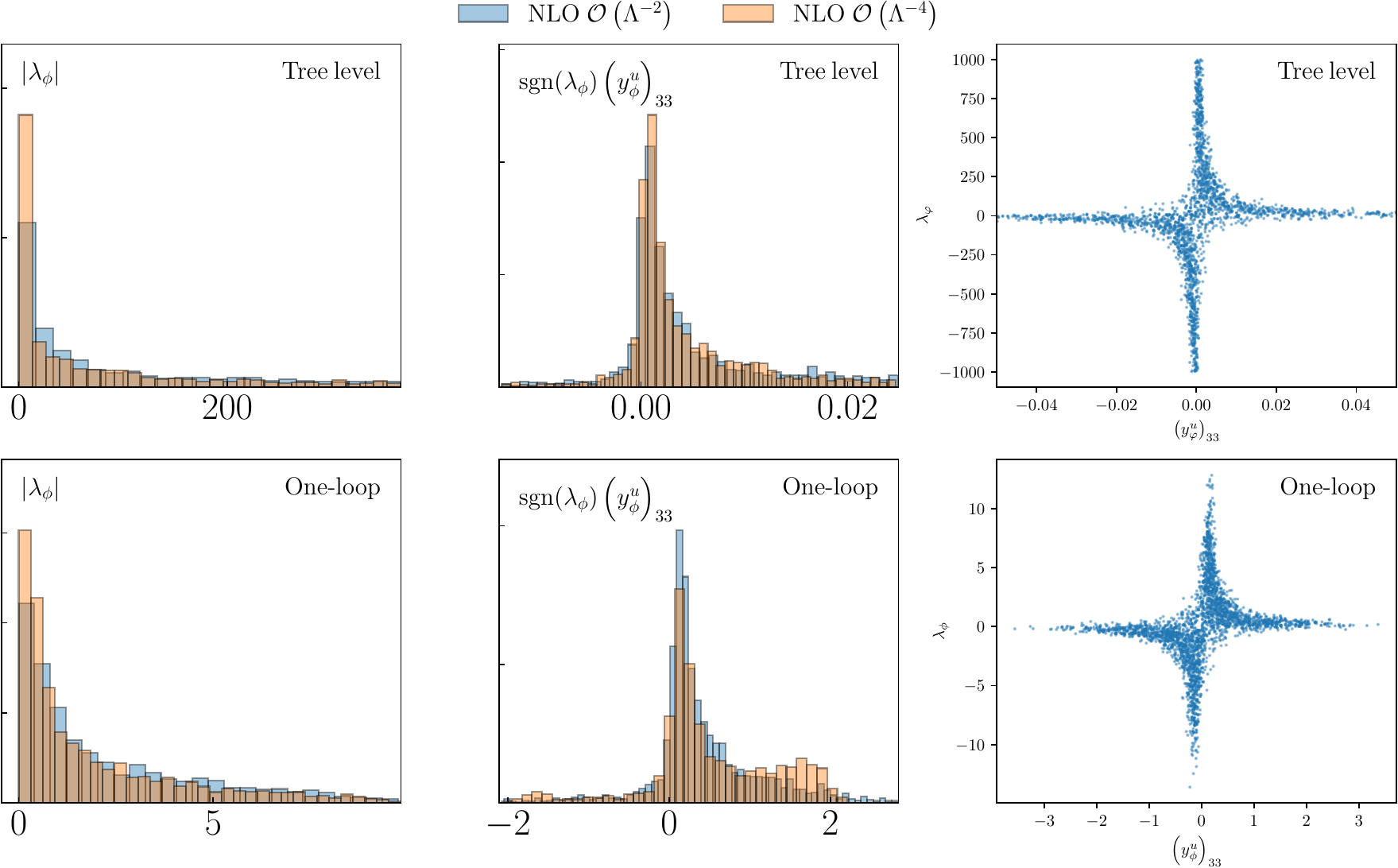}
    \caption{Marginalized posterior distributions of the two UV invariants associated
    to  the heavy scalar model $\phi$, comparing the impact of linear and quadratic EFT corrections after matching at tree level (upper panel) and at one-loop level (lower panel). 
    The rightmost panels illustrate how 
    the individual UV couplings $\lambda_\phi$ and $(y_{\phi}^u)_{33}$ are correlated, as expected
    given that only their product
    can be constrained from the data. 
    See Table~\ref{tab:cl-phi} for
    the corresponding 95\% CL intervals.
    }
    \label{fig:tree_vs_loop}
\end{figure}

The heavy scalar models that we have discussed so far 
and listed in Table~\ref{tab:cl-heavy-scalar-fermion}
depend on a single UV invariant.
On the other hand, as discussed
in Sect.~\ref{subsec:uv_invariants}, the heavy scalar $\phi$ model depends on two
different UV couplings, $\lambda_\phi$ and $(y_{\phi}^u)_{33}$, resulting in two independent invariants.
We present the corresponding $95\%$ CL intervals from tree-level 
matching in the upper part of Table~\ref{tab:cl-phi} and their distributions
in Fig.~\ref{fig:tree_vs_loop}. 
One finds that this model exhibits a degeneracy along 
the $\big(y_{\phi}^{u}\big)_{33} = 0$ direction, meaning that $\lambda_\phi$ can only be constrained whenever $\big(y_{\phi}^u\big)_{33} \ne 0$. This feature can be traced back to the tree-level matching relations in Eq.~\eqref{eq:matching_result_example_tree} with $\big(y_{\phi}^{d,e}\big)_{33}=0$ and the fact that there is no observable sensitive to  the $c_{\varphi}$
operator in the {\sc SMEFiT} dataset as well as the fact that the data does not prefer a non-zero $\big(y_{\phi}^{u}\big)_{33}$.
As we discuss below, this flat direction
is lifted once one-loop corrections to the matching
relations are accounted for.
As opposed to the UV-invariants in Table~\ref{tab:cl-heavy-scalar-fermion}, for this model the constrained UV-invariant is not positive-definite.

\begin{figure}[t]
    \centering
    \includegraphics[width=\linewidth]{figures/Granada_posteriors_HO_vs_NHO_NLO_scalars.pdf}
    \caption{Posterior distributions
    associated to the UV-invariants in the one-particle heavy scalar
 models listed in the upper part of Table~\ref{tab:cl-heavy-scalar-fermion}
 and
   obtained from tree-level matching.
   Note that all UV-invariants
   for the considered models
   are positive-definite. 
   We
   compare the
   results
   based on linear
   and quadratic corrections 
   in the EFT expansion, in both cases
   with NLO QCD accuracy.
   To facilitate visualisation, the posterior distributions are cut after they have dropped to $5\%$ of its maximum, though
   this does not affect the calculation
   of the bounds listed in Table~\ref{tab:cl-heavy-scalar-fermion}.
   }\label{fig:heavy_scalar_quad}
\end{figure}

\begin{figure}[t]
    \centering
    \includegraphics[width=\linewidth]{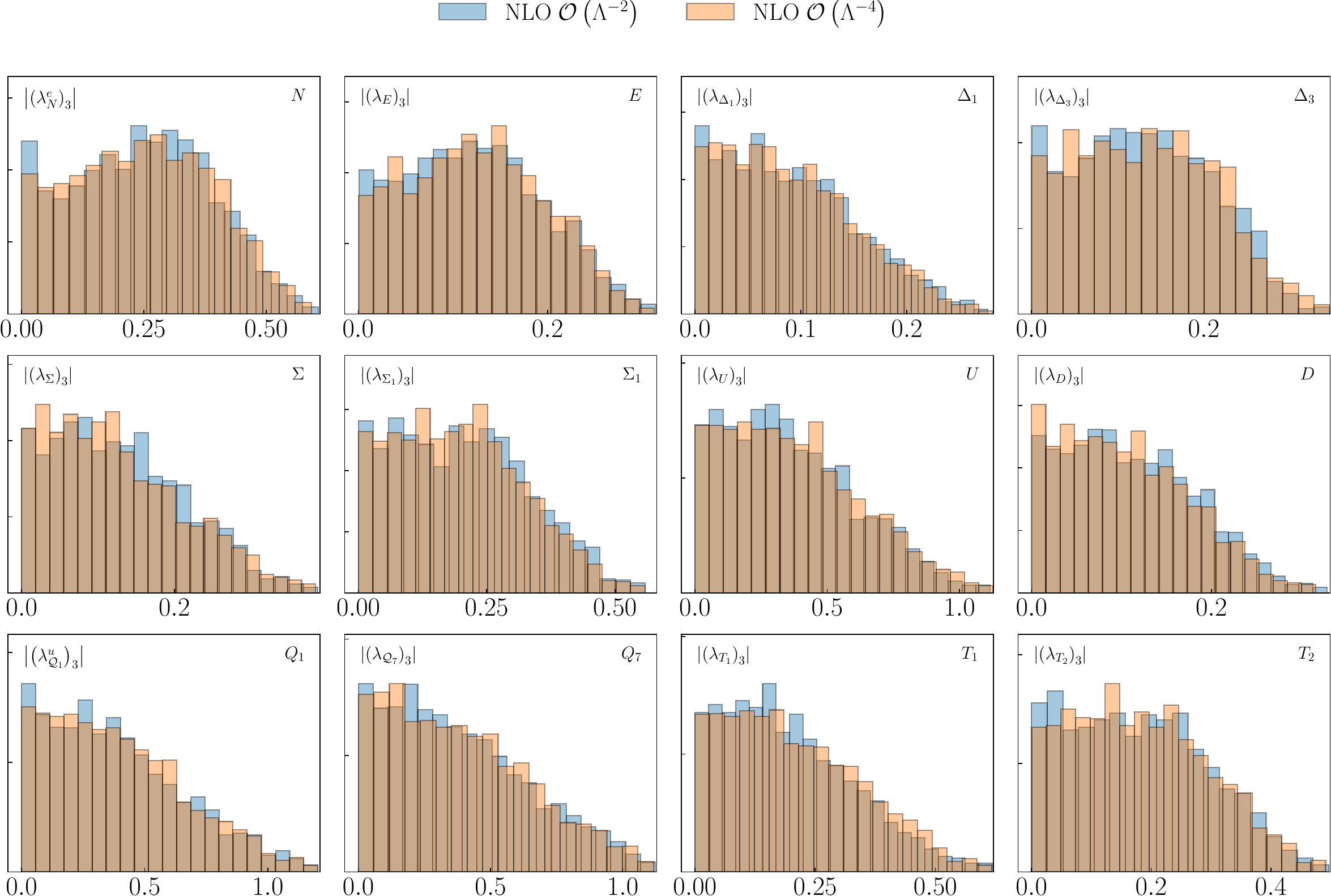}
    \caption{Same as
    Fig.~\ref{fig:heavy_scalar_quad}
    for the one-particle models
    composed of heavy vector-like
    fermions}
    \label{fig:heavy_vfermion_nlo}
\end{figure}

\paragraph{Heavy fermions.}
Following the discussion
of the results for the single-particle
BSM extensions with heavy scalars,
we move to the corresponding
analysis involving the heavy vector-like fermions listed in Table~\ref{tab:UV_particles}.
We provide their 95\% CL bounds
in the lower part of Table~\ref{tab:cl-heavy-scalar-fermion}, with the corresponding posterior distributions shown in Fig.~\ref{fig:heavy_vfermion_nlo}. 
All the (positive-definite) UV-invariants
can be constrained from the fit, leaving no flat
directions, and are consistent
with the SM hypothesis.
One observes how the constraints
achieved in all the heavy fermion models are in general
similar, with differences no larger
than a factor 4
depending on the specific gauge representation. Additionally, all the bounds are $\mcO(0.1)$, indicating that current data probes weakly-coupled fermions with masses around $1$~TeV.

For these heavy fermion models, 
differences between fits carried
out at the linear and quadratic levels
in the EFT expansion are minimal. 
These reason is that these UV scenarios
are largely constrained by the precisely measured EWPOs from LEP and SLC,
composed by processes for which quadratic EFT corrections
are very small~\cite{ewpos}.
The same considerations apply
to the stability with respect to higher-order
QCD corrections, which are negligible
for the EWPOs.

\begin{figure}[t]
    \centering
    \includegraphics[width=\linewidth]{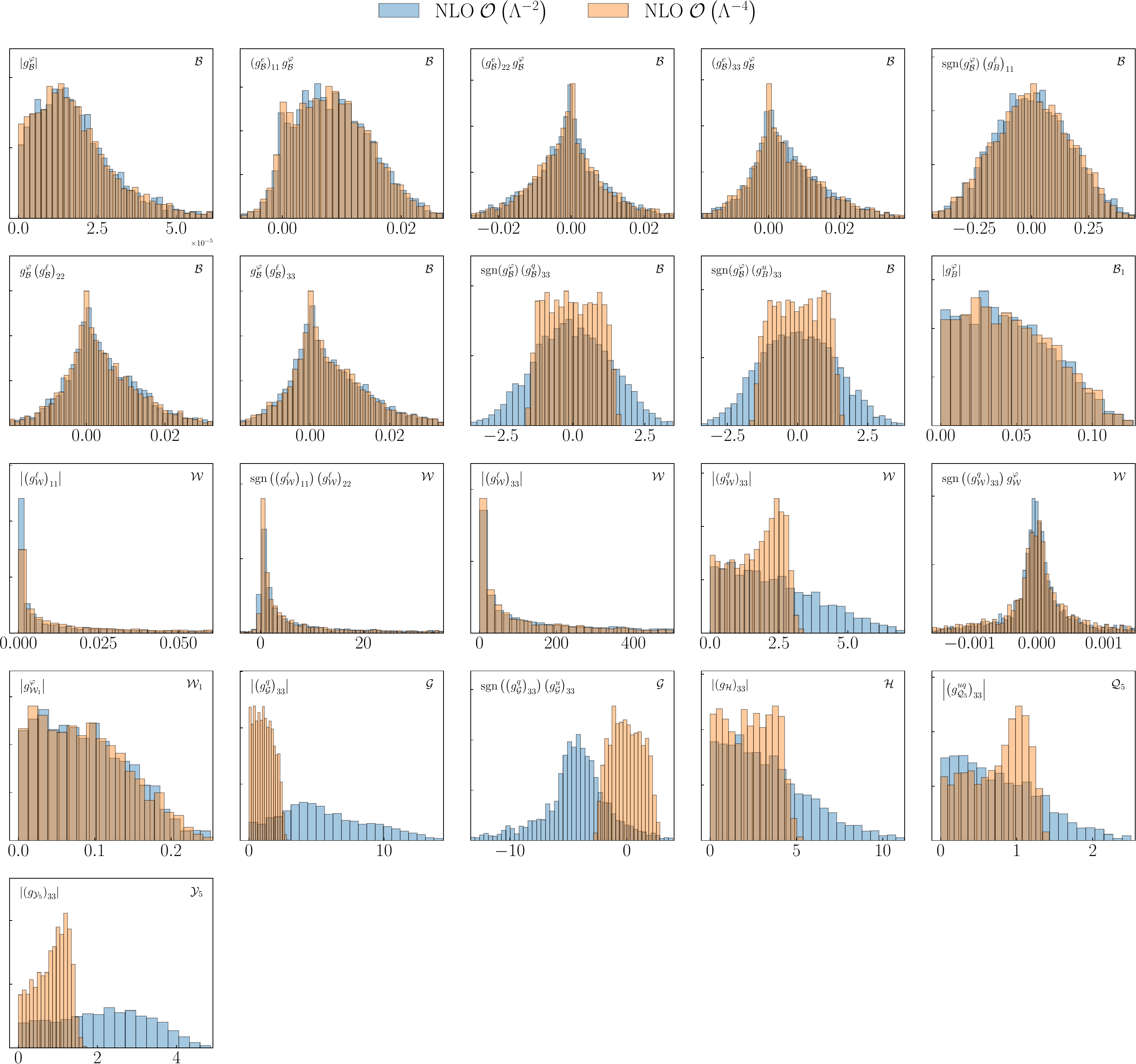}
    \caption{Same as
    Fig.~\ref{fig:heavy_scalar_quad}
    for the one-particle models
    composed of heavy vector-like
    bosons}
    \label{fig:heavy_vbosons_nlo}
\end{figure}

\paragraph{Heavy vector bosons.}
The last category of single-particle models
listed in Table~\ref{tab:UV_particles}
is the one composed of heavy vector bosons.
We provide their
$95\%$ marginalised CL intervals in Table~\ref{tab:cl-heavy-boson}, with the corresponding posterior distributions in Fig.~\ref{fig:heavy_vbosons_nlo} in which we compare linear and quadratic EFT corrections at NLO QCD.

 A first observation is that most of the heavy vector models depend on multiple UV-invariants. To be specific, the $\mathcal{B}$, $\mathcal{W}$ and $\mathcal{G}$ vector models have associated 9, 5 and 2 UV-invariants respectively, while the other vector models are characterised by a single UV-invariant. For all models considered, results are consistent with the SM scenario, corresponding to vanishing UV parameters. 
 
Regarding model $\mathcal{B}$, its results can be understood as follows. First of all, we observe a strong increase in sensitivity at the quadratic level in the last two invariants, $\mathrm{sgn}\lp g_{\mathcal{B}}^\varphi\rp \lp g_{\mathcal{B}}^q\rp_{33}$ and $\mathrm{sgn}\lp g_{\mathcal{B}}^\varphi\rp \lp g_{\mathcal{B}}^u\rp_{33}$. Indeed, they generate the four-heavy operators $c_{tt}^{(1)}$ and $c_{QQ}^{(1)}$, respectively, which are characterised by large quadratic corrections. 
The other UV-invariants in the $\mathcal{B}$ model generate operators sensitive to the EWPOs which are strongly constrained from LEP data leading to strong bounds in all cases. For example, the operators  $c_{\varphi\Box}$ and $c_{\varphi D}$ are generated by $g_{\mathcal{B}}^\varphi$,
\be
\frac{c_{\varphi\Box}}{\Lambda^2} = \frac{1}{2}\frac{\lp g_{\mathcal{B}}^\varphi\rp^2}{m_{\mathcal{B}}^2} \, ,\qquad  \frac{ c_{\varphi D} }{\Lambda^2} = -2 \frac{\lp g_{\mathcal{B}}^\varphi\rp^2}{m_{\mathcal{B}}^2}\, ,
\ee
and thus provide strong bounds on the invariant $\left|g_{\mathcal{B}}^\varphi\right|$. Furthermore, the invariant $\lp g_{\mathcal{B}}^e\rp_{ii} g_{\mathcal{B}}^\varphi$ is sensitive to the leptonic Yukawa operators $c_{\varphi (e, \mu, \tau)}$ and therefore gets well constrained by LEP data as well. Yet another example related to the EWPOs is the invariant $g_{\mathcal{B}}^\varphi \lp g_{\mathcal{B}}^\ell\rp_{ii}$ that generates the operators $c_{\varphi\ell_2}$ and $c_{\varphi\ell_3}$ for $i=1,2$, respectively. Finally, $\lp g_{\mathcal{B}}^\ell\rp_{11}$ generates $\lp c_{\ell\ell}\rp_{1111}$ and thus gets constrained by Bhabha scattering. For this model, we have chosen the UV-invariants such that they agree with the constrained directions. Had we built them in the same way that for other models, it would be explicit that there are 5 poorly constrained directions along $|\big(g_{\mathcal{B}}^{e}\big)_{11}|$, $|\big(g_{\mathcal{B}}^{e}\big)_{22}|$, $|\big(g_{\mathcal{B}}^{e}\big)_{33}|$, $|\big(g_{\mathcal{B}}^{\ell}\big)_{22}|$, and $|\big(g_{\mathcal{B}}^{\ell}\big)_{33}|$. The model $\mathcal{B}_1$ is only sensitive to operators that can be constrained via EWPOs, hence no improved sensitivity is observed after adding quadratic EFT effects.

Moving on to model $\mathcal{W}$, we observe two flat directions along $\lp g_{\mathcal{W}}^{\ell}\rp_{33}$ and $\mathrm{sgn}\left(\left(g_{\mathcal{W}}^\ell\right)_{11}\right)\left(g_{\mathcal{W}}^\ell\right)_{22}$. In the matching relations, the UV coupling $\lp g_{\mathcal{W}}^{\ell}\rp_{33}$ enters exclusively in a product together with $g_{\mathcal{W}}^\varphi$. Now, $g_{\mathcal{W}}^\varphi$ already gets strongly constrained via other independent relations to $c_{\varphi\Box}, c_{b\varphi}, c_{t \varphi}$ and $c_{\tau, \varphi}$. As a result, the UV invariant $\left|\lp g_{\mathcal{W}}^{\ell}\rp_{33}\right|$ is left essentially unconstrained as no other matching relation exists to disentangle $ \lp g_{\mathcal{W}}^{\ell}\rp_{33}$ from $g_{\mathcal{W}}^\varphi$.
A similar argument holds for the second flat direction we observe along $\mathrm{sgn}\left(\left(g_{\mathcal{W}}^\ell\right)_{11}\right)\left(g_{\mathcal{W}}^\ell\right)_{22}$. The UV parameter $\lp g_{\mathcal{W}}\rp_{22}$ only enters as a product with either $\lp g_W^\ell\rp_{11}$ or $g_W^\varphi$, both of which already get constrained via other independent matching relations, e.g. $\lp c_{\ell\ell}\rp_{1111} \sim \lp g_{\mathcal{W}}^\ell\rp^2$ and the aforementioned reason in case of $g_{\mathcal{W}}^\varphi$. In fact, these bounds can be considered meaningless since they are of the order or above the perturbative limit $4\pi$ and well in excess of more refined perturbative unitarity bounds~\cite{Barducci:2023lqx}. 
The four-heavy operators $c_{QQ}^{(1, 8)}$ are responsible for the increased sensitivity we observe in $\lp g_{\mathcal{W}}^q\rp_{33}$ after including quadratic EFT corrections.

Finally, concerning the models $\mathcal{G}$, $\mathcal{H}$, $\mathcal{Q}_5$ and $\mathcal{Y}_5$, we observe a significant tightening of the bounds after quadratic EFT corrections are accounted for. This is entirely due to their sensitivity to the four-heavy operators, as can be seen in Table \ref{tab:heavy_scalar_vboson_sens}.

\begin{table}[t]
\begin{center}
\scriptsize

\renewcommand{\arraystretch}{1.6}\begin{tabular}{c|c|c|c|c|c}
Model & UV invariants & LO $\mathcal{O}\left(\Lambda^{-2}\right)$ &LO $\mathcal{O}\left(\Lambda^{-4}\right)$ & NLO $\mathcal{O}\left(\Lambda^{-2}\right)$ & NLO $\mathcal{O}\left(\Lambda^{-4}\right)$ \\
\toprule
\multirow{9}{*}{$\mathcal{B}$} & $\left|g_{\mathcal{B}}^\varphi\right|$ & [0, $4.4\cdot 10^{-5}$] & [0, $4.4\cdot 10^{-5}$] & [0, $4.6\cdot 10^{-5}$] & [0, $4.6\cdot 10^{-5}$] \\
 & $\left(g_{\mathcal{B}}^e\right)_{11} g_{\mathcal{B}}^\varphi$ & [-0.0033, 0.021] & [-0.0028, 0.021] & [-0.0022, 0.022] & [-0.0032, 0.021]  \\
 & $\left(g_{\mathcal{B}}^e\right)_{22} g_{\mathcal{B}}^\varphi$ & [-0.026, 0.022] & [-0.026, 0.022] & [-0.025, 0.025] & [-0.025, 0.025]  \\
 & $\left(g_{\mathcal{B}}^e\right)_{33} g_{\mathcal{B}}^\varphi$ & [-0.015, 0.030] & [-0.015, 0.029] & [-0.016, 0.029] & [-0.014, 0.030]  \\
 & $\mathrm{sgn}(g_{\mathcal{B}}^\varphi) \left(g_B^\ell\right)_{11}$ & [-0.32, 0.33] & [-0.33, 0.32] & [-0.32, 0.33] & [-0.32, 0.30]  \\
 & $g_{\mathcal{B}}^\varphi \left(g_{\mathcal{B}}^\ell \right)_{22}$ & [-0.016, 0.027] & [-0.016, 0.027] & [-0.016, 0.028] & [-0.016, 0.029]  \\
 & $g_{\mathcal{B}}^\varphi \left(g_{\mathcal{B}}^\ell \right)_{33}$ & [-0.014, 0.028] & [-0.015, 0.027] & [-0.014, 0.029] & [-0.015, 0.028]  \\
 & $\mathrm{sgn}(g_\mathcal{B}^\varphi)\left(g_\mathcal{B}^q\right)_{33}$ & [-3.1, 3.0] & [-1.4, 1.5] & [-2.4, 2.6] & [-1.4, 1.4]  \\
 & $\mathrm{sgn}(g_\mathcal{B}^\varphi)\left(g_B^u\right)_{33}$ & [-3.2, 2.9] & [-1.5, 1.4] & [-2.5, 2.6] & [-1.4, 1.4]  \\
 \midrule
$\mathcal{B}_1$ & $\left|g_B^\varphi\right|$ & [0, 0.099] & [0, 0.098] & [0, 0.098] & [0, 0.099] \\
\midrule
\multirow{5}{*}{$\mathcal{W}$}& $\left|\left(g_{\mathcal{W}}^\ell\right)_{11}\right|$ & [0, 0.33] & [0, 0.32] & [0, 0.28] & [0, 0.34] \\
 & $\mathrm{sgn}\left(\left(g_{\mathcal{W}}^\ell\right)_{11}\right)\left(g_{\mathcal{W}}^\ell\right)_{22}$ & [-34, $4.6\cdot 10^{2}$] & [-20, $5.3 \cdot 10^{2}$] & [-17, $7.0\cdot 10^{2}$] & [-21, $3.6\cdot 10^{1}$]  \\
 & $\left|\left(g_{\mathcal{W}}^\ell\right)_{33}\right|$ & [0, $7.9\cdot 10^{2}$] & [0, $7.8\cdot 10^{2}$] & [0, $8.0\cdot 10^{2}$] & [0, $7.7\cdot 10^{2}$]  \\
 & $\left|\left(g_{\mathcal{W}}^q\right)_{33}\right|$ & [0, 6.4] & [0, 3.1] & [0, 5.6] & [0, 2.9]  \\
 & $\mathrm{sgn}\left(\left(g_{\mathcal{W}}^q\right)_{33}\right)g_{\mathcal{W}}^\varphi$ & [-0.022, 0.019] & [-0.024, 0.017] & [-0.026, 0.020] & [-0.024, 0.020]  \\
\midrule
\multirow{2}{*}{$\mathcal{G}$}& $\left|\left(g_{\mathcal{G}}^q\right)_{33}\right|$ & [0, 12] & [0, 2.5] & [0, 12] & [0, 2.3] \\
 & $\mathrm{sgn}\left(\left(g_{\mathcal{G}}^q\right)_{33}\right)\left(g_{\mathcal{G}}^u\right)_{33}$ & [-11, 1.8] & [-2.3, 2.6] & [-11, 2.2] & [-2.2, 2.4]  \\
 \midrule
$\mathcal{H}$ & $\left|\left(g_{\mathcal{H}}\right)_{33}\right|$ & [0, 8.0] & [0, 4.3] & [0, 8.0] & [0, 4.4] \\
\midrule
$\mathcal{Q}_5$ & $\left|\left(g_{\mathcal{Q}_5}^{uq}\right)_{33}\right|$ & [0, 2.1] & [0.026, 1.4] & [0, 1.8] & [0, 1.3] \\
\midrule
$\mathcal{Y}_5$ & $\left|\left(g_{\mathcal{Y}_5}\right)_{33}\right|$ & [0.046, 4.5] & [0.031, 1.6] & [0, 4.0] & [0.053, 1.4] \\
\bottomrule
\end{tabular}
\end{center}
\caption{Same as Table~\ref{tab:cl-heavy-scalar-fermion}
for the UV invariants associated
to the one-particle heavy vector boson extensions
}
\label{tab:cl-heavy-boson}
\end{table}


\paragraph{Comparison to the {\sc\small Fitmaker} results.}
The global SMEFT analysis carried
out in~\cite{Ellis:2020unq} by the
{\sc\small Fitmaker} collaboration
also provided bounds for selected single-particle
extensions of the SM obtained
from tree-level matching. 
Specifically, they provide
results for  21 different models 
including heavy scalars, fermions, and vector bosons.
These UV models were highlighted in red
in Tables~\ref{tab:heavy_scalar_vboson_sens}
and~\ref{tab:heavy_vfermion_sens}.

As further discussed in App.~\ref{app:UV_details}, the {\sc\small Fitmaker} models are assumed to couple in a flavour-universal way to the SM fields.
This flavour assumption is inconsistent
with the corresponding one adopted by {\sc SMEFiT} and leads in most models to the generation of  non-vanishing EFT
coefficients which are not part of our baseline fitting basis. 
Furthermore, in some scenarios,
like in the heavy scalar $\phi$ model, 
they lead to coefficients with flavour indices breaking the {\sc SMEFiT} flavour symmetries. 
For this reason, in order to benchmark
our results for the single-particle BSM
extensions with those of~\cite{Ellis:2020unq}, we adopt their same matching relations and ignore the effects of EFT coefficients not included in our fit.

\begin{figure}[t]
    \centering
    \includegraphics[width=0.99\textwidth]{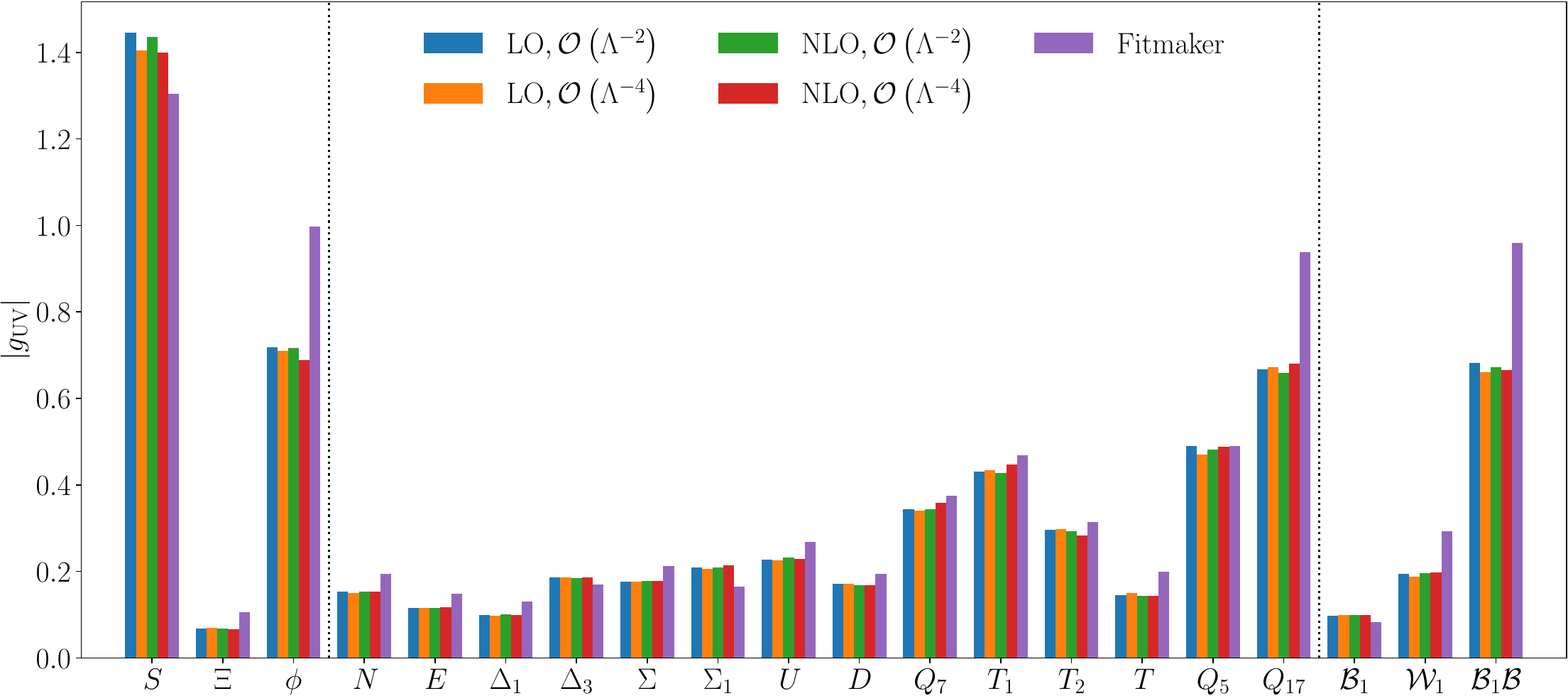}
    \caption{Upper $95\%$ CL bounds on the absolute value of the UV couplings $g_{\rm UV}$ obtained for the single-particle extensions 
    considered in~\cite{Ellis:2020unq}
    compared to the results of this work. 
    A heavy particle mass of $m_{\rm UV}=1$ TeV
    is assumed. 
    We display sequentially the results
    for the scalar, fermion, and vector particles separated by vertically dashed lines.
    }
    \label{fig:fitmaker_vs_smefit_all}
\end{figure}

We provide in Fig.~\ref{fig:fitmaker_vs_smefit_all}
a comparison between
the upper $95\%$ CL bounds on the absolute value of the UV couplings $g_{\rm UV}$  for the single-particle extensions  obtained in~\cite{Ellis:2020unq}
with the results of this work, where  a heavy particle mass of $m_{\rm UV}=1$ TeV
    is assumed for all models. 
We find satisfactory agreement for most of the models considered, with residual differences
between the {\sc\small FitMaker} bounds
(based on a linear EFT fit at LO)
and our own ones with the same theory settings explained by different choices
of the input dataset and of their statistical
treatment.

Non-negligible differences are
instead observed for the heavy
scalar model $\phi$, the heavy
fermion model $Q_{17}$,
and the heavy vector models
$\mathcal{W}_1$ and $\mathcal{B}_1\mathcal{B}$.
Differences for the $\phi$ model
are explained by the fact that the
{\sc\small SMEFiT} basis includes four-heavy operators generated by this model, such
as $c_{Qt}^1$ and $c_{Qt}^8$ (Table~\ref{tab:heavy_scalar_vboson_sens}), constrained
by $t\bar{t}b\bar{b}$ 
and $t\bar{t}t\bar{t}$ cross-sections 
and which
are absent from the  {\sc\small Fitmaker} analysis.
Differences for the fermionic model $Q_{17}$ 
as well as for the  heavy vector models
$\mathcal{W}_1$ and $\mathcal{B}_1\mathcal{B}$ originate from the inclusion of different Higgs production datasets in the fit.

From the comparison in Fig.~\ref{fig:fitmaker_vs_smefit_all}, one
further observes that the change in sensitivity after including either NLO QCD or quadratic EFT corrections is mild, indicating that for these models the dominant sensitivity on the UV couplings
is already provided by the
linear EFT predictions with LO accuracy. 
In this context, one should note that a $20\%$ improvement at the level of the Wilson coefficients $c_i$ corresponds to only a $10\%$ enhancement at the level of the UV parameters $g_{\mathrm{UV}}$ given that the typical matching relations have the form $c_i \sim g_{\mathrm{UV}}^2$.

All in all, we conclude that this comparison with {\sc\small FitMaker}
is successful and provides a non-trivial
validation of our new pipeline
enabling direct fits of UV coupligs 
within the {\sc\small SMEFiT}
framework.

\subsection{Multi-particle models matched at tree level}

Moving beyond one-particle models,
we now study  UV-completions
of the SM which include multiple heavy
particles.
Specifically, we consider a UV model 
which includes two heavy vector-like
fermions, $Q_1$ and $Q_7$,
and a heavy vector-boson, $\mathcal{W}$, see Table~\ref{tab:UV_particles} for their
quantum numbers and gauge group
representation.
In  the case of equal masses and couplings
for the two heavy fermions, this model
satisfies
custodial symmetry~\cite{Marzocca:2020jze}.
The two heavy fermions
$Q_1$ and $Q_7$ generate the same two operators, namely $c_{t\varphi}$ and $c_{\varphi t}$. 
A contribution to the top Yukawa operator $c_{t\varphi}$ is also generated by 
the heavy vector-boson $\mathcal{W}$, introducing an interesting interplay between the quark bidoublets on the one hand and the neutral vector triplet on the other hand. 
As indicated in Table~\ref{tab:heavy_scalar_vboson_sens},
several other operators in addition
to $c_{t\varphi}$ are generated
when integrating out
the heavy vector boson $\mathcal{W}$.

It should be emphasized that
we make this choice of 
multi-particle model for illustrative
purposes as well as to compare
with the benchmark studies of~\cite{Marzocca:2020jze}, and that
results for any other combination
of the heavy BSM particles listed
in Table~\ref{tab:UV_particles} 
can easily be obtained within our approach.
The only limitations are 
that the number of UV couplings
must remain smaller than the number
of EFT coefficients entering the analysis,
and that the input data used in the global
fit exhibits sensitivity to the matched
UV model.

\begin{figure}[t]
    \centering
    \includegraphics[width=\linewidth]{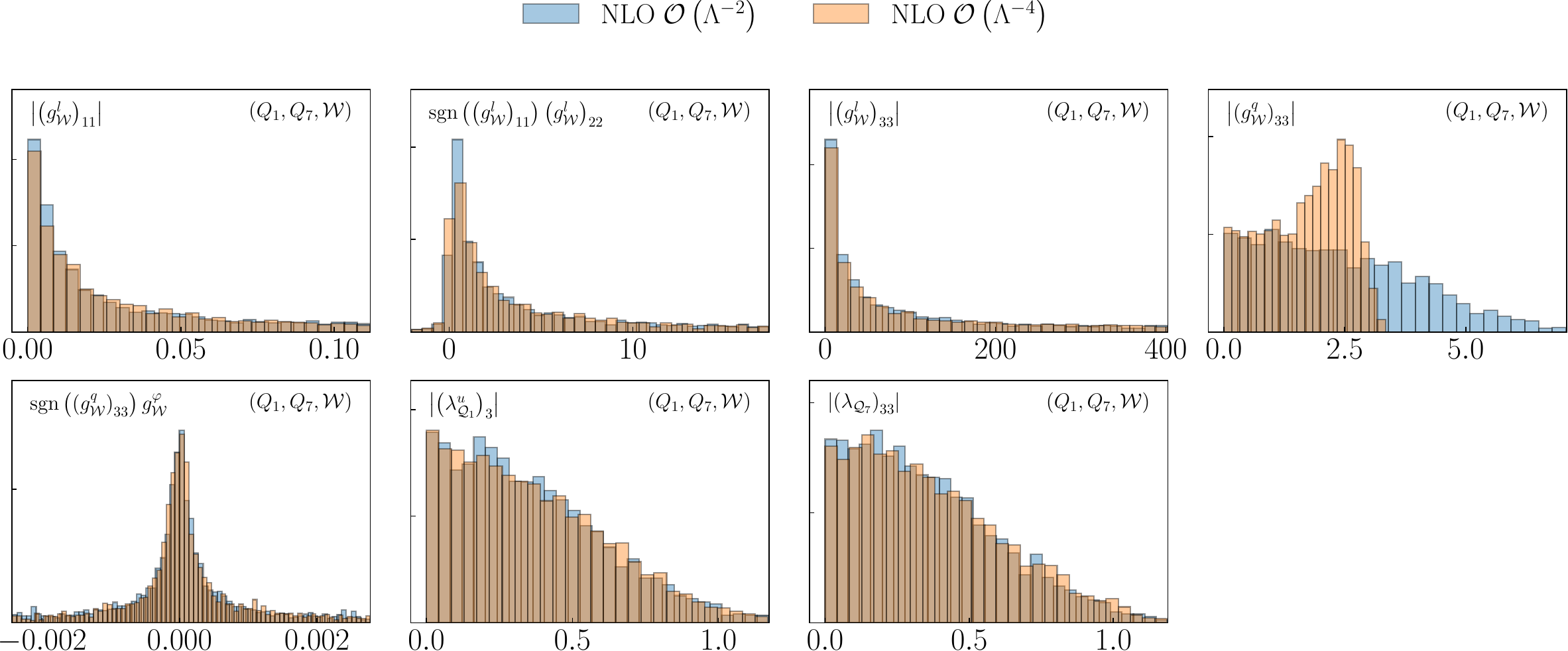}
    \caption{The posterior distributions of the UV invariants in the
three-particle model consisting
of two heavy fermions $Q_1,Q_7$ and a heavy
vector boson $\mathcal{W}$, see
also  see Table~\ref{tab:cl-mp}, 
comparing the impact of linear (blue) versus quadratic (orange) corrections in the EFT expansion.
}
    \label{fig:mp-posteriors}
\end{figure}

We provide in Table~\ref{tab:cl-mp} the $95\%$ CL intervals for the UV invariants associated to this three-particle model. A common
value of the heavy mass, 
 $m_{Q_1}=m_{Q_7}=
m_\mathcal{W} = 1$ TeV, is assumed.
As in the case of the one-particle models,
we compare results at the linear and quadratic EFT level and at LO and NLO in the QCD expansion.
The associated posterior distributions
for the UV invariants
comparing the impact of linear  versus quadratic EFT corrections are shown in Fig.~\ref{fig:mp-posteriors}.
On the one hand, one notices an  increase in sensitivity at the quadratic level in case of $(g_{\mathcal{W}}^{q})_{33}$, which is
consistent with the results of 
the one-particle model
analysis shown in Table~\ref{tab:cl-heavy-boson}. 
On the other hand, for some of the UV-invariants in the
model, such as $|(g_{\mathcal{W}}^l)_{11}|$,
the bounds become looser once quadratic EFT
corrections are accounted for, presumably due to the appearance of a second minimum
in the marginalised $\chi^2$ profiles.
For this specific model, it turns out that one has a quasi-flat direction
in the UV-invariant $\left|\left(g_{\mathcal{W}}^l\right)_{33}\right|$.

\begin{table}[t]
\begin{center}
\scriptsize
\renewcommand{\arraystretch}{1.5}\begin{tabular}{c|c|c|c|c|c}
\toprule
Model & UV invariants & LO $\mathcal{O}\left(\Lambda^{-2}\right)$ &LO $\mathcal{O}\left(\Lambda^{-4}\right)$ & NLO $\mathcal{O}\left(\Lambda^{-2}\right)$ & NLO $\mathcal{O}\left(\Lambda^{-4}\right)$ \\
\midrule
\multirow{7}{*}{$(Q_1,Q_7,\mathcal{W})$} & $\left|\left(g_{\mathcal{W}}^l\right)_{11}\right|$ & [0, 0.43] & [0, 0.31] & [0, 0.36] & [0, 0.28] \\
 & $\mathrm{sgn}\left(\left(g_{\mathcal{W}}^l\right)_{11}\right)\left(g_{\mathcal{W}}^l\right)_{22}$ & [-3.3, 40] & [-3.9, 49] & [-4.7, 53] & [-4.8, 49]  \\
 & $\left|\left(g_{\mathcal{W}}^l\right)_{33}\right|$ & [0, $7.7\cdot 10^{2}$] & [0, $7.9\cdot 10^{2}$] & [0, $7.4\cdot 10^{2}$] & [0, $7.6\cdot 10^{2}$]  \\
 & $\left|\left(g_{\mathcal{W}}^q\right)_{33}\right|$ & [0, 6.5] & [0.011, 3.1] & [0, 5.4] & [0.014, 2.9]  \\
 & $\mathrm{sgn}\left(\left(g_{\mathcal{W}}^q\right)_{33}\right)g_{\mathcal{W}}^\varphi$ & [-0.013, 0.012] & [-0.014, 0.019] & [-0.013, 0.0099] & [-0.014, 0.011]  \\
 & $\left|\left(\lambda_{\mathcal{Q}_1}^u\right)_3\right|$ & [0, 0.87] & [0, 0.88] & [0, 0.86] & [0, 0.86]  \\
 & $\left|\left(\lambda_{\mathcal{Q}_7}\right)_{33}\right|$ & [0, 0.88] & [0, 0.87] & [0, 0.84] & [0, 0.87]  \\
\bottomrule
\end{tabular}
\end{center}
\caption{Same as Table~\ref{tab:cl-heavy-scalar-fermion}
for the UV invariants associated
to the three-particle model consisting
of two heavy fermions $Q_1,Q_7$ and a heavy
vector boson $\mathcal{W}$ obtained from
tree-level matching.
A common value of the heavy mass is assumed for
the three particles, $m_{Q_1}=m_{Q_7}=
m_\mathcal{W} = 1$ TeV.
See Fig.~\ref{fig:mp-different-masses} for the corresponding
results in a scenario with $m_{Q_1} \ne m_{Q_7} \ne
m_\mathcal{W}$. 
For this model we have a flat direction
in the UV-invariant $\left|\left(g_{\mathcal{W}}^l\right)_{33}\right|$.
}
\label{tab:cl-mp}
\end{table}

\begin{figure}[t]
    \centering
    \includegraphics[width=\linewidth]{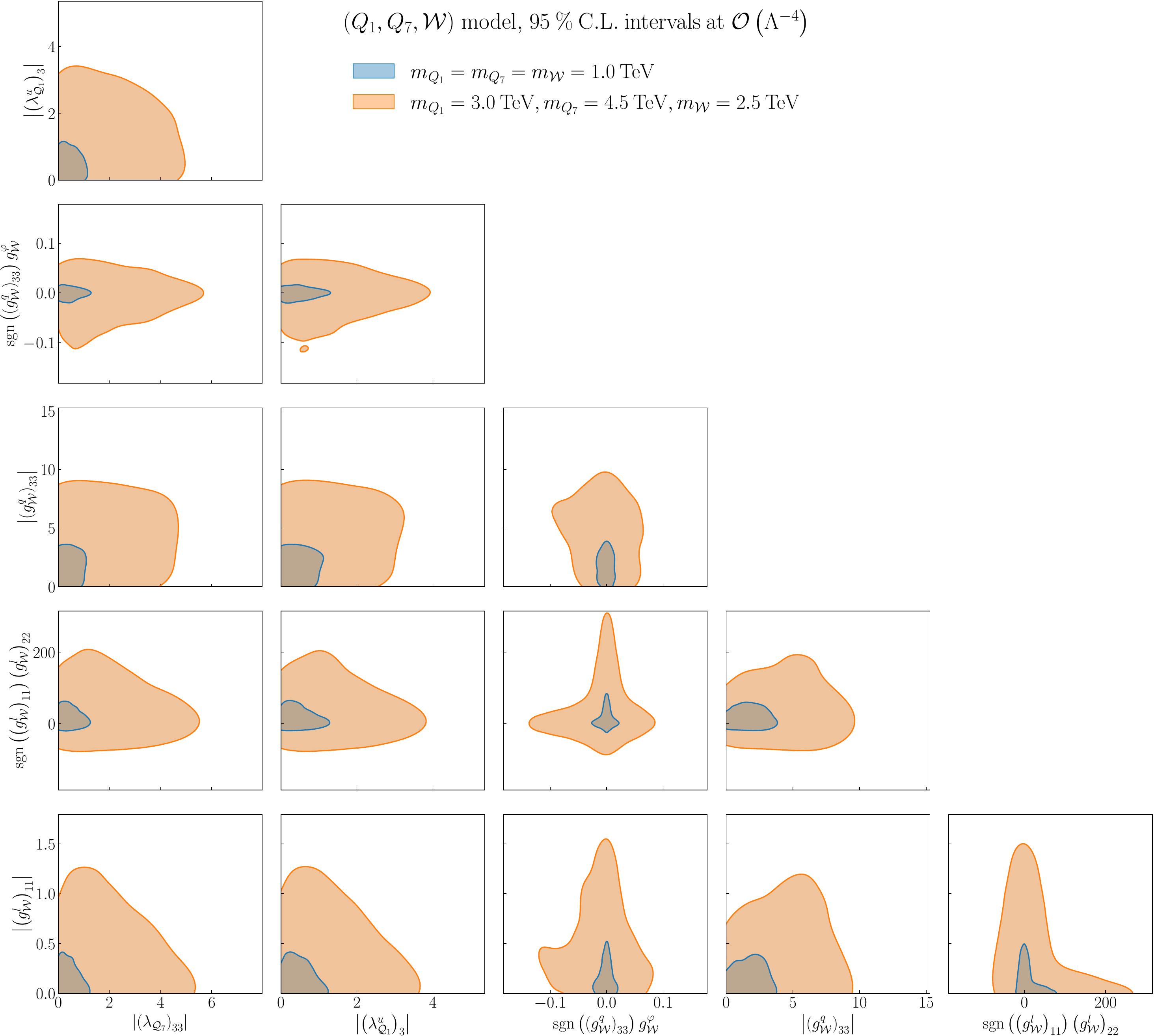}
    \caption{Pair-wise marginalised 
95\% contours for the UV invariants
associated
to the three-particle model consisting
of two heavy fermions, $Q_1$ and $Q_7$, and the heavy
vector boson $\mathcal{W}$, see also Fig.~\ref{fig:mp-posteriors} and Table~\ref{tab:cl-mp}
for the corresponding single-invariant posterior distributions and 95\% CL bounds.
We compare results for a common
value of the heavy mass, $m_{Q_1}=m_{Q_7}=
m_\mathcal{W} = 1$ TeV, with another
scenario
with different heavy masses,
$m_{Q_1}=3$ TeV, $m_{Q_7}=4.5$ TeV,
and $m_\mathcal{W}=2.5$ TeV. 
The EFT fit is carried out at NLO QCD
with $\mathcal{O}\lp \Lambda^{-4}\rp$
corrections included.
As compared to the list of invariants
in Table~\ref{tab:cl-mp}, we exclude
$\left|\left(g_{\mathcal{W}}^l\right)_{33}\right|$ given that it is essentially
unconstrained from the fitted data.
}
    \label{fig:mp-different-masses}
\end{figure}

\begin{figure}[t]
    \centering
    \includegraphics[width=\linewidth]{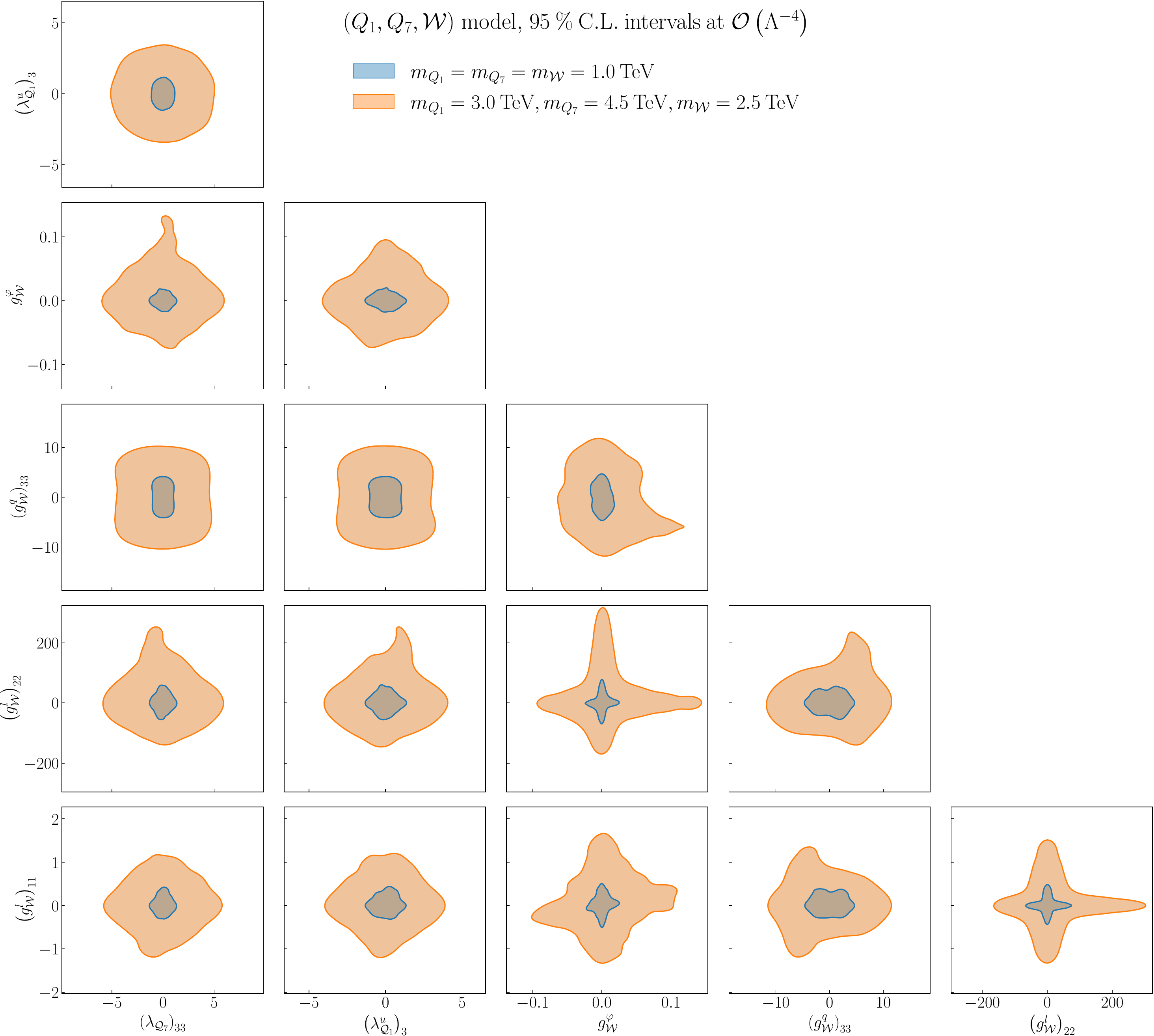}
    \caption{Same as Fig.~\ref{fig:mp-different-masses} in the case of
    the original UV couplings.
}
    \label{fig:mp-different-masses-uv-param}
\end{figure}

While the results
of Table~\ref{tab:cl-heavy-boson} assume
a common value of the heavy particle masses,
it is trivial to extend them
to different masses for each
of the different particles in the model.
This way, one can assess the dependence
of the UV-coupling fit results
on the assumptions of the heavy particle
masses.
With this motivation, Fig.~\ref{fig:mp-different-masses} and Fig.~\ref{fig:mp-different-masses-uv-param} display pair-wise marginalised 95\% contours for the relevant UV invariants and the original Lagrangian parameters in the model respectively, see Fig.~\ref{fig:mp-posteriors}
for the corresponding single-invariant posterior distributions. 
The baseline results with a common mass of
1 TeV are compared to 
a scenario
with three different heavy masses,
$m_{Q_1}=3$ TeV, $m_{Q_7}=4.5$ TeV,
and $m_\mathcal{W}=2.5$ TeV. 
We exclude
$\left|\left(g_{\mathcal{W}}^\ell\right)_{33}\right|$ from this comparison, given that it is essentially
unconstrained from the fitted data.

From the comparison in  Fig.~\ref{fig:mp-different-masses} one observes, as expected, that assuming heavier BSM particles
leads to looser constraints on the
UV couplings.
Taking into account that different terms in the matching relations scale with the heavy particle
masses in a different manner, it
is not possible in general to rescale
the bounds obtained from the equal mass
scenario to another with different heavy
masses.
Nevertheless, given that in a single-particle extension we know that bounds worsen by
a factor $\lp m_{\rm UV}^{*}/m_{\rm UV}\rp^2$
if the heavy particle mass is increased
from $m_{\rm UV}$ up to $m_{\rm UV}^*$,
one can expect that in this three-particle scenario
the bounds are degraded by an amount between $\sim 5$ and $\sim 20$ depending on the specific UV coupling, a estimate which is consistent with the results in Fig.~\ref{fig:mp-different-masses}.
This comparison also highlights how
for very heavy particles we lose all
sensitivity and the global EFT fit cannot provide competitive constraints 
on the UV model parameters.

\subsection{Single-particle models matched at one-loop}
\label{subsec:results_oneloop}
All results presented so far relied on tree-level matching relations. 
We now present results for UV-coupling fits
in the case of matching relations obtained at the one-loop level, and study their effect on the UV parameter space as compared
to results based on tree-level matching.
As discussed in Sect.~\ref{subsec:oneloop}, for
this purpose we can also deploy {\sc\small MatchMakerEFT}~interfaced to {\sc\small SMEFiT} via \matchTOfit~in order to obtain one-loop matching relations suitable for their use in the fit in an automated way.
%
In its current incarnation, this pipeline
enables using any
of the single-particle heavy scalar
and heavy fermion models (and combinations
thereof) listed
in Table~\ref{tab:UV_particles} 
when matched at the one-loop level.
We note here that the automation of the one-loop matching for  heavy vector bosons has not been 
achieved yet.
For concreteness, here
we present results based on the one-loop matching of the heavy scalar $\phi$ defined
by Eq.~(\ref{eq:Lag_UV_full_Varphi_model}), and also discussed in Sect.~\ref{subsec:heavy-scalar}, and the heavy fermions $T_1$ and $T_2$. We compare its bounds to those previously obtained from the tree-level matching analysis. 

Table~\ref{tab:cl-phi} compares
the 95\% CL bounds obtained for the two UV invariants in the $\phi$ model following either
tree-level or one-loop matching relations,
with the associated marginalised posterior
distributions shown in Fig.~\ref{fig:tree_vs_loop}.
In contrast to the bounds obtained at tree-level, one-loop corrections enable lifting the flat direction along the $|\lambda_\phi|$ invariant.
This effect is a consequence of the operators $\mathcal{O}_{\varphi \Box}, \mathcal{O}_{b\varphi}$ and $\mathcal{O}_{t\varphi}$, which receive additional one-loop matching contributions resulting in independent constraints on $\lambda_\phi$.
Specifically, the one-loop matching relations
for the Wilson coefficients
associated to $\mathcal{O}_{\varphi \Box}$ and $\mathcal{O}_{t\varphi}$ read

\begin{align}
    \label{eq:matching-1loop-lambda}
    \frac{c_{\varphi\Box}}{\Lambda^2} = & -\frac{g_1^4}{7680\pi^2}\frac{1}{m_\phi^2}-\frac{g_2^4}{2560\pi^2}\frac{1}{m_\phi^2}-\frac{3}{32\pi^2} \frac{\lambda_{\phi}^2}{m_\phi^2}, \nonumber \\
    \frac{c_{t\varphi}}{\Lambda^2} = & -\frac{\lambda_\phi\, \big(y_{\phi}^{u}\big)_{33}}{m_\phi^2} - \frac{g_{2}^4 y_{t}^{\text {SM}}}{3840 \pi^2}\frac{1}{m_\phi^2} + \frac{y_{t}^{\text {SM}}}{16\pi^2 }\frac{\lambda_\phi^2 }{m_\phi^2} +\frac{(4 \left(y_{b}^{\text{SM}} \right)^2 - 13 \left(y_{t}^{\text{SM}}\right)^2 )}{64\pi^2}\frac{\lambda_\phi\, \big(y_{\phi}^{u}\big)_{33}}{m_\phi^2} \\
    & - \lp 12 \lambda_{\varphi}^{\text {SM}} + \left(y_{b}^{\text{SM}} \right)^2 - 11 \left(y_{t}^{\text{SM}} \right)^2 \rp \frac{y_{t}^{\text {SM}} }{64 \pi^2} \frac{\big(y_{\phi}^{u}\big)_{33}^2}{m_\phi^2} + \frac{3 }{128\pi^2}\frac{\lambda_\phi\, \big(y_{\phi}^{u}\big)_{33}^3}{m_\phi^2}\, ,\nonumber
\end{align}

\noindent
where $\lambda_{\varphi}^{\text{SM}}$ is the quartic Higgs self-coupling in the SM.
In Eq.~\eqref{eq:matching-1loop-lambda}, 
all terms with a factor $1/\pi^2$ 
arise from one-loop corrections, indicating that
$ c_{\varphi\Box}$ is entirely loop-generated
while for $ c_{t\varphi}$
the tree-level matching relation is simply
$c_{t\varphi}=-\lambda_\phi\, \big(y_{\phi}^{u}\big)_{33}/m_\phi^2$.
This additional  dependence on $\lambda_\phi$
arising from one-loop corrections
is hence responsible for closing the tree-level
flat direction.
We have also  checked that the bounds resulting from one-parameter SMEFT fits of the Wilson
coefficients $c_{\varphi\Box}$ and $c_{t\varphi}$ to the same datasets lead to similar bounds on $|\lambda_\phi|$ as compared to those reported in Table~\ref{tab:one_loop_matching_phi} after using 
the relations in Eq.~(\ref{eq:matching-1loop-lambda}) 

On the other hand, as a consequence of one-loop corrections to the matching relations
one also observes a degradation
in the sensitivity 
to the UV-invariant
$\mathrm{sgn}(\lambda_{\phi})\big(y_{\phi}^u\big)_{33}$.
The reason is that the parameter region
around  arbitrarily small values of
the coupling $\big(y_{\phi}^{u}\big)_{33}$ is disfavoured now, translating into a flattening of the distribution in $\big(y_{\phi}^{u}\big)_{33}$ as observed in the rightmost panels of Fig.~\ref{fig:tree_vs_loop}. 
These results showcase that one-loop corrections
bring in not only precision but also
accuracy, in that the stronger bound
on this UV-invariant at tree-level was a consequence of a flat direction which
is lifted as soon as perturbative
corrections are accounted for.

Interestingly, the middle panels of Fig.~\ref{fig:tree_vs_loop} also indicate the appearance of a double-peak structure in the distribution of $\mathrm{sgn}(\lambda_\phi) \big(y_{\phi}^{u}\big)_{33}$ at quadratic order in the EFT expansion which is absent in case of tree-level matching.
Such structure is associated to a second
minimum in the $\chi^2$ favouring
non-zero UV couplings, and may
be related to cancellations between
different terms in the matching relations.\footnote{Such cancellations may arise in EFT fits whenever linear and quadratic corrections become comparable.}
For one-loop matching relations
such as those displayed in Fig.~\ref{eq:matching-1loop-lambda},
the minimised figure of merit
will in general be a complicated higher-order
polynomial in the UV couplings, and
in particular 
in the presence of quadratic EFT
corrections, the $\chi^2(\mathbf{c})$
will be a quartic form of the Wilson
coefficients.
Therefore, for the specific case of 
the heavy scalar model, the $\chi^2(\mathbf{g})$ expressed in terms
of the UV couplings will include 
terms of up to $\mathcal{O}\lp \lambda_\phi^8\rp$
and $\mathcal{O}\lp \big( y_{\phi}^{u}\big)^{16}_{33}\rp$,
see Eqns~(\ref{eq:matching-1loop-lambda}) and~(\ref{eq:oneloop_logs_example}),
as well as the various cross-terms.
The minima structure 
of $\chi^2(\mathbf{g})$ can then only be resolved
by means of a numerical analysis
such as the one enabled by our strategy.

\begin{figure}
    \centering
    \includegraphics[width=.9\linewidth]{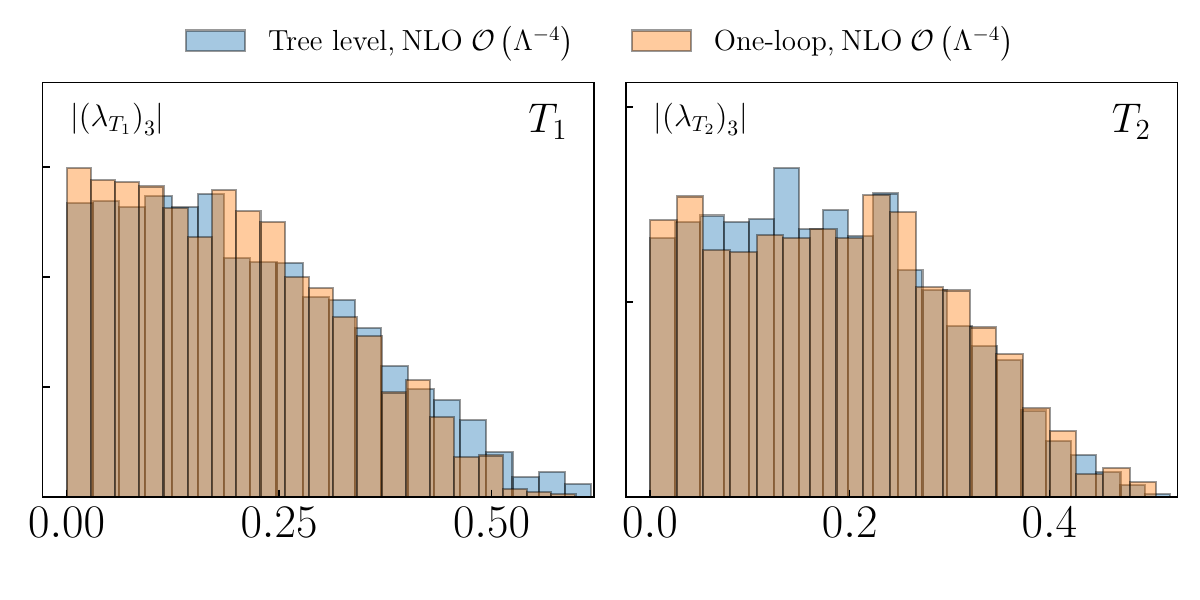}
    \caption{Posterior distributions associated to the UV-invariants in the one-particle heavy fermion models $T_1$ and $T_2$, comparing the effect of tree and one-loop level matching. 
    In both cases, we display results based on quadratic corrections in the EFT expansion and at NLO QCD accuracy.}
    \label{fig:tree_vs_loop_T12}
\end{figure}

Regarding heavy fermions matched at one-loop, we include in Fig.~\ref{fig:tree_vs_loop_T12} a comparison of the effect of one-loop matching as compared to only tree-level for the models $T_1$ and $T_2$. We notice a slightly increased sensitivity thanks to the one-loop matching effects, with this additional sensitivity entering through the extra operators generated at one-loop. Indeed, these two models generated only four operators at tree-level, while at one-loop they generate all but two of the operators in our fitting basis. However, since the main constraint on these models come from EWPOs, the improvement on the bounds is relatively small. We stress that this limited impact of one-loop matching is not a general result, but limited to these vector-like fermion models.

The above discussion with the models $\phi$, $T_1$ and $T_2$ as examples demonstrates that one-loop corrections to the SMEFT matching relations provide non-trivial information on the parameter space of the considered
UV models. 
 Loop corrections not only  modify the sensitivity to UV couplings already constrained by tree-level,  but 
can also lift flat or quasi-flat  directions.
This result further motivates
ongoing efforts to increase
the perturbative accuracy
of EFT matching relations. 

\section{Summary and outlook}
\label{sec:summary}
In this work we have presented and validated
a general strategy to scrutinize the parameters of BSM models  by using the SMEFT
as an intermediate bridge to the experimental
data.
By combining {\sc\small MatchMakerEFT} (for the matching between UV models and the EFT) with {\sc\small match2fit} (for the conversion of the matching results to fit run card) and {\sc\small SMEFiT} (for the comparison with data
and derivation of posterior distributions), this
pipeline enables to constrain the masses and
couplings of any UV model that can be matched
to the SMEFT either at tree-level or at one-loop.
While in this work we adopt {\sc\small MatchMakerEFT}, our approach is fully
general and any other matching framework could be adopted.

This flexible pipeline results
in the automation of the bound-setting
procedure on UV-complete models that can be matched to the SMEFT.
To illustrate its reach,
we apply it to derive constraints on a wide
range of single-particle
models extending the SM with heavy scalars,
fermions, or vector bosons with different
quantum numbers.
We also consider a three-particle
model which combines two heavy vector-like
fermions with a vector boson.
While most of the results presented arise
from tree-level matching relations, 
our approach applies equally well to one-loop matching results as shown by our results with one heavy scalar doublet model and two vector-like fermion models.

All the tools used in this analysis are 
made publicly available, ensuring the reproducibility
of our results and enabling their extension to 
analyse other user-defined BSM scenarios.
In addition to a new version of {\sc\small SMEFiT}
with the option of performing fits in the parameter space of UV couplings, we
release the Mathematica package \matchTOfit\,
whose goal is to streamline the interface
between EFT matching and fitting codes.
In its current incarnation, \matchTOfit\,
connects {\sc\small MatchMakerEFT}
with {\sc SMEFiT}, but could be extended
to other combinations in the future.
The availability of this pipeline opens up therefore the possibility of easily including SMEFT-derived constraints in upcoming model-building efforts.

Several directions could be considered for future work generalising the results presented here. An obvious one is the automation of the UV invariants computation at one-loop and testing our techniques with one-loop matched vector boson models, once the matching results for the latter class of models become available. The generalisation of the flavour symmetry assumed for the fit would also allow to fit more general models consistently.

As we have shown in the case
of the heavy scalar model,  one-loop matching
allows us to bound directions in the UV parameter space which are
flat after tree-level matching. This can be crucial in resolving degeneracies among models, which could be further helped by including  dimension-8 operators and their positivity bounds in the fit~\cite{Zhang:2021eeo}.
While a bottom-up determination of dimension-8 operators from data is hampered by their large number
in the absence of extra assumptions, specific UV models only generate a relatively small number
of dimension-8 operators, facilitating
their integration in the SMEFT fit. 

One could also consider accounting for RGE effects between the cutoff scale given by the UV model and the scale at which the global fit is performed. 
The latter could be combined with the inclusion of RGE effects in the cross sections entering the fit~\cite{Aoude:2022aro}.
We are currently working on the interfacing of \matchTOfit~with existing codes that solve the SMEFT RGE~\cite{Fuentes-Martin:2020zaz,DiNoi:2022ejg} and on the inclusion of the results in~\cite{Aoude:2022aro}.

Finally, it would be beneficial to consider
more flexible flavour symmetries which
are automatically consistent with the fitting
code, something which
would anyway be required for a full extension
of multi-particle models to one-loop matching.

All in all, our results provide valuable
insights on ongoing model-building efforts by using the 
SMEFT to connect them with the experimental data,
hence providing complementary information to that derived
from direct searches for new heavy particles at the
LHC and elsewhere.
The pipeline presented
in this work brings closer one of the original
goals of the EFT program at the LHC, namely
bypassing the need to directly compare
the predictions of UV-complete models
with the experimental measurements
and instead using the SMEFT
as the bridge connecting UV physics with data.

\section*{Acknowledgments}
We acknowledge useful discussions with J. Santiago, M. Chala, K. Mimasu, C. Severi, F. Maltoni, L. Mantani, T. Giani, J. Pagès, A. Thomsen, and Y. Oda.
A.~R. thanks the High Energy Theory Group of Universidad de Granada and Theory Group of Nikhef for their hospitality during early stages of this work.
The work of
A.~R. and E.~V. is supported by the European Research Council (ERC)
under the European Union’s Horizon 2020 research and innovation programme (Grant
agreement No. 949451)
and  by a Royal Society University Research Fellowship through grant URF/R1/201553.
The work of J.~t.~H. and G.~M is supported by the Dutch Research Council (NWO).
The work of J.~R. is supported by the Dutch Research Council (NWO) and by the Netherlands eScience Center.


\appendix

\section{Baseline SMEFiT global analysis}
\label{app:ewpos}

The {\sc\small SMEFiT} framework provides a flexible and robust open-source tool to constrain the SMEFT
parameter space using the information provided by experimental data from energy-frontier particle
colliders such as the LHC as well as from
lower-energy experiments.

 Given a statistical model defined by a likelihood
 $L(\boldsymbol{c})$, depending on  parameters $\boldsymbol{c}$ specifying the 
 theoretical predictions 
 for the observables entering the fit, and the
 corresponding dataset $\mathcal{D}$, {\sc\small SMEFiT}  infers the posterior distribution $P(\boldsymbol{c}|\mathcal{D})$ by means
 of sampling techniques~\cite{Feroz:2013hea, Feroz:2007kg}.
 {\sc\small SMEFiT} supports the inclusion of quadratic EFT corrections, can work up to NLO precision in the QCD expansion,
 admits general user-defined flavour assumptions and theoretical restrictions, and does
 not have limitations on the number or type of dimension-six operators that can be considered.

Theoretical restrictions on the EFT parameter space can be imposed as follows.
Given a set of $N$ fit coefficients $\{c_1,\dots, c_N\}$,
{\sc\small SMEFiT} allows the user to impose relations of the general form,
\be
\label{eq:constraints_coeffs_app}
\sum_{i} a_i \lp c_1\rp ^{n_{1,i}} \dots \lp c_N\rp ^{n_{N,i}}=0 \, ,
\ee
where the sum runs over all the possible combinations of the coefficients and $n_{j,i}$ are real numbers such that $\sum_{j} n_{j,i}$ is equal to a constant to preserve the correct dimensionality for all $i$. 
For instance, the relationship between the
Wilson coefficients in Eq.~(\ref{eq:rel_UV_tree_2}) arising from the 
heavy scalar model matched at tree-level can be
rewritten in the same format as Eq.~(\ref{eq:constraints_coeffs_app}), namely
\be
\label{eq:constraints_coeffs_app_2}
\lp c_{qd}^{(1)}\rp_{3333} \lp \lp c_{u\varphi}\rp_{33} \rp^2   - \lp c_{qu}^{(1)} \rp _{3333}  \lp  \lp c_{d\varphi}\rp_{33} \rp^2  = 0
\ee
These relations have to be imposed on a case-by-case basis via a run card and the user must declare which are the independent coefficients to be fitted. 
This functionality can be used to either relate the fit parameters among them,
as in Eq.~(\ref{eq:constraints_coeffs_app_2}), or to perform a mapping to a different
set of degrees of freedom.
The latter feature allows one to perform with {\sc\small SMEFiT} the fits on the UV model parameters as presented in this work.

As mentioned in Sect.~\ref{sec:smefit}, in comparison to~\cite{Giani:2023gfq}, which in turn updates the global SMEFT analysis
of Higgs, top quark, and diboson data presented in~\cite{Ethier:2021bye}, here we 
implement in {\sc\small SMEFiT} the information provided by the legacy electroweak
precision observables (EWPOs) from LEP and SLAC~\cite{ALEPH:2005ab} in an exact manner.
Previously, the constraints provided by these EWPOs in the SMEFT parameter space were accounted in an approximate manner, assuming EWPOs were measured with vanishing uncertainties.
This assumption introduced additional relations between a subset of the Wilson coefficients.
While such approximation served the purposes of interpreting LHC measurements, it is not appropriate
in the context of matching to UV-complete models given that  in many BSM scenarios the EWPOs
provide the dominant constraints~\cite{Ellis:2020unq}.
The new implementation of EWPOs in {\sc\small SMEFiT} will be discussed in
a separate publication~\cite{ewpos}
and here we provide
for completeness a short summary.

\begin{figure}[t]
    \centering
    \includegraphics[width=0.90\linewidth]{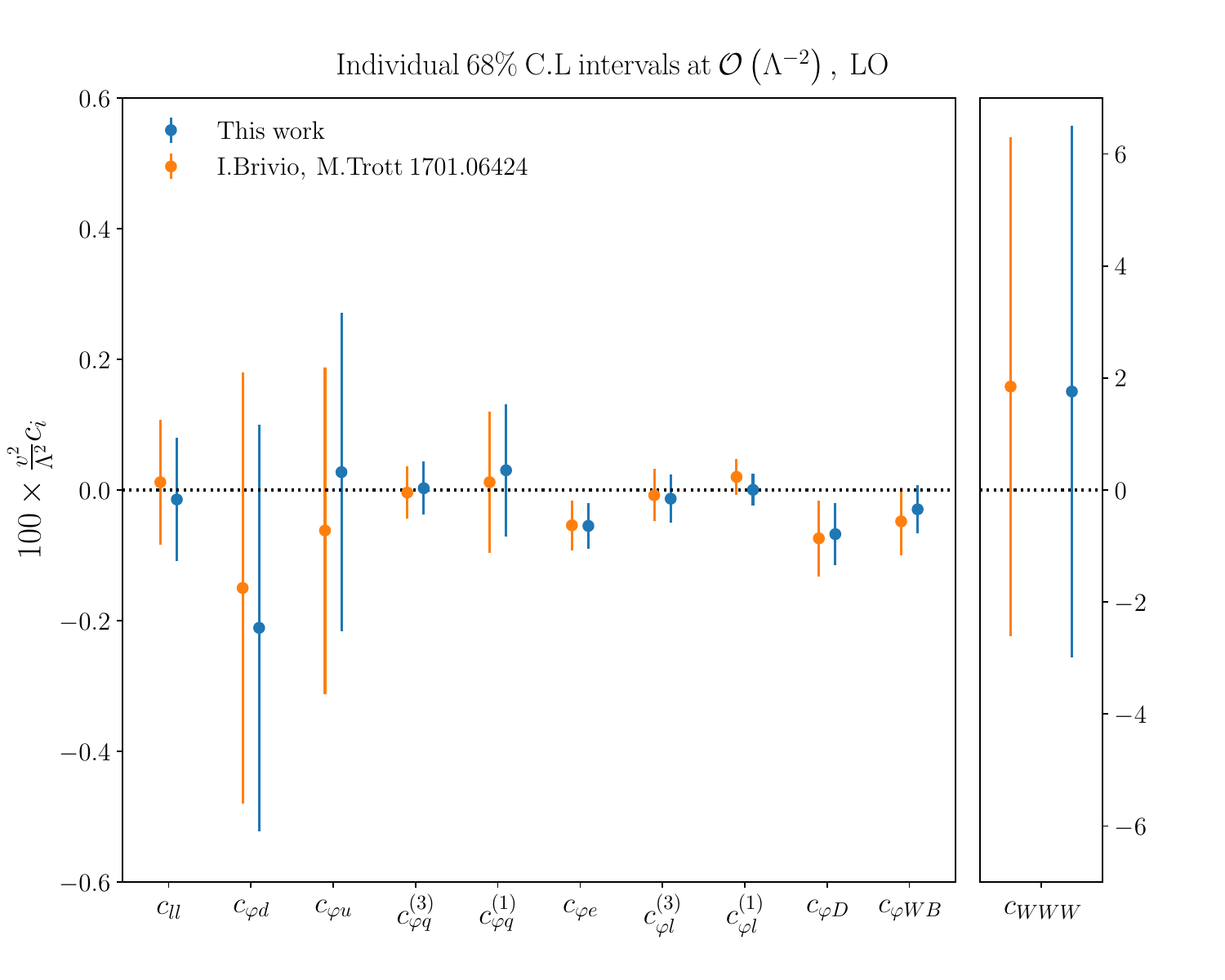}
    \caption{Comparison of a {\sc\small SMEFiT}-based analysis
of the electroweak precision observables from LEP against the corresponding analysis from~\cite{Brivio:2017bnu}
based on the same dataset.
We display bounds at the $68\%$ CL for one-parameter fits using linear EFT and LO calculations.
In this comparison, we adopt the convention of~\cite{Brivio:2017bnu} for the
triple gauge operator $O_{WWW}$, which differs from the SMEFT@NLO one by a negative sign. 
For consistency with the settings
adopted in~\cite{Brivio:2017bnu}, the triple
gauge operator $c_{WWW}$ has been fitted to a subset
of the available LEP data composed by only 4 bins.
}
    \label{fig:ewpo_benchmark}
\end{figure}

Experimental data and SM predictions
for the legacy EWPOs from LEP and SLAC are taken from~\cite{ALEPH:2005ab}.
%
Linear and quadratic SMEFT corrections to these observables arising from dimension-six operators have been computed with {\sc\small mg5\_aMC@NLO} interfaced to SMEFT@NLO.
Our implementation of the  EWPOs has been benchmarked by comparing with the results
of~\cite{Brivio:2017bnu}, assuming the same theoretical settings and choice of input dataset. 
Specifically, for this benchmarking exercise we use the $m_W$ electroweak input scheme, linear EFT corrections, LO cross-sections, and assume flavour universality.
Furthermore, we adopt the same convention on triple gauge operator $O_{WWW}$, which differs from SMEFT@NLO by a negative sign.
Fig.~\ref{fig:ewpo_benchmark} displays the bounds at the $68\%$ CL for one-parameter fits
in the two studies.
Good agreement between our implementation and that of~\cite{Brivio:2017bnu} is obtained for all operators.
Residual differences can be explained due to the inclusion of flavour-sensitive observables from LEP which are absent in~\cite{Brivio:2017bnu}.

Taking into account this updated EWPOs
implementation, when compared to the results of~\cite{Ethier:2021bye,Giani:2023gfq} the present global SMEFT analysis used
to constrain UV models
contains
$14$ additional Wilson coefficients $\mathbf{c}$ to be constrained from the fit for a total of 50 independent parameters.
The  new degrees of freedom are constrained not only by the EWPOs, but also top quark and diboson measurements.
For this reason,
we have recomputed EFT cross-sections for all the processes where such operators are relevant.
As in the original analysis~\cite{Ethier:2021bye}, we use {\sc\small mg5\_aMC@NLO}
interfaced to SMEFT@NLO to evaluate linear and quadratic EFT corrections which are augmented with  NLO QCD perturbative corrections whenever available.
To avoid overlap between datasets entering the PDF and EFT fits~\cite{Kassabov:2023hbm,Greljo:2021kvv}, we use NNPDF4.0 NNLO no-top~\cite{NNPDF:2021njg} as input PDF set.
The values of the input electroweak parameters are taken to be $m_W = 80.387$ GeV, $m_Z =  91.1876$ GeV and $G_F = 1.1663787\cdot 10^{-5}$ $\mathrm{GeV}^{-2}$.
Finally, the same flavour assumptions as in~\cite{Ethier:2021bye} are used i.e. a U(2)$_q\times$ U(2)$_u\times $U(3)$_d\times [$U(1)$_\ell\times $U(1)$_e]^3$ symmetry in the quark and lepton sectors. 

\begin{figure}[htbp]
    \centering
\includegraphics[width=0.88\linewidth]{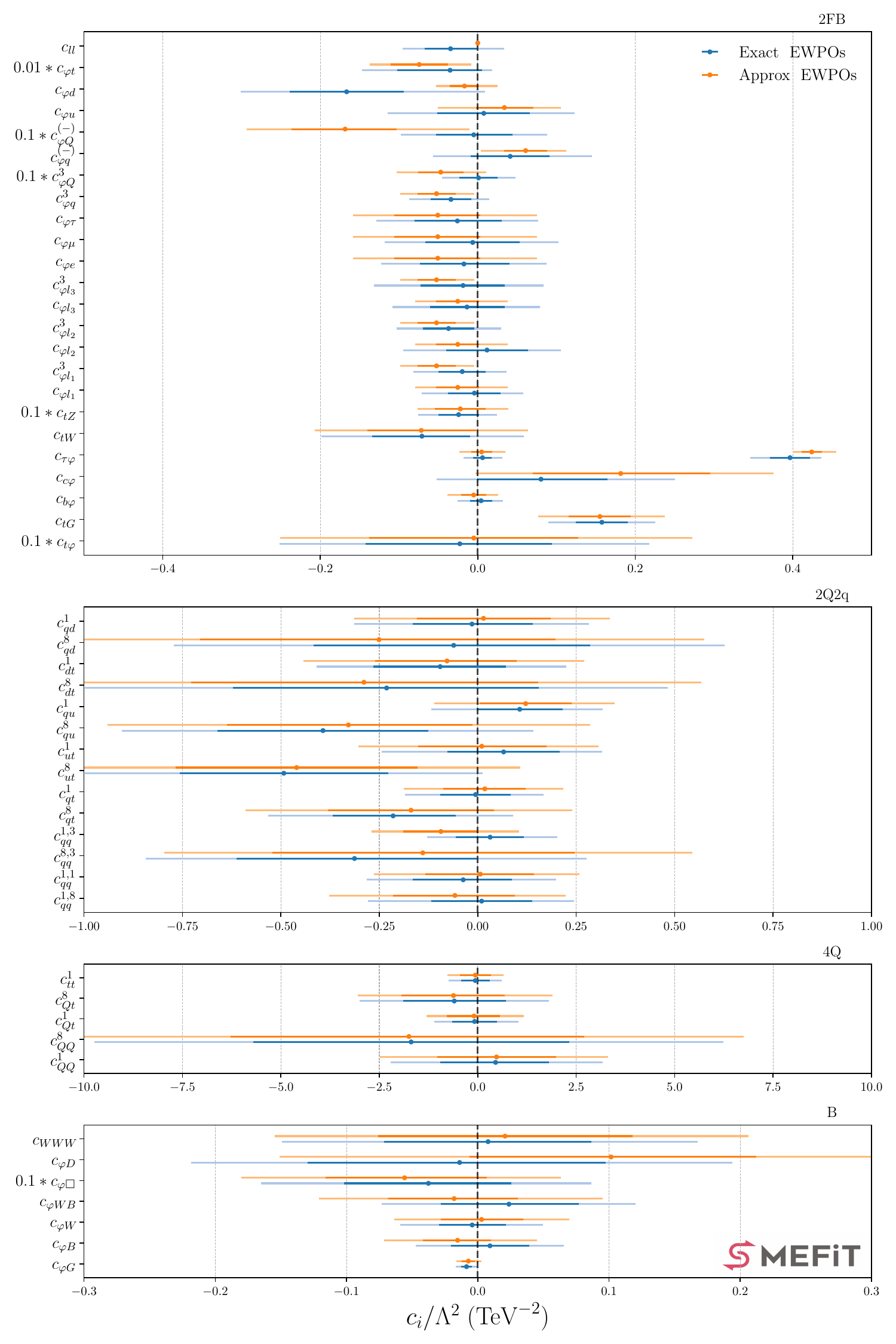}
    \caption{The 
    results of the global {\sc\small SMEFiT} analysis, comparing
    the approximated~\cite{Ethier:2021bye,Giani:2023gfq} and exact~\cite{ewpos}
    implementation of the EWPOs.
    For each of the Wilson coefficients entering the fit,
    we indicate the best-fit results and the  marginalised 68\% CL (95\% CL) intervals
    with thicker (thinner) lines.
    The fit labelled as ``Approx EWPOs"
    differs slightly from that of~\cite{Giani:2023gfq}
    due to the use of 
    improved theory predictions, see text
    for more details.
    Operators are grouped by type: from top to bottom, we show two-fermion-two-boson (2F2B),  two-heavy-two-light (2Q2q), four-heavy quarks (4Q), and purely bosonic (B) operators.
    }
    \label{fig:bounds_smefit30_vs_smefit20}
\end{figure}

In addition to Fig.~\ref{fig:post_smefit30_vs_smefit20-label},
the impact of the exact implementation of EWPOs 
in the SMEFT fit is also illustrated in Fig.~\ref{fig:bounds_smefit30_vs_smefit20},
which compares the 68\% and 95\% CL
 marginalised bounds on the Wilson coefficients 
 obtained with the approximate
 and exact implementations.
 As mentioned before,
 the results
labelled as ``Approx EWPOs'' differ slightly from 
those in~\cite{Giani:2023gfq} due to the use of improved EFT theory predictions.
In general one observes good agreement
between the 
results of the global EFT fit with
approximate and exact 
implementation of the EWPOs, with small
differences whose origin is discussed in~\cite{ewpos}.


\section{The {\sc match2fit}~package}
\label{app:mmeft2smefit}
\label{app:mmeft2smefit-conceptual-description}

As mentioned in the introduction,
several public tools to carry out
global analyses of experimental data
in the SMEFT framework have been presented.
Some of these tools also include the
option to carry out parameter scans
in the space of pre-defined UV models matched to SMEFT. 
While the matching procedure
between UV models and the SMEFT coefficients has been automated at tree-level and partially at 1-loop level, currently 
there is no streamlined method enabling
the interface of the
output of these matching codes with
SMEFT fitting frameworks.
Such interface would allow  
using the SMEFT to impose bounds on the
parameter space of general, user-defined
UV-complete models.

In this appendix we describe
the {\sc\small match2fit}~package used in this work
to interface an EFT automated matching code,
specifically {\sc\small MatchMakerEFT}, with
a global fitting analysis framework, in this  case
{\sc\small SMEFiT}.
The current public version of this package is available on \href{https://github.com/arossia94/match2fit}{ Github~\faicon{github} }.
We emphasize that the novel functionality of \matchTOfit~is the automated translation of matching results between the formats used by the matching and fitting codes.
This could be extended to connect different SMEFT-related codes. 
We provide here a succinct introduction
to this code, and point
the interested reader to the 
\href{https://github.com/arossia94/match2fit/blob/main/match2fit_Users_Manual.pdf}{user's manual}
for more details.

The package has two different working modes. 
The first one reads the matching output in the format used by {\sc\small MatchMakerEFT}~and parses it to the format of run cards that can be fed into \smefit.
In this mode, the mandatory inputs are the location of the file containing the matching results and a numerical value (in TeV) for the mass of the heavy particle. 
Optionally, one can also specify a name for the UV model and define a ``collection'', 
a set of UV models with common characteristics. 
Depending on the executed function, the code will print the run card for a scan on the UV parameters with or without the accompanying file that defines the UV invariants (see Sect.~\ref{subsec:uv_invariants}).

The second mode combines the first one with a wrapper for {\sc\small MatchMakerEFT}~to perform the matching of a given UV model to SMEFT. 
This model allows the user to go from the files that define the UV model to the run cards for \smefit~in a single command.
It is also possible to just perform the matching without producing the run cards for SMEFiT.
The input required in this mode is the one that {\sc\small MatchMakerEFT}~needs to describe the heavy particles that will be integrated out.  More precisely, it needs the \verb|.fr|, \verb|.red| and \verb|.gauge| files, and optionally the \verb|.red| file. The \verb|.fr| file is just a FeynRules file that defines the heavy particle(s), its (their) free Lagrangian and its (their) interactions with the SM. 
The SM and SMEFT models are included in {\sc\small MatchMakerEFT}~and can be used
in this context without further modification. 
See~\cite{Carmona:2021xtq} for more details on how to write the \verb|.fr| file and what the other files must contain.

The {\sc\small match2fit} code consists of
a Wolfram Mathematica Package designed and tested in version $12.1$ or later. 
To unlock its full functionality, this package requires a working installation of {\sc\small MatchMakerEFT}.

\paragraph{Main commands.}

In order to highlight its usage, we 
describe now the key commands
made available by the {\sc\small match2fit}
package.

\begin{itemize}

\item {\lstinline[emph={directory,model}]|parametersList[directory,model]|}: Both arguments should be strings. This function reads the file \verb|model.fr| in \verb|directory|, recognizes the masses and couplings of the heavy particles to be integrated out, and gives back an array with two elements. 
The first element is a list with the symbolical expression of the masses of the heavy particles. The second element is a list of the couplings defined for these heavy particles, excluding gauge couplings but not self-interactions.
    
\item {\lstinline[emph={matchResFile,looplevel}]|parametersListFromMatchingResult[matchResFile,looplevel]|}: Its first argument must be a string with the address of the file containing the matching results. It recognizes the masses and couplings of the heavy particles from those results and gives an output in the same format as \lstinline|parametersList|. This code assumes that any parameter with a name starting by \textit{m} or \textit{M} corresponds to a mass and identifies any other parameter as a coupling. If the input of \lstinline|parametersListFromMatchingResult| is the matching result obtained with the model fed into \lstinline|parametersList|, any difference in their outputs should be due to couplings (or whole particles) that do not affect the tree-level matching result. If the second argument, \lstinline|looplevel|, is equal to \lstinline|0|, \lstinline|"tree"|, or \lstinline|"Tree"|, and the input file contains 1-loop matching results, the 1-loop contributions will be ignored. To use 1-loop results, \lstinline|looplevel| must be set to \lstinline|1|, \lstinline|"loop"|, or \lstinline|"1loop"|. If the input file corresponds to tree-level matching results, there might be some spurious 1-loop pieces due to evanescent shifts (see \href{https://gitlab.com/m4103/matchmaker - eft/-/issues/12}{here}) and hence \lstinline|looplevel| must be set to \lstinline|0|, \lstinline|"tree"|, or \lstinline|"Tree"| to ensure their consistent removal.

\item {\lstinline[emph={matchResFile}]|flavourSymChecker[matchResFile,Options]|}: It takes the file with matching results specified as input and checks if those results are compatible with the \smefit~flavour symmetry, U$(2)_{q}\times$U$(2)_{u}\times$U$(3)_{d}\times(\text{U}(1)_{\ell}\times\text{U}(1)_{e})^3$. If the constraints are satisfied, it returns YES. If not, it returns NO and it prints the first WC for which it found a symmetry violation. The option \lstinline|"UVFlavourAssumption"| allows the user to specify a replacement list that can be used to apply flavour assumptions on the UV parameters. The left-hand side of the replacement rule should contain some of the UV parameters listed by \lstinline|parametersListFromMatchingResult| or \lstinline|parametersList| with or without numerical indices, according to how they appear in the matching results file. The code supports up to 4 numerical indices in a single group, i.e. couplings such as gUV, gUV$[1]$, gUV$[1,3]$ and gUV$[1,3,2,4]$ are supported, but gUV$[1][3]$ or gUV$[1][2,3]$ are not. The default value of \lstinline|"UVFlavourAssumption"| is an empty list. An example of how to set this option is,
\begin{lstlisting}
{"UVFlavourAssumption" -> {gWtiQ[i_, j_] :> KroneckerDelta[i,j] * KroneckerDelta[i, 3] gWqf[3, 3] }}
\end{lstlisting}
This function is prepared to handle tree-level matching results only. If it detects 1-loop results in the input file, it warns the user of its limitations and proceeds to ignore the 1-loop contributions. Some models matched at tree level might trigger the warning due to the presence of spurious 1-loop pieces related to the evanescent shifts (see \href{https://gitlab.com/m4103/matchmaker - eft/-/issues/12}{here}) which are consistently ignored by this function.

\item {\lstinline[emph={matchResFile}]|flavourSolver[matchResFile, Options]|}: It takes the file with matching results specified as input and tries to solve the constraints imposed by the \smefit~flavour symmetry, U$(2)_{q}\times$U$(2)_{u}\times$U$(3)_{d}\times(\text{U}(1)_{\ell}\times\text{U}(1)_{e})^3$, for the UV couplings. The running time, the number of solutions and their complexity depend on the model. It returns all found solutions. The only Option of this function is \lstinline|"UVFlavourAssumption"|, which follows the same description given in the function \lstinline|flavourSymChecker|. This function considers the SM Yukawa couplings as symbolical variables and they can be set to zero with the option \lstinline|"UVFlavourAssumption"|. As with \lstinline|flavourSymChecker|, this function is prepared to handle tree-level matching results only.

\item {\lstinline[emph={directory, model, looplevel}]|matcher[directory, model, looplevel]|}: it runs {\sc\small MatchMakerEFT}~and performs the tree-level matching without printing any run card for \smefit. It takes two strings as arguments, \emph{\lstinline|directory|} and \emph{\lstinline|model|}. The first one is the directory where the package will look for the files \verb|model.fr|, \verb|model.red| and \verb|model.gauge|. If the code does not find one of those files, it will print a warning. It does not check for the existence of \verb|model.herm|. The expected content of each of those files is specified in the documentation of {\sc\small MatchMakerEFT} ~\cite{Carmona:2021xtq}. The argument \lstinline|looplevel| decides the matching order, \lstinline|looplevel| equal to \lstinline|0|, \lstinline|"tree"|, or \lstinline|"Tree"| will produce tree-level matching, while setting it to \lstinline|1|, \lstinline|"loop"|, or \lstinline|"1loop"| will produce 1-loop level matching. {\sc\small MatchMakerEFT}~will create the folder \verb|directory/model_MM|, inside which the matching results will be stored as \verb|MatchingResult.dat|. The code will check if {\sc\small MatchMakerEFT}~reported any problem during the matching and will print a warning if so. After performing the matching, the package will remove most of the files and directories created by itself or {\sc\small MatchMakerEFT}~for the sake of tidiness. It will only leave the directory \verb|model_MM| and 2 files inside: \verb|MatchingResult.dat| and \verb|MatchingProblems.dat|.

\item{\lstinline[emph={matchResFile, mass, looplevel}]|matchResToUVscanCard[matchResFile,mass, looplevel, Options]|}: Function that reads the file with the tree-level matching results and prints the cards required for a UV scan. \emph{\lstinline|matchResFile|} must be a string with the exact address of the file that contains the matching results to be used. The format of that file should be exactly like the file \verb|MatchingResult.dat| produced by {\sc\small MatchMakerEFT}. The argument \emph{\lstinline|mass|} should be the value in TeV that the mass(es) of the UV particle(s) will be set to. \emph{\lstinline|mass|} can be one numerical value or a list of them \emph{$\lbrace m_1, ..,m_N\rbrace$}, the latter being useful in the case of a multiparticle model. The order of the masses is the one returned by \lstinline|parametersListFromMatchingResult|. If the user specifies only one numerical mass value for a multiparticle model, all the particles will be assigned the same mass. If \lstinline|parametersListFromMatchingResult| identifies $K$ masses and $N<K$, the code will assume  $m_i = m_N$ for $ N \leqslant i \leqslant K$. If $N>K$, the values $m_i$ with $K<i\leqslant N$ will be ignored.
The mass values are also printed on the card names and inside the cards. For multiple masses, the convention is to take the integer part of each value and stick them together in sequence. The third argument is \emph{\lstinline|looplevel|}, which indicates the loop order of the matching results to be parsed and it works as in \lstinline|parametersListFromMatchingResult|. When \emph{\lstinline|looplevel|} is set to \lstinline|1|, \lstinline|loop| or \lstinline|1loop|, the code chooses the matching scale as the minimum of the masses of the UV particles.
There are 3 options. The first one is \lstinline|"UVFlavourAssumption"|, which is identical to the one of the function \lstinline|flavourSymChecker|, see its description for details on this option.
The second option is \lstinline|"Collection"|, a string indicating the Collection to which the model belongs. Its default value is \lstinline|"UserCollection"|. Finally, the option \lstinline|"Model"| is a string that specifies the model name to be printed on the run cards, with default value \verb|"UserModel"|. An example of how to set these options is replacing the argument \lstinline|Options| by:

\begin{lstlisting}
{"UVFlavourAssumption" -> {lambdaT1[a_] :> lambdaT1[3]}, 
 "Collection" -> "TestMatchingCollection","Model"->"TestModel"}
\end{lstlisting}

\end{itemize}

The package also includes a function that integrates the steps of matching and printing all the run cards according to the result of said matching:

\begin{itemize}
    
\item {\lstinline[emph={directory, model, mass, looplevel}]|modelToUVscanCard[directory,model, mass, looplevel, Options]|}: Function that takes the files that define the UV model, performs the tree-level matching by running {\sc\small MatchMakerEFT}, and prints the run cards needed for a UV scan. The first two arguments are exactly like in \verb|matcher|, i.e. the program will look for the files \verb|model.fr|, \verb|model.red|, and \verb|model.gauge| in \verb|directory| and will do the tree-level matching based on them. The argument \verb|model| also defines the name of the model. \verb|mass| should be the value(s) in TeV that the mass(es) of the UV particle(s) will be set to. The handling of several mass values is equal to the function {\lstinline|matchResToUVscanCard|}. The argument \lstinline|looplevel| indicates the loop order at which the matching will be performed and that will be considered when printing the run cards. Its allowed values are as for previous functions. When \emph{\lstinline|looplevel|} is set to \lstinline|1|, \lstinline|loop| or \lstinline|1loop|, the code chooses the matching scale as the minimum of the masses of the UV particles. This function has 2 options. The first one is \lstinline|"UVFlavourAssumption"|, which is identical to the one of the function \lstinline|flavourSymChecker|, see its description for details on this option.
The second option is \lstinline|"Collection"|, a string indicating the Collection to which the model belongs. Its default value is \lstinline|"UserCollection"|. An example of how to set these options is replacing the argument \lstinline|Options| by:

\begin{lstlisting}
{"UVFlavourAssumption" -> {lambdaT1[a_] :> lambdaT1[3]}, 
 "Collection" -> "TestMatchingCollection"}
\end{lstlisting}

\end{itemize}

\paragraph{Limitations and outlook.}
%
In its current incarnation, all SM couplings are numerically evaluated when printing the run cards 
used as input in the global SMEFT fit. 
Their values are hard-coded in the package and summarised in App.~A of the {\sc\small match2fit}  user manual.
Future versions of {\sc match2fit} should allow the user to vary easily these values.

Another restriction of {\sc match2fit}
is that it does not check for the fulfilment of the \smefit~flavour assumptions automatically when printing the run cards for \smefit. This generates a degree of arbitrariness, e.g. the code uses the matching result for $(c_{\varphi q}^{(3)})_{22}$ as the result for $(c_{\varphi q}^{(3)})_{ii}$ without verifying first that that $(c_{\varphi q}^{(3)})_{22}=(c_{\varphi q}^{(3)})_{11}$.
We summarise these choices in App.~B of the user manual. 
The user should be aware of this limitation and compensate for it with the required assumptions on the UV couplings.
Support for alternative flavour structures on the Wilson coefficients will be added in the future,
to ensure full compatibility with updated versions of the \smefit~analysis and/or becoming compatible with other fitting codes.

We also note that {\sc match2fit}
assumes that all UV couplings are real and applies this assumption when interpreting the matching results. The support for complex UV couplings (and hence also of the corresponding WCs) will be added in future releases.

In the long term, {\sc match2fit} could
be extended to process matching results
provided in the WCxf format~\cite{Aebischer:2017ugx}.
This would facilitate the interface with other SMEFT matching codes such as {\sc\small CoDEx} and {\sc\small Matchete} as well as with codes implementing the Renormalisation Group Equations (RGEs) running of SMEFT operators. 

Since version \verb|1.8|, the pipeline required to perform the fit of the UV model matched to the SMEFT at 1-loop level is fully automated. However, accessory functions that might help to analyse the fit results, in particular the one that computes the UV invariants, remain limited to tree-level results. They will support 1-loop results in the future.

\section{Origin of the logarithms in 1-loop matching formulas.}
\label{app:logarithms}

The logarithms in one-loop matching expressions such as that of Eq.~(\ref{eq:oneloop_logs_example}) arise as a consequence of the RG running of both the UV couplings and the EFT coefficients.
Their appearance ensures that the matching
result is correct irrespective of the choice
of matching scale~\cite{Gherardi:2020det, Carmona:2021xtq}. 
In this appendix we revisit this point
in the explicit case of the
heavy scalar model defined by
the Lagrangian of Eq.~\eqref{eq:Lag_UV_full_Varphi_model}. 
We start with the general expression, valid for a any
 matching scale $Q$, for the Wilson coefficient $c_{qu,pqrs}^{(8)}$,
which is the most general form in flavour space of the coefficient \cQteight~ generated by this model.
 Subsequently, we evaluate this matching
 relation at two
 different scales, $Q=m_\phi$ and $Q=\mu<m_\phi$, to obtain the results for $c_{qu,pqrs}^{(8)}|_{Q=m_\phi}$ and $c_{qu,pqrs}^{(8)}|_{Q=\mu}$. We expect only the former to be free of RG logarithms.

\begin{figure}[h!]
	\centering
		\includegraphics[width=0.9\linewidth]{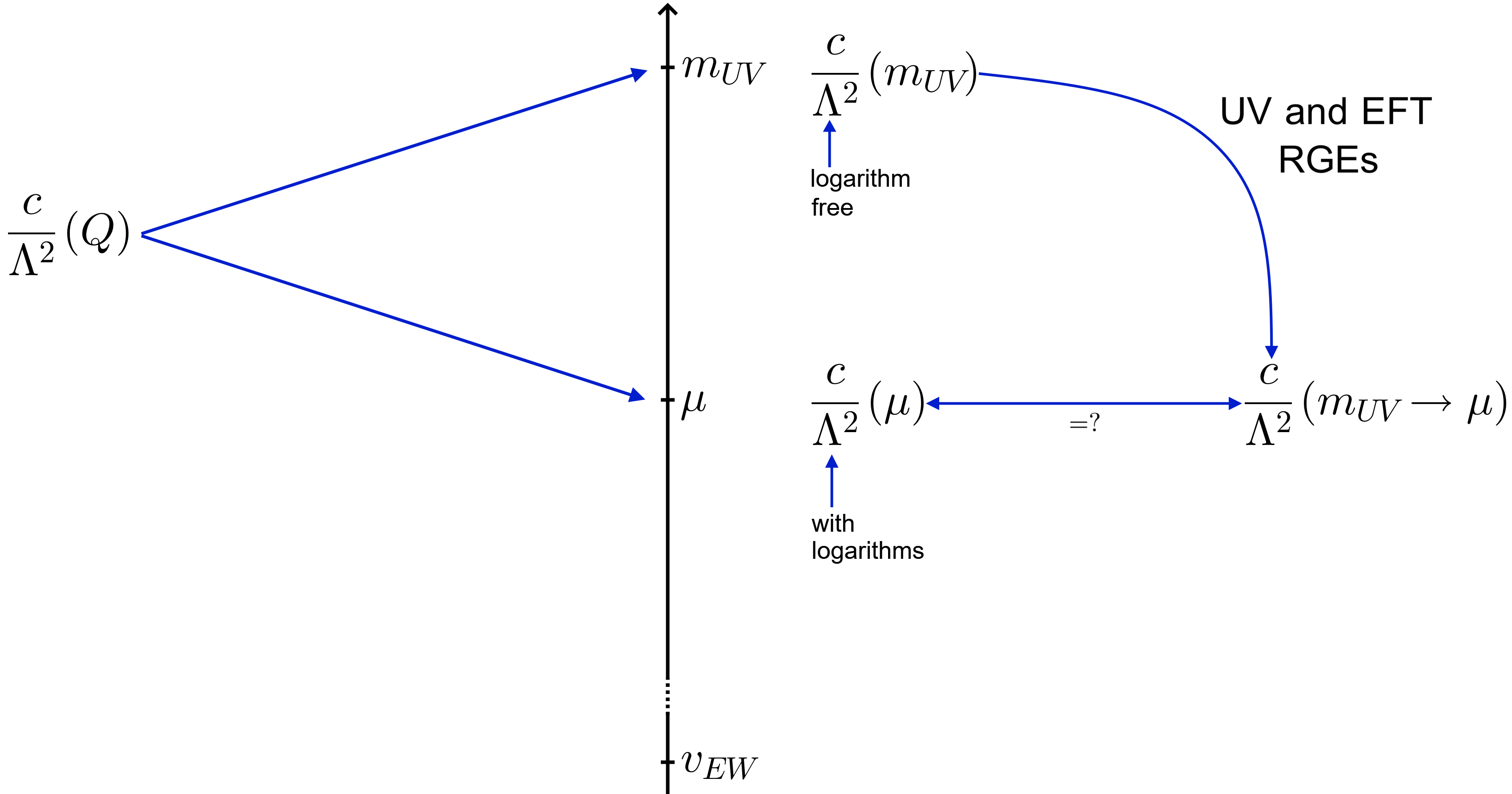}  
	\caption{Sketch of the procedure adopted
 to verify the origin of the logarithms
 arising in the one-loop matching formulas.
 The general matching expression for
 the EFT coefficient $c/\Lambda^2$ in terms of UV
 couplings is evaluated at two different scales, $Q=m_{\rm UV}$ and $Q=\mu < m_{\rm UV}$,
 and one verify whether or not the two results
 are entirely related by RGE running.
 }
	\label{fig:Log_check}
\end{figure}

Then we  use the RGEs for the Wilson
coefficients in the EFT and for the UV couplings of the heavy scalar model to evolve $c_{qu,pqrs}^{(8)}|_{Q=m_\phi}$ down to the scale $\mu$ to obtain $c_{qu,pqrs}^{(8)}|_{Q=m_\phi\to\mu}$. 
The comparison between the two calculations, $c_{qu,pqrs}^{(8)}|_{Q=m_\phi\to\mu}$ and $c_{qu,pqrs}^{(8)}|_{Q=\mu}$, will tell us whether the logarithms arising
in the general matching formula correspond to the ones generated by the RG running or not. 
This procedure, which was also adopted
in App.~C of~\cite{Carmona:2021xtq}, is sketched in Fig.~\ref{fig:Log_check}.

For illustration purposes, we perform this check separately at $\mathcal{O}\lp g_3^2\rp $ and then at $\mathcal{O}\lp g_1^2\rp$, i.e. the contributions that depends on the matching scale $Q$ will only be kept at that order. 
On the other hand, we perform it without applying any flavour assumption on the EFT coefficient nor the UV model.

\paragraph{Order $g_3^2$ validation.}
The starting point of this calculation
is the general matching result at one-loop for 
the Wilson coefficient $c_{qu,prst}^{(8)}$
in the heavy scalar model, given by
\begin{align}
	\frac{c_{qu,prst}^{(8)}}{\Lambda^2}=&-\frac{\big(y_{\phi}^{u}\big)_{pt} \big(y_{\phi}^{u}\big)_{rs}^{*}}{m_\phi^2}-\frac{g_3^2\,\delta_{s,t} \big(y_{\phi}^{d}\big)_{ir} \big(y_{\phi}^{d}\big)_{ip}^{*} }{144 \pi^2\,m_{\phi}^2} \(4-3\log\lp \frac{m_{\phi}^2}{Q^2}\rp \) \nonumber \\  
    &-\frac{g_3^2\,\delta_{p,r} \big(y_{\phi}^{u}\big)_{it} \big(y_{\phi}^{u}\big)_{is}^{*} }{72 \pi^2\,m_{\phi}^2} \(4-3\log\lp \frac{m_{\phi}^2}{Q^2}\rp \) 
    -\frac{g_3^2\,\delta_{s,t} \big(y_{\phi}^{u}\big)_{pi} \big(y_{\phi}^{u}\big)_{ri}^{*} }{144 \pi^2\,m_{\phi}^2} \(4-3\log\lp \frac{m_{\phi}^2}{Q^2}\rp\)\\
    & +\frac{g_3^2 \big(y_{\phi}^{u}\big)_{pt} \big(y_{\phi}^{u}\big)_{rs}^{*} }{48\pi^2 m_\phi^2}+\mcO\(\frac{g_3^0}{16\pi^2}\), \nonumber
\label{eq:rge_check_cqu8_1loop}
\end{align}
where $prst$ indicate the flavour indices and we omitted the one-loop contributions that do not depend on  the $g_3$ coupling.
As expected, all the logarithms appearing
in this expression are removed by setting the matching scale at the heavy scalar mass, $Q=m_\phi$.

The renormalisation group equations 
relevant to describe the scale dependence of this specific Wilscon coefficient are given by~\cite{Alonso:2013hga},
\begin{align}
	\mu\frac{d c_{qu,prst}^{(8)}}{d\mu} = \frac{\beta(c_{qu,prst}^{(8)})}{16\pi^2}= & \frac{g_3^2}{16\pi^2}\left(\frac{4}{3}\delta_{s,t} \( c_{qq,pjjr}^{(1)} + c_{qq,jrpj}^{(1)} \) +4 \delta_{s,t} \(c_{qq,pjjr}^{(3)}+c_{qq,jrpj}^{(3)}\) \right. \nonumber\\
 & +\frac{2}{3}\delta_{s,t} \left(c_{qu,prjj}^{(8)}+c_{qd,prjj}^{(8)}\right) +\frac{4}{3} \delta_{p,r} c_{qu,jjst}^{(8)} +\frac{2}{3} \delta_{p,r} c_{ud,stjj}^{(8)} \\
 &\left. + \frac{4}{3} \delta_{p,r} c_{uu,sjjt} + \frac{4}{3} \delta_{p,r} c_{uu,jtsj} - 14 c_{qu,prst}^{(8)} - 12 c_{qu,prst}^{(1)} \right), \nonumber
\end{align}

\begin{equation}
	\mu\frac{d \big(y_{\phi}^{u}\big)_{pt} }{d\mu}=\frac{\beta\left(\big(y_{\phi}^{u}\big)_{pt}\right)}{16\pi^2}=-\frac{g_3^2}{2\pi^2} \big(y_{\phi}^{u}\big)_{pt},
\end{equation}
\begin{equation}
		\mu\frac{d m_\phi}{d\mu}=0,
\end{equation}
where again we have kept only those terms proportional to $g_3^2$.
The RGE equations for the Yukawa
coupling $\big(y_{\phi}^{u}\big)_{pt}$ and the heavy mass $m_\phi$ were computed with {\sc\small RGBeta}~\cite{Thomsen:2021ncy} and cross-checked against 
the results of~\cite{Ferreira:2015rha}.
We can solve these RGEs at leading log accuracy to obtain,
\begin{equation}
	c_{qu,prst}^{(8)}\(\mu\)=c_{qu,prst}^{(8)}\(m_\phi\)+\frac{\beta(c_{qu,prst}^{(8)})}{32\pi^2}\log\(\frac{\mu^2}{m_\phi^2}\),
\end{equation}
and similarly for $\big(y_{\phi}^{u}\big)_{pt}$, while $m_\phi$ is constant at this order.
The required tree-level matching results are,
\begin{equation}
	c_{qq,ijkl}^{(1)}=c_{qq,ijkl}^{(3)}=c_{ud,ijkl}^{(8)}=c_{uu,ijkl}=0,
\end{equation}
\begin{equation}
	\frac{c_{qd,ijkl}^{(8)}}{\Lambda^2}= -\frac{\big(y_{\phi}^{d}\big)_{kj} \big(y_{\phi}^{d}\big)^{*}_{li}}{m_\phi^2},
\end{equation}
\begin{equation}
	\frac{c_{qu,ijkl}^{(1)}}{\Lambda^2} = -\frac{ \big(y_{\phi}^{u}\big)_{il} \big(y_{\phi}^{u}\big)^{*}_{jk} }{6\, m_\phi^2},
\end{equation}
which substituting in the corresponding
beta function allows one to compute $\frac{c_{qu,prst}^{(8)}}{\Lambda^2}\Big|_{Q=\mu}$ as,
\begin{align}
\label{appD:eqC9}
\frac{c_{qu,prst}^{(8)}}{\Lambda^2}\Big|_{Q=\mu}= &\frac{c_{qu,prst}^{(8)}}{\Lambda^2}\Big|_{Q=m_{\phi}} +  \frac{\beta(c_{qu,prst}^{(8)})}{32\pi^2\Lambda^2}\log\(\frac{\mu^2}{m_\phi^2}\),\nonumber\\
= & \frac{c_{qu,prst}^{(8)}}{\Lambda^2}\Big|_{{\rm tree},Q=m_{\phi}}+\frac{c_{qu,prst}^{(8)}}{\Lambda^2}\Big|_{ { \rm loop},Q=m_{\phi}}+  \frac{\beta(c_{qu,prst}^{(8)})}{32\pi^2\Lambda^2}\log\(\frac{\mu^2}{m_\phi^2}\),
\end{align}
where we have defined
\begin{equation}
	\frac{c_{qu,prst}^{(8)}}{\Lambda^2}\Big|_{{\rm loop},Q=m_{\phi}}= -\frac{g_3^2\,\delta_{s,t} \big(y_{\phi}^{d}\big)_{ir} \big( y_{\phi}^{d} \big)^{*}_{ip} }{36 \pi^2\,m_{\phi}^2} 
	-\frac{g_3^2\,\delta_{p,r} \big( y_{\phi}^{u} \big)_{it} \big( y_{\phi}^{u} \big)^{*}_{is} }{18 \pi^2\,m_{\phi}^2}
	-\frac{g_3^2\,\delta_{s,t} \big( y_{\phi}^{u} \big)_{pi} \big( y_{\phi}^{u} \big)^{*}_{ri} }{36 \pi^2\,m_{\phi}^2} + \frac{g_3^2 \big( y_{\phi}^{u} \big)_{pt} \big( y_{\phi}^{u}\big)^{*}_{rs} }{48\pi^2 m_\phi^2}.
\end{equation}
The last term in Eq.~(\ref{appD:eqC9})
is the contribution related
to the running between $Q=m_\phi$
and $Q=\mu<m_{\phi}$.

To keep everything consistently at leading-log and one-loop order, all the couplings in the previous expression should be considered as constant with energy, and hence $\lp c_{qu,prst}^{(8)}/\Lambda^2\rp\Big|_{{\rm loop},Q=m_{\phi}}$ is equal to  $\lp c_{qu,prst}^{(8)}/\Lambda^2\rp \Big|_{{\rm loop},Q=\mu}$ except for the logarithmic pieces. The same is not true for the tree-level contribution, which has to be transformed as follows.
\begin{align}
	\frac{c_{qu,prst}^{(8)}}{\Lambda^2}\Big|_{tree,Q=m_{\phi}} =& -\frac{ \big( y_{\phi}^{u} \big)_{pt} \big( y_{\phi}^{u} \big)^{*}_{rs} }{m_\phi^2}\Big|_{Q=m_\phi} = -\frac{ \big( y_{\phi}^{u}(m_\phi) \big)_{pt} \big( y_{\phi}^{u}(m_\phi) \big)_{rs}^{*} }{m_\phi^2} \nonumber\\
	=&-\frac{\( \big(y_{\phi}^{u}(\mu)\big)_{pt} - \frac{ \beta\Big(\big(y_{\phi}^{u}\big)_{pt}\Big)}{32\pi^2}\log(\frac{\mu^2}{m_\phi^2})\) \( \big(y_{\phi}^{u}(\mu)\big)_{rs} - \frac{\beta\Big(\big(y_{\phi}^{u}\big)_{rs}\Big)}{32\pi^2}\log(\frac{\mu^2}{m_\phi^2}) \)^{*} }{m_\phi^2 } \nonumber \\
	=&-\frac{\big(y_{\phi}^{u}\big)_{pt} \big(y_{\phi}^{u}\big)^{*}_{rs} }{m_\phi^2}\Big|_{Q=\mu}+\frac{\beta\Big( \big( y_{\phi}^{u} \big)_{pt} \Big) \big( y_{\phi}^{u} \big)^{*}_{rs} }{32\pi^2\,m_{\phi}^2}\log\lp \frac{\mu^2}{m_\phi^2}\rp  +\frac{ \big( y_{\phi}^{u} \big)_{pt} \beta\Big( \big(y_{\phi}^{u}\big)_{rs} \Big)^{*} }{ 32 \pi^2 \, m_{\phi}^2}\log\lp \frac{\mu^2}{m_\phi^2}\rp,
\end{align}
where in the first two lines we indicate explicitly the scale at which the UV couplings are evaluated and in the last line we have kept only the leading log terms. The first term in the last line can be identified as $\lp c_{qu,prst}^{(8)}/\Lambda^2\rp \Big|_{{\rm tree},Q=\mu}$.
Hence, now we have,
\begin{equation}
	\frac{c_{qu,prst}^{(8)}}{\Lambda^2}\Big|_{Q=\mu}= \frac{c_{qu,prst}^{(8)}}{\Lambda^2}\Big|_{tree,Q=\mu}+\frac{c_{qu,prst}^{(8)}}{\Lambda^2}\Big|_{{\rm loop},Q=m_{\phi}} +\frac{ \beta\Big( \big( y_{\phi}^{u} \big)_{pt} \Big) \big( y_{\phi}^{u} \big)^{*}_{rs} + \big( y_{\phi}^{u} \big)_{pt} \beta\Big( \big( y_{\phi}^{u} \big)_{rs} \Big)^{*} +\beta( c_{qu,prst}^{(8)})}{32\pi^2\,m_{\phi}^2}\log\lp \frac{\mu^2}{m_\phi^2}\rp.
 \nonumber
\end{equation}
Finally, one has to replace the expressions for the $\beta$ functions computed before, use the tree-level matching results inside the $\beta$ function for the Wilson coefficient, and reorder the terms. This leads to Eq.~\eqref{eq:rge_check_cqu8_1loop} evaluated at $Q=\mu$, concluding the check.

\paragraph{Order $g_1^2$ validation.}
Next we perform the same check at order $g_1^2$.
The general matching result at 1-loop for $c_{qu,prst}^{(8)}$ is
\begin{equation}
	\begin{split}
		\frac{c_{qu,prst}^{(8)}}{\Lambda^2}=-\frac{\big(y_{\phi}^{u}\big)_{pt} \big( y_{\phi}^{u} \big)^{*}_{rs} }{m_\phi^2} -\frac{25 g_1^2 \big( y_{\phi}^{u} \big)_{pt} \big(y_{\phi}^{u}\big)^{*}_{rs}  }{ 1152 \pi^2 m_{\phi}^{2} } + \mcO\(\frac{g_1^0}{16\pi^2}\),
	\end{split}
 \label{eq:rge_check_cqu8_g1}
\end{equation}
where the omitted terms are 1-loop contributions that do not depend on $g_1$. In this case, there are no logarithms in the expression. Hence, the only terms that have a dependence on the energy scale are the ones generated at tree level.

The RGE equations we need are in this case are~\cite{Alonso:2013hga,Ferreira:2015rha},
\begin{equation}
	\begin{split}
		\mu\frac{d c_{qu,prst}^{(8)}}{d\mu} =\frac{\beta(c_{qu,prst}^{(8)})}{16\pi^2}=-\frac{g_1^2}{12\pi^2}c_{qu,prst}^{(8)},
	\end{split}
\end{equation}
\begin{equation}
	\begin{split}
		\mu\frac{d \big(y_{\phi}^{u}\big)_{pt}}{d\mu}=\frac{\beta\Big(\big(y_{\phi}^{u}\big)_{pt} \Big) }{16\pi^2}=-\frac{ 17 g_1^2}{192\pi^2} \big( y_{\phi}^{u} \big)_{pt},
	\end{split}
\end{equation}
\begin{equation}
		\mu\frac{d m_\phi}{d\mu}= \frac{\beta(m_{\phi})}{16\pi^2} = -\frac{3g_1^2}{64\pi^2} m_{\phi},
\end{equation}
where we have kept only those terms proportional to $g_1^2$.
The leading-log solution reads,
\begin{equation}
	c_{qu,prst}^{(8)}\(\mu\)=c_{qu,prst}^{(8)}\(m_\phi\)+\frac{\beta(c_{qu,prst}^{(8)})}{32\pi^2} \log\(\frac{\mu^2}{m_\phi^2}\),
\end{equation}
\begin{equation}
	m^2_{\phi}\(m_\phi\)=m^2_{\phi}\( \mu \)- m_{\phi}\(\mu\)\frac{\beta(m_{\phi})}{16\pi^2} \log\(\frac{\mu^2}{m_\phi^2}\),
\end{equation}
and analogously for the UV coupling.

All the required tree-level matching results were shown in the previous subsection. We can compute $\frac{c_{qu,prst}^{(8)}}{\Lambda^2}\Big|_{Q=\mu}$ as,
\begin{align}
		\frac{c_{qu,prst}^{(8)}}{\Lambda^2}\Big|_{Q=\mu} &= \frac{c_{qu,prst}^{(8)}}{\Lambda^2}\Big|_{Q=m_{\phi}} + 
		\frac{\beta(c_{qu,prst}^{(8)})}{32\pi^2\Lambda^2}\log\(\frac{\mu^2}{m_\phi^2}\),\nonumber\\
		&= \frac{c_{qu,prst}^{(8)}}{\Lambda^2}\Big|_{{\rm tree},Q=m_{\phi}} + \frac{c_{qu,prst}^{(8)}}{\Lambda^2}\Big|_{{\rm loop},Q=m_{\phi}} +  \frac{\beta(c_{qu,prst}^{(8)})}{32\pi^2\Lambda^2}\log\(\frac{\mu^2}{m_\phi^2}\),
\end{align}
where
\begin{equation}
	\frac{c_{qu,prst}^{(8)}}{\Lambda^2}\Big|_{loop,Q=m_{\phi}}= -\frac{25 g_1^2 \big( y_{\phi}^{u} \big)_{pt} \big( y_{\phi}^{u} \big)^{*}_{rs}  }{ 1152 \pi^2 m_{\phi}^{2} }.
\end{equation}
Let us remark again that all the UV couplings in $\frac{c_{qu,prst}^{(8)}}{\Lambda^2}\Big|_{{\rm loop},Q=m_{\phi}}$ should be considered as constant with energy. 
The tree-level contribution at the two different scales are related as,
	\begin{align}
		\frac{c_{qu,prst}^{(8)}}{\Lambda^2}\Big|_{tree,Q=m_{\phi}} &= -\frac{ \big( y_{\phi}^{u} \big)_{pt} \big( y_{\phi}^{u} \big)^{*}_{rs} }{m_\phi^2}\Big|_{Q=m_\phi} = -\frac{\big( y_{\phi}^{u}(m_\phi)\big)_{pt} \big(y_{\phi}^{u}(m_\phi) \big)^{*}_{rs} }{m_\phi^2} \nonumber\\
		&=-\frac{ \( \big( y_{\phi}^{u}(\mu) \big)_{pt} - \frac{\beta\Big( \big(y_{\phi}^{u} \big)_{pt} \Big) }{32\pi^2}\log(\frac{\mu^2}{m_\phi^2})\) \( \big( y_{\phi}^{u}(\mu) \big)_{rs} - \frac{\beta\Big( \big( y_{\phi}^{u} \big)_{rs} \Big)}{32\pi^2}\log(\frac{\mu^2}{m_\phi^2}) \)^{*} }{ m^2_{\phi}\( \mu \)- m_{\phi}\(\mu\)\frac{\beta(m_{\phi})}{16\pi^2} \log\(\frac{\mu^2}{m_\phi^2}\) } \nonumber\\
		&=-\frac{ \big( y_{\phi}^{u} \big)_{pt} \big( y_{\phi}^{u} \big)^{*}_{rs} }{m_\phi^2}\Big|_{Q=\mu}+\frac{\beta\Big( \big( y_{\phi}^{u}\big)_{pt}\Big) \big( y_{\phi}^{u} \big)^{*}_{rs} +\big( y_{\phi}^{u} \big)_{pt} \beta\Big( \big(y_{\phi}^{u}\big)_{rs} \Big)^* }{ 32\pi^2\,m_{\phi}^2 }\log(\frac{\mu^2}{m_\phi^2}) \nonumber\\
   &\phantom{=} - \frac{ \big( y_{\phi}^{u} \big)_{pt} \big( y_{\phi}^{u} \big)^{*}_{rs} }{ m_\phi^2} \frac{\beta(m_{\phi})}{16\pi^2 m_{\phi}} \log\(\frac{\mu^2}{m_\phi^2}\),
	\end{align}
where we have kept only the leading-log terms. The first term in the last line is $\frac{c_{qu,prst}^{(8)}}{\Lambda^2}\Big|_{{\rm tree},Q=\mu}$.

Hence, in total we have,
\begin{align}
	\frac{c_{qu,prst}^{(8)}}{\Lambda^2}\Big|_{Q=\mu} =& \frac{c_{qu,prst}^{(8)}}{\Lambda^2}\Big|_{{\rm tree},Q=\mu} + \frac{c_{qu,prst}^{(8)}}{\Lambda^2}\Big|_{{\rm loop},Q=m_{\phi}}\nonumber\\
	&+\frac{\beta\Big(\big(y_{\phi}^{u}\big)_{pt}\Big) \big(y_{\phi}^{u}\big)^{*}_{rs} + \big(y_{\phi}^{u}\big)_{pt} \beta\Big( \big( y_{\phi}^{u} \big)_{rs} \Big)^{*} - 2 \tfrac{\beta(m_{\phi})}{m_\phi} \big( y_{\phi}^{u} \big)_{pt} \big( y_{\phi}^{u} \big)^{*}_{rs}  +\frac{m_\phi^2}{ \Lambda^2}\beta(c_{qu,prst}^{(8)})}{32\pi^2\,m_{\phi}^2}\log\lp \frac{\mu^2}{m_\phi^2}\rp \, . \nonumber
\end{align}
Replacing the $\beta$ functions computed before and reordering, one finds that the logarithmic term vanishes exactly, in perfect agreement with Eq.~\eqref{eq:rge_check_cqu8_g1}. 

Finally, let us remark the key role played by the RGEs of the UV variables, $y_{\phi}^{u}$ and $m_{\phi}$, in both checks at order $g_3^2$ and $g_1^2$. Had we not considered them, there would have been residual logarithmic effects not accounted for by the RGEs, as found in~\cite{Brivio:2021alv}. The same checks can be performed with other WCs and at all orders in the couplings.

\section{Additional details on UV models.}
\label{app:UV_details}
\label{app:uv_model_details}

In this appendix we provide complementary
information on the UV-complete scenarios
studied in this work, in particular concerning
the tree-level matched one-particle models.

For each of the one-particle
UV models considered here
and listed in Table~\ref{tab:UV_particles},
we indicate the relevant couplings
entering their Lagrangian in
Table~\ref{tab:uv_couplings_SMEFiT},
where as elsewhere in the paper
we follow the notation from~\cite{deBlas:2017xtg}.
These are the UV parameters 
which are constrained from 
the data via the matching procedure.
 These couplings have been selected such that the resulting
 Wilson coefficients at the matching scale fulfil the \smefit~flavour assumptions
 in the case of tree-level matching.
 
\begin{table}[t]
\renewcommand{\arraystretch}{1.5}
\begin{center}
\begin{tabular}{c|c||c|c||c|c}
\multicolumn{2}{c||}{\bf Scalars} & \multicolumn{2}{c||}{\bf Fermions} & \multicolumn{2}{c}{\bf Vectors} \\
\midrule
Model & UV couplings & Model & UV couplings & Model & UV couplings \\
\hline
$\mathcal{S}$ & $\kappa_{\mathcal{S}}$ & $N$ & $\left(\lambda_N^e\right)_3$ & $\mathcal{B}$ & $\left(g_B^u\right)_{33}$, $\left(g_B^q\right)_{33}$, $g_B^\varphi$, \\
 $\phi$       & $\lambda_{\phi},\,\, (y_{\phi}^u)_{33}$ & $E$ & $\left(\lambda_E\right)_3$ & & $\left(g_B^e\right)_{11}$, $\left(g_B^e\right)_{22}$, $\left(g_B^e\right)_{33}$,  \\
$\Xi$  & $\kappa_{\Xi}$ & $\Delta_1$ & $\left(\lambda_{\Delta_1}\right)_3$ & & $\left(g_B^\ell\right)_{22}$, $\left(g_B^\ell \right)_{33}$ \\
$\Xi_1$ & $\kappa_{\Xi_1}$ & $\Delta_3$ & $\left(\lambda_{\Delta_3}\right)_3$ & $\mathcal{B}_1$ & $g_{B_1}^\varphi$  \\
$\omega_1$ & $\left(y_{\omega_1}^{qq}\right)_{33}$ & $\Sigma$ & $\left(\lambda_{\Sigma}\right)_3$ & $\mathcal{W}$ &  $\left(g_{\mathcal{W}}^l\right)_{11} = 2\,\left(g_{\mathcal{W}}^l\right)_{22}$, $\left(g_{\mathcal{W}}^l\right)_{33}$ \\
$\omega_4$ & $\left(y_{\omega_4}^{uu}\right)_{33}$ & $\Sigma_1$ & $\left(\lambda_{\Sigma_1}\right)_3$ & & $g_{\mathcal{W}}^\varphi$, $\left(g_{\mathcal{W}}^q\right)_{33}$  \\
$\zeta$ & $\left( y_{\zeta}^{qq} \right)_{33}$ & $U$ & $\left(\lambda_{U}\right)_3$ & $\mathcal{W}_1$ & $g_{\mathcal{W}_1}^\varphi$ \\
$\Omega_1$ & $\left(y_{\Omega_1}^{qq}\right)_{33}$ & $D$ & $\left(\lambda_{D}\right)_3$ & $\mathcal{G}$ &  $\left(g_{\mathcal{G}}^q \right)_{33}$, $\left(g_{\mathcal{G}}^u\right)_{33}$ \\
$\Omega_4$ & $\left(y_{\Omega_4}\right)_{33}$ & $Q_1$ & $\left(\lambda_{\mathcal{Q}_1}^u\right)_3$ &  &  \\
$\Upsilon$ & $\left(y_{\Upsilon}\right)_{33}$ & $Q_7$ & $\left(\lambda_{\mathcal{Q}_7}\right)_3$ &  $\mathcal{H}$ & $\left(g_{\mathcal{H}}\right)_{33}$ \\
$\Phi$ & $\left(y_{\Phi}^{qu}\right)_{33}$ & $T_1$ & $\left(\lambda_{T_1}\right)_3$ &  $\mathcal{Q}_5$ & $\left(g_{\mathcal{Q}_5}^{uq}\right)_{33}$ \\
  &   & $T_2$ & $\left(\lambda_{T_2}\right)_3$ & $\mathcal{Y}_5$ & $\left(g_{\mathcal{Y}_5}\right)_{33}$ \\
\bottomrule
\end{tabular}
\caption{\label{tab:uv_couplings_SMEFiT} For each of the one-particle
UV models considered in this work
and described in Table~\ref{tab:UV_particles},
we list the couplings
entering their Lagrangian,
restricting ourselves to those which are consistent with the {\sc\small SMEFiT} flavour assumption after tree-level matching.
}
\end{center}
\end{table}

It is worth noting that in most models
considered, the interaction with the top quark is singled out and treated differently from the lighter quarks, consistently
with the {\sc\small SMEFiT} assumptions.
This is a well-motivated assumption and is realised in many common UV scenarios, such as in partial compositeness. 
Concerning the leptonic sector, 
the {\sc\small SMEFiT} flavour assumption allows for independent couplings to the different generations. 

This situation differs from the models considered by the {\sc\small FitMaker} collaboration~\cite{Ellis:2020unq}, in which the same heavy particles are used but the couplings are assumed to be flavour-universal.
For the case of heavy spin-one bosons, they assume couplings only to the Higgs boson where appropriate. 
The {\sc\small FitMaker} models $BB_1$ and $Q_{17}$ arise by considering the pairs of heavy particles $\lbrace \mathcal{B}, \mathcal{B}_1 \rbrace$ and $\lbrace Q_1, Q_7 \rbrace$ respectively, with degenerate masses and couplings. 
The models $T$ and $TB$ contain the heavy fermions $U$ and $Q_1$ respectively with specially rescaled couplings~\cite{Dawson:2020oco}.

Table~\ref{tab:Fitmaker_models_flavour} indicates, for the UV models
 considered in the {\sc\small FitMaker}
 study~\cite{Ellis:2020unq} and which are compared
 with our results
 in Fig.~\ref{fig:fitmaker_vs_smefit_all},
  whether they comply 
 with the flavour symmetry assumed in {\sc\small SMEFiT}, and if
 this is not the case what are the differences.
 Here ``more restrictive than {\sc\small SMEFiT}'' means that applying the {\sc\small SMEFiT} flavour symmetry induces more non-vanishing UV couplings than in the {\sc\small FitMaker} case.
For the purposes of the benchmark
comparison of Fig.~\ref{fig:fitmaker_vs_smefit_all}, the effect of 
these additional symmetry-breaking Wilson coefficients has been ignored.

\begin{table}[htbp]
	\centering
 \renewcommand{\arraystretch}{1.5}
	\begin{tabular}{c|c|c}
		$\quad\quad$ Model
  $\quad\quad$& $\quad\quad$ Compliant $\quad\quad$ & Details\tabularnewline
		\midrule
		$\mathcal{S}\, (S)$ & Yes & - \tabularnewline
		\hline
		$\phi$  & No &  $(c_{ledq})_{3333}\neq0$, $(c_{quqd})_{3333}\neq0$, $(c_{lequ})_{3333}\neq0$\tabularnewline
		\hline
		$\Xi$ & Yes & - \tabularnewline
		\hline
		$\Xi_1$ & Yes & - \tabularnewline
		\hline
		$\mathcal{B}\,(B)$ & Yes & More restrictive than {\sc\small SMEFiT}\tabularnewline
		\hline
		$\mathcal{B}_1\,(B_1)$ & Yes & - \tabularnewline
		\hline
		$\mathcal{W}\,(W)$ & Yes & More restrictive than {\sc\small SMEFiT} \tabularnewline
		\hline
		$\mathcal{W}_1\,(W_1)$ & Yes & More restrictive than {\sc\small SMEFiT}\tabularnewline
		\hline
		$N$ & No & $(c_{\varphi l}^{(1),(3)})_{ij} \neq 0$ for $i\neq j$ \tabularnewline
		\hline
		$E$ & No & $(c_{\varphi l}^{(1),(3)})_{ij} \neq 0$ for $i\neq j$ \tabularnewline
		\hline
		$T$ & Yes & - \tabularnewline
		\hline
		$\Delta_1$ & No & $(c_{\varphi e})_{ij}\neq 0$ for $i\neq j$ \tabularnewline
		\hline
		$\Delta_3$ & No & $(c_{\varphi e})_{ij}\neq 0$ for $i\neq j$ \tabularnewline
		\hline
		$\Sigma$ & No & $(c_{\varphi l}^{(1),(3)})_{ij}\neq 0$ for $i\neq j$ \tabularnewline
		\hline
		$\Sigma_1$ & No & $(c_{\varphi l}^{(1),(3)})_{ij}\neq 0$ for $i\neq j$ \tabularnewline
		\hline
		$U$ & No & $(c_{\varphi q}^{(1),(3)})_{ij}\neq 0$ and $(c_{u\varphi})_{ij}\neq 0$ for $i\neq j$ \tabularnewline
		\hline
		$D$ & No & $(c_{\varphi q}^{(1),(3)})_{ij}\neq 0$ for $i\neq j$ \tabularnewline
		\hline
		$Q_5$ & No & $(c_{\varphi d})_{ij}\neq 0$ for $i\neq j$ \tabularnewline
		\hline
		$Q_7$ & No & $(c_{\varphi u,\,u\varphi})_{ij}\neq 0$ for $i\neq j$ \tabularnewline
		\hline
		$T_1$ & No & $(c_{\varphi q}^{(1),(3)})_{ij}\neq 0$ and $(c_{u\varphi})_{ij}\neq 0$ for $i\neq j$ \tabularnewline
		\hline
		$T_2$ & No & $(c_{\varphi q}^{(1),(3)})_{ij}\neq 0$ and $(c_{u\varphi})_{ij}\neq 0$ for $i\neq j$ \tabularnewline
		\hline
        $U\,(T)$ & Yes & - \tabularnewline
        \hline
		$Q_1\,(TB)$ & No & $(c_{\varphi d})_{33}\neq (c_{\varphi d})_{11,22}$ and $(c_{\varphi ud})_{33}\neq 0$ \tabularnewline
        \hline
        $Q_{17}\,(\lbrace Q_1,\,Q_7 \rbrace)$  & Yes & - \tabularnewline
        \hline
        $BB_{1}\,(\lbrace B,\,B_1 \rbrace)$  & Yes & - \tabularnewline
  \bottomrule
	\end{tabular}
	\caption{For the UV models
 considered in the {\sc\small FitMaker}
 study and which are compared
 with the results of our analysis
 in Fig.~\ref{fig:fitmaker_vs_smefit_all},
 we indicate whether they comply 
 with the flavour symmetry assumed in {\sc\small SMEFiT}, and if
 this is not the case what are the differences.
 Here ``more restrictive than {\sc\small SMEFiT}'' means that applying the {\sc\small SMEFiT} flavour symmetry induces more non-vanishing UV couplings.
 We denote UV models
 following the notation of~\cite{deBlas:2017xtg}
 and 
    we indicate the choice in {\sc\small FitMaker} between parenthesis whenever different.
 The only model for which we use
 a different notation from~\cite{deBlas:2017xtg} is the scalar doublet model $\phi$. }
	\label{tab:Fitmaker_models_flavour}
\end{table}


\clearpage

\providecommand{\href}[2]{#2}\begingroup\raggedright\endgroup

\end{document}